\documentstyle[12pt,a4,epsf,rotate]{article}
\newcommand{\rcite}[1]{{\cite{#1}}}
\newcommand{\rref}[1]{{(\ref{#1})}}
\newcommand{\tref}[1]{{\ref{#1}}}
\newcommand{\rlabel}[1]{{\label{#1}}}
\newcommand{\rbibitem}[1]{\bibitem{#1}}
\newcommand{\be}{\begin{equation}}
\newcommand{\ee}{\end{equation}}
\newcommand{\ba}{\begin{eqnarray}}
\newcommand{\ea}{\end{eqnarray}}
\newcommand{\qvev}{\langle\overline{q}q\rangle}
\newcommand{\dis}{\displaystyle}
\newcommand{\tr}{{\rm tr}}
\newcommand{\ovpi}{\overline{\Pi}}
\newcommand{\Pids}{\Pi_{\Delta S =2}}
\begin{document}
\begin{titlepage}
\begin{flushright}
NORDITA 95/10 N,P\\
hep-ph/9502335
\end{flushright}
\vspace{2cm}

\begin{center}
{\huge\bf
Chiral Lagrangians and Nambu-Jona-Lasinio like models}\\[2cm]
Johan Bijnens\\[0.5cm]
NORDITA, Blegdamsvej 17\\
DK-2100 Copenhagen \o, Denmark
\end{center}

\vspace{2cm}
\begin{abstract}
We discuss the low-energy analysis of models involving quarks and
four-fermion couplings. The relation with QCD and with other
models of mesons and meson plus quarks at low energies is discussed.
A short description of how the heat-kernel expansion
can be used to get regularization independent information, is given.

The anomaly within this class of models and a physical
prescription to obtain the correct flavour anomaly while keeping as
much of the VMD aspects as possible is discussed. The major part is the
discussion within this framework of the order $p^4$ action and
of two and some three-point functions to all orders
in momenta and quark masses. Some results
on hadronic matrix elements are given.
\end{abstract}
\end{titlepage}
\tableofcontents
\section{Introduction}
\rlabel{intro}

The problems of dealing with the strong interaction at low and intermediate
energies are well known. At short distance we can use perturbative Quantum
Chromo Dynamics (QCD)\rcite{QCD}
but due to asymptotic freedom this can no longer be
done at low energies. The coupling constant there becomes too large.
A general method, that is, however, extremely manpower and computer intensive,
is using lattice gauge theory methods. An overview of this field can be found
in the recent lectures by Sharpe\rcite{sharpe} or in any of the
proceedings of the annual lattice conferences.

At very low energies we can use the methods of Chiral Perturbation
Theory (CHPT).
A good overview of the present state of the art here can be found in the
DA$\Phi$NE workshop report\rcite{CHPT}.
CHPT is a rigorous consequence of the symmetry pattern
in QCD and its spontaneous breaking. Both perturbative QCD and CHPT are good
theories in the sense that it is in principle possible to go to higher orders
and calculate unambiguously. The size of the higher orders also gives an
estimate of the expected accuracy of the result. A disadvantage of CHPT is
that as soon as we start going beyond lowest order, the number of free
parameters increases very rapidly, thus making calculations beyond the lowest
few orders rather impractical. We would thus like to obtain these free
parameters directly from QCD.

This has so far been rather difficult to do. The reason is that all
available approaches,like lattice QCD, QCD sum rules\rcite{QCDsum}, etc.\ ,
have problems with enforcing the correct chiral behaviour. We would also
like to understand the physics behind the numbers from the lattice
calculations in a more intuitive fashion. Therefore there is a need for
some models that interpolate between QCD and CHPT. We will require that these
models have the correct chiral symmetry behaviour.

It should be kept in mind that these are models and not QCD. The hope is
that these models will catch enough of the essential part of the behaviour
of QCD at low energies that they can be useful.
Two major classes exist, those with
higher resonances than the pseudoscalars included and staying at the
hadronic level, or those with some kind of quarks.  Both of these have their
drawbacks. In the first case there still tends to be a large number of
parameters and in the second case most models do not include
confinement.
Confinement
is treated by explicitly looking at colour singlet
observables only.
The other drawback is inherent in the use of a model. It is not
possible to systematically expand and get closer to the ``true'' answer.

We will look at models including some kind of constituent quarks. The
main
motivation is that the standard constituent quark picture
explains the hadron spectrum rather well. It has problems
when interactions have to be included. It also tends to
break chiral symmetry explicitly. Here we do not attempt to explain
the hadron spectrum but instead focus on the few lowest lying states only.

The class of models we will look at, is those where the fundamental
Lagrangian contains quarks and sometimes also explicitly meson fields.
There exists a whole set of these models of increasing sophistication.
Models that are mainly for study of the spectrum like the bag model are not
included. See \rcite{Donoghue} for a review of various aspects
of this whole area.

The lowest member of the hierarchy are the quark-loop models. Here the basic
premise is that interactions of mesons proceed only via quark loops. The
kinetic term for the mesons is added by hand. As a rule these models have
some problems with chiral symmetry. In particular pointlike couplings
of more than one meson to a quark-antiquark pair have to be added in order
to be consistent. This goes under various names like bare-quark-loop model.
A version that incorporates chiral symmetry correctly and also considers
gluons is known as the Georgi-Manohar model\rcite{GM}. Another variation
is to use the linear sigma model coupled to quarks.

The next level is what I would call improved quark-loop models. Here also
the kinetic terms of the mesons are generated by the quark loops. The
degrees of freedom corresponding to the mesons still have to be added
explicitly by hand. This leads to somewhat counterintuitive results
when calculating loops of mesons\rcite{BR}.
This class started as integrating the nonanomalous
variation of the measure under axial transformations and its most
recent member is known as the QCD effective action model\rcite{ERT}, that
reference also contains a rather exhaustive list of references to earlier work.

The third level differs from the previous in that it starts with
a Lagrangian which is purely fermionic and the hadronic fields are generated
by the model itself. The simplest models here are those that
add four-fermion interaction terms to the kinetic terms for the fermions.
These are usually known as extended Nambu-Jona-Lasinio (ENJL)\rcite{NJL1}
models. They have the
advantage of being very economical in the total number of parameters and of
generating the spontaneous breakdown of chiral symmetry by itself. The
previous class of models has the latter put in by hand.
Most of the remainder will be devoted to this class of models. A review
of the more traditional way of treating this model can be found in
\rcite{NJL2}.

The most ambitious method has been to find a chirally symmetric solution
to the Schwinger Dyson equations. These methods are typically plagued by
instabilities in the solution of the equations. In the end they tend to
be more or less like nonlocal ENJL models. They typically also have a lot of
free parameters. A recent reference is \rcite{Cahill}.

Some common features of all these models are that they contain a type of
constituent quark mass and confinement is introduced by hand. The quarks
are integrated out in favour of an effective action in terms of colourless
fields only. The analysis also assumes keeping only the leading term
in the expansion in the number of colours, $1/N_c$\rcite{revNc}, only. This is
not always explicitly stated but there are very few papers trying to go
beyond the leading term.

I will concentrate on the ENJL models since they are the simplest ones
where the spontaneous symmetry breaking and the mesonic states are
generated dynamically rather than put in by hand. Various arguments
for this model in terms of QCD exist, see \rcite{NJL2,BBR}
and section \tref{QCD}. A physics
argument for the pointlike fermion interaction is that in lattice calculations
the lowest glueball mass tends to be around 1~GeV. So correlations due
to gluons below this scale might be suppressed.

In this review no attempt was made to get a complete reference list. For this
I refer to the more standard review\rcite{NJL2} but let me give a few more
background references.
The original model\rcite{NJL1} was introduced as a simple dynamical
model to understand the pions as Goldstone bosons from the
spontaneously broken chiral symmetry as originally suggested by Nambu.
After the advent of QCD there were various attempts at deriving such a model
from QCD, see e.g. \rcite{Kleinert}. Then the model lay dormant for some
time till it was revived in the early eighties by Volkov,
Ebert and collaborators\rcite{Volkov}. At about the same time a more
theoretical argument for these models was given in \rcite{Wadia}.
A partial list of references where the phenomenological success of this
model was shown is \rcite{Kleinert} to \rcite{3g}.

There has also been some work on the NJL model on the lattice. This was
mainly concerned with the attempt of finding a continuum limit
(cut-off to infinity)\rcite{NJLlattice}.

In the mean time a parallel development took place in the derivation of the
Wess-Zumino-Witten term\rcite{WZW} from quark models\rcite{Manohar}.
This approach was then also used for the non-anomalous part of the effective
action\rcite{50} to \rcite{50c}. This can be found reviewed in \rcite{Ball}.
This was later extended to include gluonic effects\rcite{ERT} and
applied to nonleptonic weak matrix elements \rcite{AP1,raf1}, the higher order
``anomalous'' effective action \rcite{15} and the $\pi^+-\pi^0$ mass
difference\rcite{BR}. The requirement of propagating pseudoscalars that
was found in the last reference provided an extra reason to go to purely
fermionic models including chiral symmetry breaking and possibly gluonic
effects.

In \rcite{BBR} the first step was taken by a low-energy expansion analysis
of the extended Nambu-Jona-lasinio model. This was then extended to
all orders in momenta for two-point functions in the chiral limit
in \rcite{BRZ} and with non-zero current quark masses in \rcite{BP2}.
Some work along similar lines can be found in \rcite{Vogl,shakin} and
\rcite{bernard} but without the emphasis on regularization
independence.
In \rcite{BRZ} also the $\pi^+-\pi^0$ mass difference was analyzed.
The extensions to three point functions can be found in \rcite{BP2}
and the extensions due to the anomaly were discussed in \rcite{BP1}
and used for the $\pi^0\gamma^*\gamma^*$ vertex in \rcite{BP2}.
This work was then extended to the $B_K$ parameter\rcite{BP3,BP4}.
A low-energy analysis of more vector and axial-vector meson processes
was also performed \rcite{Ximo}
and the application to the muon $g-2$ discussed\rcite{eduardo1}.
In addition several talks about this work have been
given \rcite{eduardo2,eduardo3,dafne,brijuni,qcd94}.
It is this series of work that is
reviewed in this Physics Reports.

The report is organized as follows. In section \tref{QCD} we discuss
the Nambu-Jona-Lasinio model, its various extensions and its connection
with QCD. In Sect. \tref{spontaneous} the occurrence of spontaneous
chiral symmetry breaking is discussed. The next section, \tref{hadronic},
 is a short
overview of the low-energy hadronic Lagrangians whose parameters we will try to
understand in the context of the ENJL model.
The relation of ENJL to other models is discussed in the next section.
The regularization method used and the arguments behind
the regularization independence of some of the results are
given in section \tref{regularization}.
Sect. \tref{anomaly}
discusses the implementation of the QCD anomalous Ward identities within this
framework. This is essentially the discussion given in \rcite{BP1}.
Then we reach the main results reviewed here.

The low-energy expansion analysis is in Sect. \tref{p4analysis},
the extension to all orders in momenta and
quark masses in the next section, while some three-point functions
are discussed in Sect. \tref{threep}. Here there is also a more general
discussion of the emergence of vector meson dominance (VMD) and a more general
meson dominance in this class of models.

Then we give a short overview of the results for nonleptonic matrix elements
obtained so far. These calculations are among the most nontrivial uses
of the ENJL model performed so far, Sect. \tref{matrixelem}. In the last
section, we briefly recapitulate the main conclusions.
The appendices contain the derivation of the ward identities at one-loop
to all orders in momenta and masses and the explicit expressions for some
of the one-loop functions needed. For a review of
the heat kernel expansion I refer to \rcite{Ball} and to \rcite{ERT,AP1}
and \rcite{BBR} for the specific notation used in this report.

\section{The Nambu-Jona-Lasinio model and its possible connection
with QCD}
\rlabel{QCD}

In this section the arguments for the ENJL model as a low-energy
approximation to the QCD Lagrangian are discussed. The different ways of
looking at this model are also presented from a QCD viewpoint.

The QCD Lagrangian is given by
\ba
\rlabel{LQCD}
{\cal L}_{QCD}(x) &=& {\cal L}^o_{QCD} + \bar
q\gamma^{\mu}(v_{\mu}+\gamma_5a_{\mu})q - \bar q(s-i\gamma_5p)q,
\nonumber\\
{\cal L}^0_{QCD} &=& - {1\over 4} \sum^{8}_{a=1} G^{(a)}_{\mu\nu}
G^{(a)\mu\nu} + i\bar q \gamma^{\mu}(\partial_{\mu} + i G_{\mu})q\ .
\ea
We restrict ourselves here to low energies so the quarks are the up,
down and strange quarks. $q = (\overline{q}\gamma_0)^{\dag}$ and
$\overline{q}= (\overline{u}\ \overline{d}\ \overline{s})$.
The gluons in \rref{LQCD} are given by
the gluon field matrix in the fundamental
$SU(N_c=3)_{colour}$
representation,
\be
G_{\mu} \equiv g_s \sum^{N_c^2-1}_{a=1} {\lambda^{(a)}\over 2}
G^{(a)}_{\mu} (x)
\ee
with $G^{(a)}_{\mu\nu}$ the gluon field strength tensor
\be
G^{(a)}_{\mu\nu} = \partial_{\mu} G^{(a)}_{\nu} - \partial_{\nu}
G^{(a)}_{\mu} - g_s f_{abc} G^{(b)}_{\mu} G^{(c)}_{\nu},
\ee
and $g_s$ the colour
coupling constant $(\alpha_s = g^2_s/4\pi)$.
The fields $v_\mu$, $a_\mu$, $s$ and $p$ are $3\times 3$ matrices in flavour
space and denote respectively vector, axial-vector, scalar and pseudoscalar
external fields.

The coupling constant in this Lagrangian decreases with increasing
energy scales.
This is known as asymptotic freedom and is the reason why at short-distances
we can use QCD perturbation theory. The other side
of the coin is that at long distances
the coupling constant becomes strong and leads to nonperturbative physics.
This is generally known as infrared slavery and is probably also responsible
for the phenomenon of confinement.
The QCD Lagrangian and a review book can be found in Ref. \rcite{QCD}.

The Lagrangian in \rref{LQCD} has a large classical symmetry.
There is of course the gauge symmetry $SU(N_c)$. In addition the different
quark flavours are conserved leading to a $U(1)_V^3$ symmetry.
The latter is increased to a $U(3)_V= U(1)_V\times SU(3)_V$ symmetry if the
three quark masses become equal. In addition for a zero quark mass
there is an additional $U(1)_A$ for that flavour since for zero
quark mass the Lagrangian does not couple the left and right handed
combinations. For the case of zero quark masses the full classical symmetry of
the lagrangian becomes
\be
\rlabel{QCDsym}
SU(N_c)_{\rm local}\times \left(SU(3)_L\times SU(3)_R
\times U(1)_V\times U(1)_A\right)_{\rm global}\ .
\ee
Not all of these symmetries survive quantization. The $U(1)_A$ is explicitly
broken by quantum effects\footnote{The $SU(3)_L\times SU(3)_R$ is also
broken by the anomaly. These breaking effects are, however, not directly
coupled to the strong interaction so they do not prevent the use of
these symmetries in the same way as happens for the $U(1)_A$.}.
This effect is known as the anomaly.

The introduction of the external fields $v_\mu$ and $a_\mu$ allows for
the global symmetries to be made local. The explicit transformations
of the different fields are:
\ba\rlabel{12}
q_L \rightarrow g_L (x) q_L
\,\,\,&\hbox{and}&\,\,\,
q_R \rightarrow g_R (x) q_R,
\nonumber\\
l_{\mu} \equiv v_{\mu} - a_{\mu} &\rightarrow&
g_L l_{\mu} g_L^{\dagger} + i g_L
\partial_{\mu} g_L^{\dagger},
\nonumber\\
r_{\mu} \equiv v_{\mu} + a_{\mu} &\rightarrow&
g_R r_{\mu} g_R^{\dagger} + i g_R
\partial_{\mu} g_R^{\dagger},
\ea
and
\be\rlabel{15}
s+ip \rightarrow g_R (s+ip) g_L^{\dagger}\ .
\ee
Here $g_l, g_R \in SU(3)_L\times SU(3)_R$. The $U(1)_A$ is not a full
symmetry.
The quark masses are included in the Lagrangian via the scalar external
field $s$
\be
\rlabel{incmass}
s = \left(
\begin{array}{ccc}m_u & & \\ & m_d & \\ & & m_s\end{array}\right)
+ s_{\rm external}\ .
\ee

The number of colours, which is equal to three in the physical world, we
have left free in order to use it as an expansion parameter\rcite{revNc}.
We work in an expansion in inverse powers of the number
of colours, $1/N_c$, where at the same
time the QCD coupling constant is scaled so that $N_c \alpha_S$
remains constant in the large $N_c$ limit. For a review of methods used
to prove things in QCD, see \rcite{revNc}.

At low energies the coupling constant becomes strong and we cannot
simply do a perturbation series in the above Lagrangian.
The objects we will use to obtain physical observables is the
generating functional of Green's functions of the vector, axial-vector,
scalar and pseudoscalar external fields, $v,a,s,p$:
\be
\rlabel{Genfunc}
e^{i\Gamma(v,a,s,p)} = \left<0 \left\vert T \exp\left( i\int d^4 x
{\cal L}_{QCD}(x) \right) \right\vert 0\right>\ .
\ee
The generating functional can be calculated in the path-integral formalism
as:
\ba
\rlabel{QCDpath}
e^{i\Gamma(v,a,s,p)} &=&
{1\over Z} \int {\cal D}G_{\mu} {\cal D}\bar{q} {\cal
D}q \; \exp \left (i \int d^4 x {\cal L}_{\hbox{QCD}}
(q,\bar{q},G;v,a,s,p) \right )
\nonumber\\
&=& {1\over Z}\int {\cal D}G_{\mu} \hbox{exp}\left (-i\int d^4 x {1\over 4}
G^{(a)}_{\mu\nu} G^{(a)\mu\nu}\right ) {\cal D}\bar{q} {\cal
D}q \hbox{exp} \left (i \int d^4 x
\bar{q} i Dq \right ),
\ea
where $D$ denotes the Dirac operator
\be
D = \gamma^{\mu}(\partial_{\mu} + i G_{\mu}) - i\gamma^{\mu}(v_{\mu} +
\gamma_5 a_{\mu}) + i(s-i\gamma_5 p).
\ee

This generating functional is sufficient to be known at zero quark mass,
since once it is known for zero quark mass, the identification
of Eq. \rref{incmass} makes sure it is known for nonzero quark masses.
So if we know it as an expansion in external fields, we also know it
as an expansion in quark masses and external fields. This is the basic
premise underlying the formulation of Chiral Perturbation Theory
of Gasser and Leutwyler\rcite{GL1,GL}.

The main assumption underlying the approach described in this report is
to write the generating functional of Eq. \rref{Genfunc} in a different way.
At very low energies this can be done using Chiral Perturbation Theory:
\be
\rlabel{CHPTpath}
e^{i\Gamma(v,a,s,p)} =
{1\over Z}\int {\cal D}U\exp\left(i\int d^4 x {\cal L}_{\rm CHPT}\right)\ .
\ee
For an explanation of the symbols I refer to Sect. \tref{hadronic}.

This form of the generating functional can be used at low energies at the
price of introducing a relatively large number of free parameters. We would
therefore like to find an alternative way that can also be applied
at low to intermediate energies and has fewer parameters.
At present this involves making more assumptions about the low- and
intermediate energy behaviour of QCD than is inherent in using \rref{CHPTpath}.
One main approach is essentially to rewrite the generating functional in
a functional integral form where the underlying degrees of freedom are still
the quarks. There are various variations on this approach but we
will replace \rref{QCDpath} by:
\be
\rlabel{NJLpath}
e^{i\Gamma(v,a,s,p)} =
{1\over Z} \int ``{\cal D}G_{\mu}'' {\cal D}\bar{q} {\cal
D}q \; \exp \left (i \int d^4 x {\cal L}_{\rm ENJL}
(q,\bar{q},G;v,a,s,p) \right )\ .
\ee
Here the integral over the gluonic degrees of freedom is either absent
or only over low-energy gluons, see below. The Lagrangian in \rref{NJLpath}
is given by
\be\rlabel{LENJL}
{\cal L}_{QCD} \to {\cal L}_{QCD}^{\Lambda_{\chi}} +
{\cal L}^{S,P}_{NJL} +
{\cal L}^{V,A}_{NJL} +O\left({1\over \Lambda_{\chi}^{4}}\right ),
\ee
with
\be\rlabel{LSP}
{\cal L}^{S,P}_{NJL} =
{8\pi^2 G_S(\Lambda_{\chi}) \over N_c \Lambda_{\chi}^2}
\sum_{a,b}(\bar q_R^a q_L^b)(\bar q_L^b q_R^a)
\ee
and
\be\rlabel{LVA}
{\cal L}^{V,A}_{NJL} = -
{8\pi^2 G_V(\Lambda_{\chi}) \over N_c \Lambda_{\chi}^2}
\sum_{a,b}\left[(\bar q_L^a \gamma^{\mu} q_L^b)
(\bar q_L^b \gamma_{\mu} q_L^a)
+ (L \to R) \right].
\ee
The couplings $G_S$ and $G_V$ are dimensionless quantities. For later
convenience we also introduce the abbreviations
\be
\rlabel{gsgv}
g_S = {4\pi^2 G_S(\Lambda_{\chi}) \over N_c \Lambda_{\chi}^2}
\qquad {\rm and}\qquad
g_V =
{8\pi^2 G_V(\Lambda_{\chi}) \over N_c \Lambda_{\chi}^2}\ .
\ee
Notice that in sections \tref{hadronic} and \tref{p4analysis} the symbol
$g_V$ is also used for the vector-two-pseudoscalar coupling.
In principle the extra couplings in \rref{LENJL} should be calculable
in QCD as a function of $\Lambda_\chi$ and the QCD couplings.
In practice this requires knowledge of the nonperturbative domain of QCD
and we will determine all of the new parameters involved empirically.

In the mean time it might be useful to see how this type of interaction
could originate in QCD. This is illustrated in Fig. \tref{figure1}.
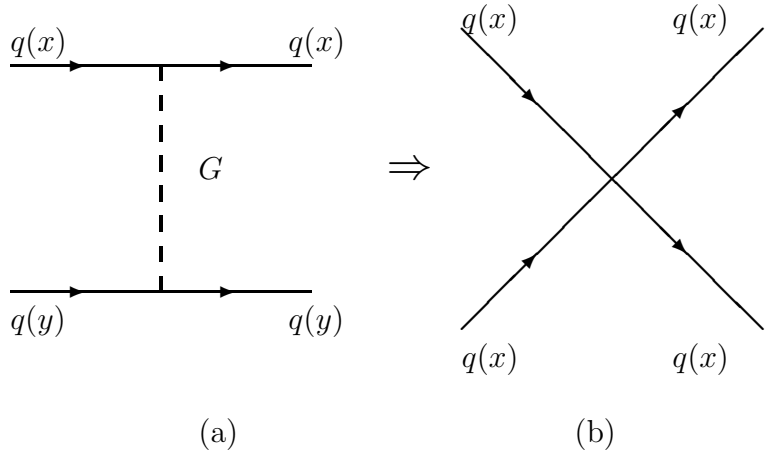
\begin{figure}[htb]
\unitlength 1cm
\begin{picture}(10,7)(-7,-3.5)
\thicklines
\put(-5,1.5){\vector(1,0){1}}
\put(-4,1.5){\vector(1,0){2}}
\put(-2,1.5){\line(1,0){1}}
\put(-5,-1.5){\vector(1,0){1}}
\put(-4,-1.5){\vector(1,0){2}}
\put(-2,-1.5){\line(1,0){1}}
\multiput(-3,1.5)(0,-0.4){8}{\line(0,-1){0.2}}
\put(-2.5,0){$G$}
\put(-5,-2){$q(y)$}
\put(-1.3,-2){$q(y)$}
\put(-5,1.7){$q(x)$}
\put(-1.3,1.7){$q(x)$}
\put(-2.5,-3.5){(a)}

\put(0,0){\Large $\Rightarrow$}

\put(1,-2){\vector(1,1){1}}
\put(2,-1){\vector(1,1){2}}
\put(4,1){\line(1,1){1}}
\put(1,2){\vector(1,-1){1}}
\put(2,1){\vector(1,-1){2}}
\put(4,-1){\line(1,-1){1}}
\put(1,2){$q(x)$}
\put(3.8,2){$q(x)$}
\put(1,-2.5){$q(x)$}
\put(3.8,-2.5){$q(x)$}
\put(2.5,-3.5){(b)}


\end{picture}

\caption{(a) Conventional one-gluon exchange between two quark vertices in
QCD.
(b) Local effective four-quark interaction emerging
from (a) with the replacement in eq.~\protect{\rref{greplace}}.}

\rlabel{figure1}
\end{figure}
In Fig. \tref{figure1}a the one-gluon-exchange interaction between quarks
is shown. If we replace the propagator by it's short-distance part only
we obtain a point like interaction. This is done via the regulator
replacement,
\be
\rlabel{greplace}
\frac{1}{Q^2} \rightarrow
\int_0^{1/\Lambda_\chi^2}d\tau\;e^{-\tau Q^2}\ ,
\ee
in the gluon propagator.
This leads in the leading $1/N_c$ limit to terms of the form \rref{LSP}
and \rref{LVA} with the constraint
\be
\rlabel{GSGV4}
G_S = 4 G_V =\frac{\alpha_S}{\pi}N_c\ .
\ee
This perturbative estimate of the extra couplings is of course only
valid at short distances where perturbative QCD can be applied.
A reliable calculation from QCD would require knowledge of all the higher
orders. In particular, the anomalous dimensions of the two operators
\rref{LSP} and \rref{LVA} are different so QCD can lead to different
predictions for these operators already at the leading order in $1/N_c$.
The constraint $G_S = 4 G_V$ has also other possible origins. In particular
if we want to understand $SU(6)$ in the baryon sector there should be
spin independence of the constituent quark couplings. This leads
precisely to this constraint.
The couplings $G_S$ and $G_V$ are ${\cal O}(1)$ in the large-$N_c$ limit.
This can also be seen in Eq. \rref{GSGV4}.

In general we could think of the Lagrangian of ENJL \rref{LENJL} as being
rooted in QCD by taking \rref{QCDpath} and performing the integral over gluons.
The resulting effective action can then be expanded in terms of local
operators of quark fields. Stopping at dimension 6 and leading order in
the number of colours the Lagrangian is then precisely of the form \rref{LENJL}
but without any gluonic degrees of freedom.
This is the standard picture of the ENJL model. An alternative view
is that we integrate out the short-distance part of gluons and quarks
and again expand the resulting effective action in local operators
leaving only the leading terms in $N_c$ and dimensions.
This again leads to a Lagrangian of the type \rref{LENJL} but this time with
low-energy gluons. Several ways of looking at these gluons are possible but
they are certainly not treatable as perturbative gluons. We will treat them
as a way to describe the gluonic effects on the vacuum, i.e., we only
keep their effects via the vacuum expectation values of gluonic operators.
This is the point of view as taken in Ref. \rcite{ERT}. One of the results
of the work reviewed here is that in the end the effects due to this
gluonic vacuum expectation values are surprisingly small.

Some alternative arguments on the basis of renormalons and QCD sum rules
also exits\rcite{Zakharov}. These arguments lead to the constraint
\be
G_V = 0\;.
\ee

There is fact some work done on extensions of the Nambu-Jona-Lasinio model
including higher order terms. Examples are the nonlocal
NJL-models\rcite{NJLnon}, Schwinger-Dyson type approaches\rcite{Cahill} and
models with some explicit higher order terms\rcite{Andrianov,Petronzio}.
These are terms that are suppressed by higher powers of $1/\Lambda_\chi$.

Here we will keep only the first terms in order to keep the number of
parameters down to a reasonable level.

It should be emphasized that this model does not include confinement. We will
circumvent this problem by only looking at observables that are explicitly
colour singlets. The intermediate lines can in principle go on-shell above
a certain energy. Mostly we will avoid this problem by working in the
domain of Euclidean momenta and then doing an extrapolation to the Minkowski
domain using Chiral Perturbation Theory. The latter method is
especially important in the treatment of nonleptonic decays in
Sect. \tref{matrixelem}.

We work at the leading order in $1/N_c$ throughout. At this order,
as remarked above, the effects of $U(1)_A$ breaking due to the anomaly are
absent. The other effects of the anomaly are still present like
the two-photon decay of the $\pi^0$. The underlying cause of this difference
is that the strong coupling constant $\alpha_S$ also goes to zero in the
large $N_c$ limit while the electromagnetic coupling does not.
One effect of this limit is that nonet symmetry becomes exact, i.e., there
is also a light pseudoscalar in the singlet channel or the $\eta'$ is
also light. Some discussions about effective lagrangians including
the anomalous effect of $U(1)_A$ breaking can be found in\rcite{Divecchia}.
A way of treating it in the context of the ENJL-model has been reviewed
in \rcite{NJL2}.

The presence of the extra pointlike interactions in \rref{LENJL} has in fact
some interesting consequences for the anomalous sector\rcite{BP1}. This is
described in Sect. \tref{anomaly}.

One more remark is needed here. We always implicitly assume that the quarks
in \rref{QCDpath} and \rref{NJLpath} are identical. I.e. there are no
other couplings of the external fields $v_\mu$, $a_\mu$, $s$ and $p$
present. In the nonlocal models the presence of extra terms is already
required by the chiral symmetry. This assumption should also be kept in
mind when judging the results from the ENJL model.

\section{Spontaneous Chiral Symmetry Breaking in the NJL model}
\rlabel{spontaneous}

The original paper of Nambu and Jona-Lasinio\rcite{NJL1} was in fact written to
show the pion as a Goldstone boson and to provide an explicit model of
spontaneous chiral symmetry breaking.
All evidence point towards a spontaneous breaking of the axial symmetry
by quark vacuum expectation values, $\qvev$, in QCD. In the large $N_c$
limit there exists a proof of this by Coleman and Witten\rcite{ColemanWitten}.
Lattice gauge theory also finds agreement with this scenario
\rcite{Latticeqvev} and a recent
reevaluation of $\qvev$ in Finite Energy Sum Rules\rcite{BPR} also gave a
value consistent with the standard scenario.

In the Nambu-Jona-Lasinio model we first have to calculate the fermion
propagator to leading order in $1/N_c$. This can be done via the
Schwinger-Dyson resummation of graphs depicted in Fig. \tref{figure2}.
\begin{figure}[htb]
\unitlength 1cm
\begin{picture}(10,3)(-7,-0.5)
\thicklines
\put(-5,0){\vector(1,0){1}}
\put(-4,0){\line(1,0){1}}
\put(-2.5,-0.1){=}
\put(4,0){\line(1,0){1}}
\put(4,0.75){\circle{1.5}}
\put(4,0){\circle*{0.2}}
\thinlines
\put(3,0){\line(1,0){1}}
\put(2.5,-0.1){+}
\put(0,0){\vector(1,0){1}}
\put(1,0){\line(1,0){1}}
\end{picture}
\caption{Schwinger--Dyson equation for the quark propagator, which leads
to the gap equation in eq. \protect{\rref{gap}}.}
\rlabel{figure2}
\end{figure}
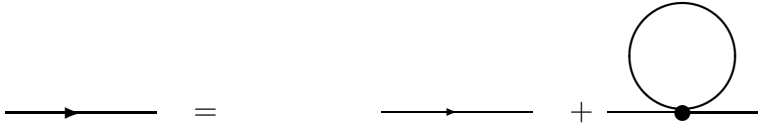
There is no wave function renormalization to this order in $1/N_c$
and the mass can be self-consistently determined from the Schwinger-Dyson
equation.
This leads to the condition
\ba
\rlabel{gap}
M_i &=& m_i - g_S \langle 0| :\overline{q}_i q_i : | 0\rangle \, ,
\\
\rlabel{VEV}
\langle 0 | : \overline{q}_i q_i : | 0 \rangle \equiv
\langle  \overline{q}_i q_i \rangle=\qvev_i &=& - N_c 4 M_i
\int \frac{{\rm d}^4p}{(2\pi)^4}
\frac{i}{p^2-M_i^2} \nonumber\\
&=& - \frac{\dis N_c}{\dis 16 \pi^2}
4 M_i^3  \Gamma\left(-1, \epsilon_i\right) \, .
\ea
Also to this order in $N_c$, the constituent quark mass of flavour $i$,
$M_i$ is independent of the momenta and only a function of $G_S$,
$\Lambda_\chi$ and $m_i$, the current mass of the $i$th flavour quark.
The dependence on $G_S$ is via $g_S$ defined in \rref{gsgv}. It is not
dependent on $G_V$. The $\Gamma$ function in Eq. \rref{VEV} is
a consequence of our regularization scheme (see Sect. \tref{regularization}).

The scalar quark-antiquark one-point function
(quark condensate)  obtains a non-trivial nonzero value. This nonzero value
breaks chiral symmetry spontaneously leading to the occurrence of a nonet
of pseudoscalar Goldstone bosons.

The dependence on the current quark-mass is somewhat obscured
 in eq. \rref{VEV}.
The quantity $\epsilon_i$ appearing in
\rref{VEV} is $M_i^2/\Lambda_\chi^2$. In figure \ref{figure3} we have plotted
the dependence of $M_i$ on $G_S$ for various values of $m_i$
and $\Lambda_\chi = 1.160$ GeV.
\begin{figure}
\epsfxsize=12cm\epsfbox{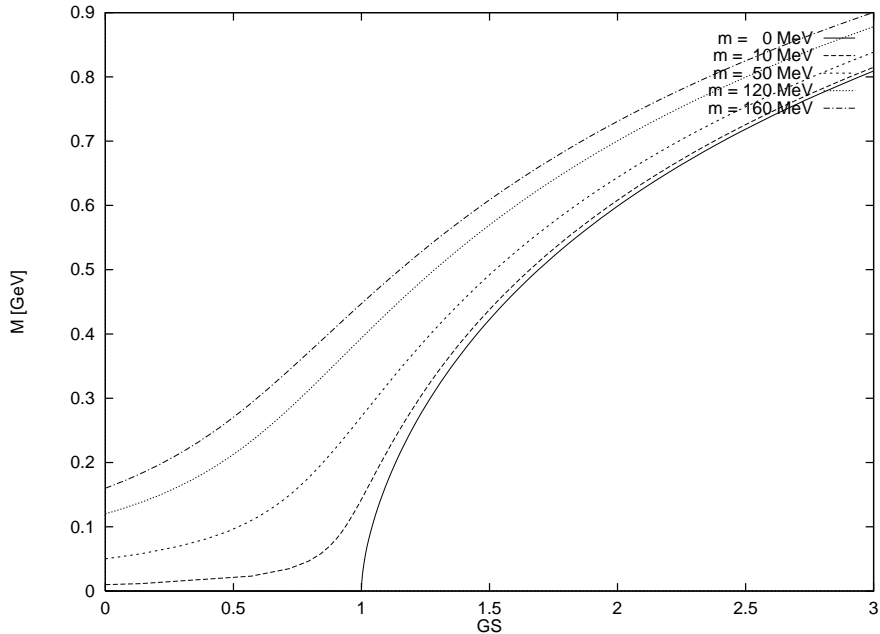}
\caption{Plot of the dependence of the constituent quark mass $M_i$ as a
function of $G_S$ for several values of $m_i$}
\rlabel{figure3}
\end{figure}
It can be seen that the value of $M_i$ for small $m_i$ converges smoothly
towards the value in the chiral limit for the spontaneously broken phase.
This is an indication that an expansion in the quark masses
as Chiral Perturbation Theory assumes for QCD is also valid in this model.
However, it can also be seen that the validity of this expansion
breaks down quickly and for $m_i \simeq 200$ MeV we already
have $2 M_i \simeq \Lambda_\chi$.
We note that the ratio of vacuum expectation values
for light quark flavours increases with increasing
current quark mass
at $p^2=0$ in this model and starts to saturate
for $m_i > 200$ MeV. In standard $\chi$PT this ratio is taken
to be 1 at lowest order
and its  behaviour with the current quark mass
is governed (at ${\cal O} (p^4)$) by the following combination
of coupling constants $2 L_8 + H_2$ \rcite{GL} in the
large $N_c$ limit. Expanding \rref{VEV} in powers of $m_i$ thus gives a
prediction for this combination of parameters, see Sect. \tref{p4analysis}.

In effect, the inclusion of gluonic corrections for this case is known
to order $\langle G^3\rangle$, see Ref. \rcite{BBR}.

\section{Low Energy Hadronic Lagrangians}
\rlabel{hadronic}

As discussed in Sect. \tref{spontaneous}
the $SU(3)_L \times SU(3)_R$ symmetry in flavour space is expected
to be spontaneously broken
down to $SU(3)_V$ in QCD. According to
Goldstone's theorem, there appears then an octet of
massless pseudoscalar
particles $(\pi,K,\eta)$. The fields of these particles
can be conveniently collected
in a $3 \times 3$ unitary matrix $U(\Phi)$ with
det$U =1$. Under local chiral transformations
\be
\rlabel{18}
U(x) \to g_R U(x) g_L^{\dagger}.
\ee
Whenever necessary,
a useful parametrization for $U(\Phi)$, which
we shall adopt, is
\be
\rlabel{19}
U(\Phi) = \hbox{exp} \left( -i \sqrt 2 {\Phi(x) \over f_0} \right)\;,
\ee
where $f_0 \simeq f_{\pi} = 93.2\, MeV$
and ($\stackrel{\rightarrow}
{\lambda}$ are Gell-Mann's
$SU(3)$ matrices with $\tr\lambda_a \lambda_b=2\delta_{ab}$)
\be
\rlabel{20}
\Phi(x) = {\stackrel{\rightarrow}{\lambda} \over \sqrt 2}.
\stackrel{\rightarrow}{\Phi}(x) =
\left( \matrix {{ \pi^0 \over \sqrt 2}+{\eta \over \sqrt 6}&\pi^+&K^+\cr
                         \pi^-&{-\pi^0 \over \sqrt 2}+{\eta \over \sqrt
 6}&K^0\cr
                 K^-&\overline{K}^0&-2{\eta \over \sqrt 6}\cr} \right).
\ee
The $0^-$ octet $\Phi(x)$ is
the ground state of the QCD hadronic spectrum.
There is a mass gap from the ground
state to the first massive multiplets with
$1^-$, $1^+$ and $0^+$ quantum numbers.
The basic idea of the effective chiral
Lagrangian approach is that, in order to describe physics of the strong
interactions at low energies, it may prove
more convenient to replace QCD by
an effective field theory which directly
involves the pseudoscalar $0^-$ octet
fields; and, perhaps, the fields of the first massive multiplets
$1^-$, $1^+$ and $0^+$ as well.
Since we work here in the leading order in $1/N_c$ we have to
add the singlet components as well. In particular we have to add to $\Phi$
\be
\Phi \rightarrow \Phi + \frac{1}{\dis \sqrt{3}}\left(
\begin{array}{ccc}\eta' & & \\ & \eta' & \\ & & \eta'\end{array}\right)\,.
\ee

The chiral symmetry of the underlying QCD
theory implies that $\Gamma(v,a,s,p)$
in eq.~\rref{Genfunc}
admits a low-energy representation
\begin{displaymath}
e^{i \Gamma(v,a,s,p)} = {1\over Z}\int
{\cal D}U {\cal D}S {\cal D}V_{\mu} {\cal D}A_{\mu}
e^{i\int d^4x {\cal L}^R_{eff}(U,S,V_{\mu},A_{\mu};v,a,s,p)}
\end{displaymath}
\be\label{21}
= {1\over Z}\int {\cal D}U e^{i\int d^4x {\cal L}_{eff} (U;v,a,s,p)}\;,
\ee
where the fields $S(x)$, $V_{\mu}(x)$ and $A_{\mu}(x)$ are those
associated with
the lowest massive scalar, vector and axial-vector particle
states of the hadronic spectrum. Both ${\cal L}^R_{eff}$ and
${\cal L}_{eff}$ are local Lagrangians, which contain in principle
an infinite number of terms.
The hope is that, for energies sufficiently small with respect
to the spontaneous
chiral symmetry breaking scale $\Lambda_{\chi}$,
the restriction of ${\cal L}^R_{eff}$ and/or ${\cal L}_{eff}$
to a few terms with the lowest
chiral dimension should provide a sufficiently
accurate description of the low-energy physics.
The success of this approach at the phenomenological level is
by now confirmed by many examples\footnote{For recent reviews see e.g.
Refs.~\rcite{46} to \rcite{eckerrep}.}.
We will later derive the effective Lagrangians
${\cal L}^R_{eff}$ and
${\cal L}_{eff}$ from the Nambu--Jona-Lasinio
cut-off version of QCD.

Let us now briefly summarize
what is known at present   about
the low-energy mesonic Lagrangians ${\cal L}^R_{eff}$ and
${\cal L}_{eff}$ from the chiral invariance properties of
${\cal L}_{QCD}$ alone.

The terms in ${\cal L}_{eff}$ with the lowest chiral dimension, i.e.
$O(p^2)$, are
\be
\rlabel{22}
{\cal L}_{eff}^{(2)} = {1\over 4} f^2_0 \Big \{
\tr D_{\mu} U D^{\mu} U^{\dagger} +
\tr(\chi U^{\dagger} + U^{\dagger} \chi)\Big
 \}
\ee
where $D_{\mu}$ denotes the covariant derivative
\be
\rlabel{23}
D_{\mu} U =\partial_{\mu} U -i(v_{\mu} + a_{\mu})U
+iU(v_{\mu} - a_{\mu})
\ee
\be
\rlabel{24}
\chi = 2B_0(s(x)+ip(x)).
\ee
The constants $f_0$ and $B_0$ are not fixed by chiral symmetry
requirements. The constant $f_0$
can be obtained from $\pi \to \mu \nu$ decay,
 and
it is the same which appears in the
normalization of the pseudoscalar field
matrix $U(\Phi)$ in \rref{20}, i.e.
\be
\rlabel{25}
f_0 \simeq f_{\pi} = 93.2\, MeV.
\ee
The constant $B_0$ is related to the vacuum expectation value

\be\label{26}
\langle 0\vert \bar {q} q
\vert 0 \rangle_{\vert q=u,d,s} = -f_0^2 B_0(1+O({\cal M})).
\ee

The terms in ${\cal L}_{eff}$ of $O(p^4)$ are also known. They have
been classified by Gasser and Leutwyler \rcite{GL}:\footnote{
There are more terms in principle because of the presence of the singlet
component as well. These have all zero coefficients at the leading order
in $1/N_c$.}
\be
\rlabel{27}
\matrix{
{\cal L}_{eff}^{(4)}(x) =&L_1(\tr D_{\mu} U^{\dagger} D^{\mu} U)^2 +
L_2 \tr(
D_{\mu} U^{\dagger} D_{\nu} U)\tr(D^{\mu} U^{\dagger} D^{\nu} U)\hfill\cr
\cr
\hphantom{{\cal L}^{(4)}(x)}&+ L_3 \tr(D_{\mu} U^{\dagger}
D^{\mu} U D_{\nu} U^{\dagger}
D^{\nu} U)+L_4 \tr( D_{\mu} U^{\dagger} D^{\mu} U)
\tr (\chi^{\dagger} U  +
U^{\dagger}\chi)\hfill\cr \cr
 \hphantom{{\cal L}^{(4)}(x)}&+ L_5 \tr[D_{\mu}
U^{\dagger} D^{\mu} U (\chi^{\dagger} U + U^{\dagger} \chi)]+ L_6
 [\tr(\chi^{\dagger} U + U^{\dagger} \chi)]^2  \hfill\cr
\cr
\hphantom{{\cal L}^{(4)}(x)}&+ L_7 [\tr(\chi U
- U^{\dagger} \chi)]^2 + L_8
\tr(\chi^{\dagger} U \chi^{\dagger} U + \chi U^{\dagger} \chi
U^{\dagger})\hfill\cr\cr
\hphantom{{\cal L}^{(4)}(x)}
&-iL_9 \tr(F^{\mu\nu}_R D_{\mu} U D_{\nu} U^{\dagger}
 +
F^{\mu\nu}_L D_{\mu} U^{\dagger} D_{\nu} U) + L_{10} \tr(U^{\dagger}
 F^{\mu\nu}_R U
F_{L\mu\nu})\hfill\cr \cr
\hphantom{{\cal L}^{(4)}(x)}&+ H_1 \tr(F^{\mu\nu}_R F_{R\mu\nu} +
F^{\mu\nu}_L F_{L\mu\nu}) + H_2 \tr(\chi^{\dagger}\chi),\hfill
\cr}
\ee
where $F_{L \mu \nu}$ and $F_{R \mu \nu}$ are the external
field-strength tensors
\ba
\rlabel{28}
F_{L \mu \nu} &=& \partial_{\mu} l_{\nu} - \partial_{\nu} l_{\mu}
- i [l_{\mu},l_{\nu}]
\nonumber\\
F_{R \mu \nu} &=& \partial_{\mu} r_{\nu} - \partial_{\nu} r_{\mu}
- i [r_{\mu},r_{\nu}]
\ea
associated with
the external left ($l_{\mu}$) and right ($r_{\mu}$)
field sources
\be
\rlabel{30}
l_{\mu} = v_{\mu} - a_{\mu}{\,\,\,},\;\;\;r_{\mu} = v_{\mu} + a_{\mu}.
\ee
The constants
$L_i$ and $H_i$ are again not fixed by chiral symmetry
requirements.
The $L_i$'s were phenomenologically determined in Ref.~\rcite{GL}.
Since then, $L_{1,2,3}$
have been fixed more accurately using data from $K_{l4}$
\rcite{kl4}.
The phenomenological values of the $L_i$'s that will be relevant for a
comparison  with our calculations, at a renormalization scale
$\mu = M_{\rho} =770\, MeV$, are collected in the first column of
Table \tref{table2}.

By contrast with ${\cal L}_{eff}$, which only has pseudoscalar fields as
physical degrees of freedom, the Lagrangian ${\cal L}^R_{eff}$ involves
chiral couplings of fields of massive $1^-$, $1^+$ and $0^+$ states to
the Goldstone fields. The general method to construct these couplings was
described a long time ago in Ref.~\rcite{CCWZ}.
An explicit construction of the
couplings for $1^-$, $1^+$ and $0^+$
fields can be found in Ref.~\rcite{36}. As
discussed in Ref.~\rcite{37},
the choice of fields to describe chiral invariant
couplings involving spin-1 particles is
not unique and, when the vector modes
are integrated out, leads to ambiguities
in the context of chiral perturbation
theory to $O(p^4)$ and higher. As shown in \rcite{37},
these ambiguities are, however,
removed when consistency with the short-distance behaviour of QCD is
incorporated. The effective Lagrangian which
we shall choose here to describe
vector couplings corresponds to the so-called model II in Ref.~\rcite{37}.

In the NJL model it is of course obvious that the different representations for
the meson fields should be identical since the original model is formulated
in terms of fermions only. The choice of fields for the mesons is purely
a matter of choice during the calculation.

The wanted ingredient for a non-linear representation of the chiral
$SU(3)_L \times SU(3)_R \equiv G$ group when
dealing with matter fields is the
compensating $SU(3)_V$ transformation $h(\Phi,g_{L,R})$
which appears under
the action of the chiral group $G$ on the coset
representative $\xi(\Phi)$ of
the $G/SU(3)_V$ manifold, i.e.
\be
\rlabel{31}
\xi(\Phi) \to g_R \xi(\Phi) h^{\dagger}(\Phi,g_{L,R})
= h(\Phi,g_{L,R}) \xi(\Phi) g_L^{\dagger}\;,
\ee
where $\xi(\Phi) \xi(\Phi) = U$ in the
chosen gauge. This defines
the $3 \times 3$ matrix representation of the
induced $SU(3)_V$ transformation.
Denoting the various matter $SU(3)_V$ multiplets by $R$ (octet) and $R_1$
(singlets), the non-linear realization of $G$ is given by
\ba
\rlabel{32}
R \to h(\Phi,g_{L,R}) R h^{\dagger}(\Phi,g_{L,R})
\nonumber\\
\rlabel{33}
R_1 \to R_1,
\ea
with the usual matrix notation for the octet
\be\rlabel{34}
R = {1\over \sqrt 2} \sum^8_{i=1} \lambda^{(i)} R^{(i)}.
\ee
The vector field matrix $V^{\mu}(x)$ representing the $SU(3)_V$
octet of $1^-$ particles;
the axial-vector field
matrix $A^{\mu}(x)$ representing $SU(3)_V$ octet of $1^+$
 particles;
and the scalar field matrix $S(x)$  representing $SU(3)_V$ octet of $0^+$
 particles
are chosen to transform like $R$ in eq.~(\tref{32}),
i.e. ($h \equiv h(\Phi,g_{L,R})$):
\be\rlabel{35}
V_{\mu} \to h V_{\mu} h^{\dagger}\,\,;\qquad
A_{\mu} \to h A_{\mu} h^{\dagger}\,\,;\qquad
S \to h S h^{\dagger}.
\ee

The procedure to construct now the lowest-order chiral Lagrangian
${\cal L}^R_{eff}$ is to write down all possible
invariant couplings to first
non-trivial order in the chiral expansion,
which are linear in the $R$ fields,
and to add of course the corresponding invariant kinetic couplings. It is
convenient for this purpose
to first set the list of possible tensor structures
involving the $R$ fields, which transform like $R$ in
eq.~\rref{32} under the action
of the chiral group $G$.
Since the non-linear realization of $G$ on the octet
field $R$ is local, one is led to define a covariant derivative
\be\rlabel{36}
d_{\mu}R = \partial_{\mu}R + [\Gamma_{\mu},R]   \;,
\ee
with a connection
\be\rlabel{37}
\Gamma_{\mu} =
{1 \over 2}\{\xi^{\dagger}[\partial_{\mu}-i(v_{\mu}+a_{\mu})]\xi
+ \xi[\partial_{\mu}-i(v_{\mu}-a_{\mu})]\xi^{\dagger}\}
\ee
ensuring the transformation property
\be
\rlabel{38}
d_{\mu}R \to h d_{\mu}R h^{\dagger}.
\ee
We can then define vector
and axial-vector field strength tensors
\be\rlabel{39}
V_{\mu\nu}=d_{\mu}V_{\nu}-d_{\nu}V_{\mu}\,\,\,{\hbox{and}}\,\,\,
A_{\mu\nu}=d_{\mu}A_{\nu}-d_{\nu}A_{\mu}\;,
\ee
which also transform like $R$, i.e.
\be\rlabel{40}
V_{\mu\nu} \to h V_{\mu\nu} h^{\dagger}
\,\,\,{\hbox{and}}\,\,\,
A_{\mu\nu} \to h A_{\mu\nu} h^{\dagger}.
\ee

There is a complementary list of terms that can be constructed
with the coset  representative
$\xi(\Phi)$ and which transform homogeneously,
i.e. like $R$ in \rref{32}. If
we restrict ourselves
to terms of $O(p^2)$ at most, here is the list:
\be
\rlabel{41}
\xi_{\mu} = i\{\xi^{\dagger}[\partial_{\mu}-i(v_{\mu}+a_{\mu})]\xi
- \xi[\partial_{\mu}-i(v_{\mu}-a_{\mu})]\xi^{\dagger}\}
= i\xi^{\dagger} D_{\mu} U \xi^{\dagger}=\xi_{\mu}^{\dagger},
\ee
\be
\rlabel{42}
\xi_{\mu}\xi_{\nu} \,\,\,{\hbox{and}}\,\,\, d_{\mu}\xi_{\nu},
\ee
\be
\rlabel{43}
\chi_{\pm} = \xi^{\dagger}\chi\xi^{\dagger} \pm \xi\chi^{\dagger}\xi,
\ee
\be
\rlabel{44}
f_{\mu\nu}^{\pm} = \xi F_{L \mu \nu}\xi^{\dagger} \pm
 \xi^{\dagger}F_{R \mu
 \nu}\xi.
\ee
Notice that $\Gamma_{\mu}$ in \rref{37}
does not transform homogeneously, but
rather like an $SU(3)_V$ Yang--Mills field, i.e.
\be\rlabel{45}
\Gamma_{\mu} \to h \Gamma_{\mu} h^{\dagger} + h\partial_{\mu} h^{\dagger} .
\ee

The most general Lagrangian ${\cal L}^R_{eff}$ to lowest non-trivial order in
the chiral expansion is then obtained by adding to ${\cal L}^{(2)}_{eff}$ in
 eq.~\rref{22} the
scalar Lagrangian
\be\rlabel{46}
{\cal L}^S = {1 \over 2} \tr\left(d_{\mu}S d^{\mu}S - M_S^2 S^2\right)
+ c_m \tr\left(S \chi^+\right) +
c_d \tr\left(S \xi_{\mu} \xi^{\mu}\right);
\ee
the vector Lagrangian
\be\rlabel{47}
{\cal L}^V = -{1 \over 4} \tr\left(V_{\mu \nu} V^{\mu \nu} - 2M_V^2
V_{\mu}V^{\mu}\right) -{1 \over 2 \sqrt 2} \left[f_V\tr\left(
V_{\mu \nu} {f^{(+)}}^{\mu \nu}\right)
+  i g_V\tr\left(V_{\mu \nu}[\xi^{\mu},\xi^{\nu}]\right)\right]+
\cdots~,
\ee
and the axial-vector Lagrangian
\be\rlabel{48}
{\cal L}^A = -{1 \over 4} \tr\left(A_{\mu \nu} A^{\mu \nu} -
2M_A^2 A_{\mu}A^{\mu}\right)
-{1 \over 2 \sqrt 2} f_A \tr\left(A_{\mu \nu} {f^{(-)}}^{\mu \nu}\right)+
\cdots ~.
\ee
The dots in ${\cal L}^V$ and ${\cal L}^A$ stand for other $O(p^3)$
 couplings which
involve the vector field $V^{\mu}$ and
axial-vector field $A^{\mu}$ instead of  the field-strength
tensors $V_{\mu\nu}$ and $A_{\mu\nu}$. They have been classified in Ref.
 \rcite{37}. As discussed
there, they play no
role in the determination of the $O(p^4)$ $L_i$ couplings
 when
the vector and axial-vector fields are integrated out.

The masses $M_V$, $M_S$ and $M_A$
and the coupling constants $c_m$, $c_d$,
 $f_V$, $g_V$ and $f_A$
are not fixed by chiral symmetry requirements. They can be determined
 phenomenologically, as
was done in Ref.~\rcite{36}.
Since later on we shall calculate masses and
 couplings only in the chiral limit, we
identify $M_V$, $M_S$ and $M_A$ to those of non-strange particles of the
corresponding multiplets, i.e.

\be\rlabel{49}
M_V=M_{\rho}=770\,MeV\,\,\, ; \,\,\, M_S=M_{a_0}=983\,MeV
\ee
and
\be\rlabel{50}
M_A=M_{a_1}=1260 \pm 30\,MeV.
\ee
The couplings $f_V$ and $g_V$ can then be
determined from the decays
 $\rho^0 \to e^+ e^-$
and $\rho \to \pi \pi$ respectively, with the result
\be\rlabel{51}
\vert f_V \vert = 0.20\qquad{\hbox{and}}\qquad\vert g_V \vert = 0.090.
\ee
The decay $a_1 \to \pi \gamma$ fixes the coupling $f_A$ to
\be\rlabel{52}
\vert f_A \vert = 0.097 \pm 0.022,
\ee
where the error is due to the experimental error in the determination
of the partial width, $\Gamma(a_1 \to \pi \gamma) = (640 \pm 246)
\,keV$. For the scalar couplings $c_m$ and $c_d$, the decay rate $a_0 \to
\eta \pi$ only fixes the linear combination \rcite{36}
\be\rlabel{53}
\left\vert c_d +
{2m_{\pi}^2 \over M_{a_0}^2 - m_{\eta}^2 - m_{\pi}^2} c_m \right\vert =
 (34.3 \pm 3.3)\,MeV.
\ee
In confronting these results with
theoretical predictions, one should
keep in mind that they have not been corrected for the effects of
chiral loop contributions.

In addition in order to describe vector interactions beyond those that can
be described by the above terms there are more terms possible. These do
however not contribute to CHPT coefficients of order $p^4$ when integrated
out. For a list of these terms see Ref. \rcite{Ximo}.

I will now shortly review the different ways vector mesons tend to be
implemented. A review can be found in \rcite{Meissner}.

There is the way of gauging the $U(3)_L\times U(3)_R$ symmetry by a set
of vector meson fields, $L_\mu$ and $R_\mu$. These can be given a mass term
without breaking the local symmetry by introducing the external fields
$l_\mu$ and $r_\mu$ defined above. To the Yang-Mills Lagrangian and the
lowest order Lagrangian for the pseudoscalar mesons, with $L_\mu$ and $R_\mu$
in the covariant derivative now, we add a term of the form
\be
\rlabel{massYM}
-\frac{1}{2}m_0^2\; \tr \left[ \left(L_\mu - \frac{1}{g}l_\mu\right)^2
+\left(R_\mu-\frac{1}{g}r_\mu\right)^2 \right]\,.
\ee
The mass $m_0$ corresponds to the vector meson mass in the chiral limit
and the axial-vector mass becomes different due to a partial Higgs mechanism,
the field $L_\mu-R_\mu$ mixes with the pseudoscalars. Including vector
mesons only in this formalism requires sending the ``bare'' pion decay constant
to infinity. This is often referred to as the gauged Yang-Mills formulation.

A variation on the Yang-Mills principle is the hidden gauge formalism
\rcite{Bando}. This formalism also allows for only the vector mesons to
be included. There are more free parameters here than in the previous
formalism at first sight but if one allows for higher order terms in
both formalisms they are fully identical. This was proven in \rcite{Bando}.
Removing the axial vector mesons from the simplest gauged Yang-Mills version
leads to the hidden gauge version (vectors only) with the extra constant
$a=1$. The usual VMD requirement has $a=2$. This version, $a=1$,
also corresponds
to Weinbergs original formulation of an effective Lagrangian for vectors and
pions\rcite{Weinbergvector}.

One can also include the vector mesons in the general form as described by
Callan et al.\rcite{CCWZ}. This is the formulation described earlier in
this section. There is also a version possible where $L_\mu$ and $R_\mu$
transform linearly under the chiral symmetry. This is the version
that the ENJL model ends up with most simply.

The last version is to use antisymmetric tensor fields to describe the
(axial-)vector mesons. This was the formulation chosen in\rcite{36,GL1}.
This can be related to the other approaches by choosing the field strength
rather than the bare field as the interpolating fields for the vectors.

All of these formalisms can lead to identical physics by introducing extra
pointlike pion couplings and higher order couplings as well. As such it is a
matter of taste which version one chooses. Some of them tends to require
fewer additional pointlike pion couplings. This tends to be true mostly
for the Yang-Mills like versions. See \rcite{37} for the analysis to
order $p^4$.
In the sector involving $\varepsilon_{\mu\nu\alpha\beta}$ this tends
not to be so simple\rcite{Petronzio2}.

\section{Relation to other models}
\rlabel{othermodels}

As discussed in the introduction there are several variations
on the theme of effective Lagrangians with quarks and mesons.
In this section we describe how the ENJL model is related to the
other approaches.
This is an extended version of the discussion in \rcite{BBR}. The relation
with the Georgi-Manohar model and in particular the discussion about the
pion-quark coupling can be found in \rcite{Perisraf}.

For this comparison we first introduce a version that includes both
bosonic and fermionic fields in the Lagrangian.
Following the standard procedure
of introducing auxiliary fields, we
rearrange the Nambu--Jona-Lasinio
cut-off version of the QCD Lagrangian in
an equivalent Lagrangian which is only quadratic in the quark fields. For this
purpose, we introduce three complex $3\times 3$
auxiliary field matrices
$M(x)$, $L_{\mu}(x)$ and $R_{\mu}(x)$; the so-called collective field
variables, which under the chiral group $G$ transform as
\be\rlabel{54}
M \to g_R M g_L^{\dagger}
\ee
\be\rlabel{55}
L_{\mu}\to g_L L_{\mu} g_L^{\dagger}\,\,\,{\hbox{and}}\,\,\,R_{\mu}\to g_R
 R_{\mu} g_R^{\dagger}.
\ee
We can then write the following identities:
\begin{displaymath}
 \exp \,i\int d^4x {\cal L}_{NJL}^{S,P}(x) =\hbox{\hskip5cm}
\end{displaymath}
\be\rlabel{56}
\int {\cal D} M  \,  \exp \,i\int d^4x  \left\{ -\left(\bar q_L
 M^{\dagger} q_R
+ \, h.c. \right) - {N_c \Lambda_{\chi}^2 \over 8\pi^2
 G_S(\Lambda_{\chi})}\tr(M^{\dagger} M)\right\};
\ee
and
\begin{displaymath}
 {\exp} \,i\int d^4x {\cal L}_{NJL}^{V,A}(x) = \hbox{\hskip5cm}
\end{displaymath}
\be\rlabel{57}
\int {\cal D}L_{\mu}{\cal D}R_{\mu}  \exp  \,i\int d^4x
\left\{
\left[ \bar q_L \gamma^{\mu} L_{\mu} q_L + {N_c \Lambda_{\chi}^2 \over 8\pi^2
 G_V(\Lambda_{\chi})}
{1 \over4}\tr L^{\mu}L_{\mu}\right] + (L \to R)\right\},
\ee
where ${\cal L}_{NJL}^{S,P}(x)$ and ${\cal L}_{NJL}^{V,A}(x)$ are
the four-fermion  Lagrangians in (\ref{LSP}) and (\ref{LVA}).

By polar decomposition
\be\rlabel{58}
M = U \tilde H = \xi H \xi,
\ee
with $U$ unitary and $\tilde H$ (and $H$) Hermitian. From the
transformation laws of $M$ and
$\xi$ in eqs.~(\ref{54}) and (\ref{31}),
it follows that $H$ transforms homogeneously, i.e.
\be\rlabel{59}
H \to h(\Phi,g_{L,R}) H h^{\dagger}(\Phi,g_{L,R}).
\ee
The path integral measure in eq.~(\ref{56})
can then also be written as
\begin{displaymath}
\exp \,\,i\int d^4x {\cal L}_{NJL}^{S,P}(x) = \hbox{\hskip5cm}
\end{displaymath}
\be\rlabel{60}
\int {\cal D} \xi {\cal D} H \, \exp\,\, i\int d^4x  \left\{
-\left(\bar q_L \xi^{\dagger} H \xi^{\dagger} q_R + \bar q_R \xi H \xi
 q_L\right)
- {N_c \Lambda_{\chi}^2 \over 8\pi^2 G_S(\Lambda_{\chi})}\tr H^2\right\}.
\ee
We are interested in the effective action
 $\Gamma_{eff}(H,\xi,L_{\mu},R_{\mu};v,a,s,p)$
defined in terms of the new auxiliary fields $H$,$\xi$,$L_{\mu}$, $R_{\mu}$;
and in the presence of the external field sources $v_{\mu}$, $a_{\mu}$,$s$ and
$p$, i.e.
\begin{displaymath}
e^{i \Gamma_{eff}(H,\xi,L_{\mu},R_{\mu};v,a,s,p)} =
{1 \over Z}
\int {\cal D}G_{\mu} \exp \left (-i \int d^4 x{1\over 4}G^{(a)}_{\mu\nu}
G^{(a)\mu\nu}\right)
\hbox{\hskip3cm}
\end{displaymath}
\begin{displaymath}
\times
\exp \,i \int d^4 x \left\{
{N_c\Lambda_{\chi}^2 \over 8\pi^2 G_V(\Lambda_{\chi})}{1
 \over4}[\tr(L^{\mu}L_{\mu})+
\tr(R^{\mu}R_{\mu})]
- {N_c \Lambda_{\chi}^2 \over 8\pi^2 G_S(\Lambda_{\chi})}
\tr H^2\right\}
\end{displaymath}
\be
\rlabel{61}
\times
\int {\cal D}\bar{q}i {\cal D}q \;
\exp\,\, i \int d^4 x \left\{
\bar{q}{\cal D}_{QCD} q +
\bar q_L \gamma^{\mu}L_{\mu} q_L +
\bar q_R \gamma^{\mu}R_{\mu} q_R
-\left(\bar q_L \xi^{\dagger} H \xi^{\dagger} q_R + \bar q_R \xi H \xi
 q_L\right) \right\},
\ee
with ${\cal D}_{QCD}$ the QCD Dirac operator:
\be\rlabel{62}
{\cal D}_{QCD} = \gamma^{\mu} (\partial_{\mu} + i G_{\mu})
- i \gamma^{\mu}(v_{\mu}+\gamma_5a_{\mu}) + i(s-i\gamma_5p).
\ee
The integrand is now quadratic in the fermion fields.

Here we can easily see how when we integrate out the quarks we will end
up with different implementations of the vector fields.
The fields $L_\mu$ and $R_\mu$ correspond to the linear version discussed
in the previous section.
We can decouple the external fields $l_\mu$ and $r_\mu$ by doing a shift
of the auxiliary vector fields
\be
L(R)_\mu \rightarrow L'(R')_\mu = L(R)_\mu  + l(r)_\mu\,.
\ee
Notice that this leads to precisely the type of mass term added in the
gauged Yang-Mills vector description and the $L'_\mu$ and $R'_\mu$
transform nonlinearly as gauge bosons under the chiral group.
The relation with the CCWZ version will be given in section \tref{p4analysis}.

In principle we could also choose various versions for the scalars
and pseudoscalars by the various choices possible for the matrix
$M$. Two possibilities are shown in \rref{58}. $H$ transforms in the CCWZ
fashion\rcite{CCWZ} while $\tilde{H}$ transforms as a purely lefthanded
scalar.

Most of the other quark-meson models described in the introduction
are models containing quarks and pseudoscalars only.
The QCD effective action model\rcite{ERT} follows simply by setting
\be
\rlabel{meanfield}
L_\mu = R_\mu = 0\qquad {\rm and}\qquad
H = \langle H\rangle= M_Q = -g_S\qvev\,,
\ee
where $\qvev$ is the quark vacuum expectation value derived in section
\tref{spontaneous}. The advantage of the present approach is that the
spontaneous symmetry breaking that was added by hand in that model is now
generated spontaneously. The approximations \rref{meanfield}
will be referred to later as the
mean field approximations.

The Georgi-Manohar model\rcite{GM} requires a little more work to obtain.
Here there is an additional free parameter, $g_A$, the axial-coupling
of the pseudoscalars to the constituent quarks. There have been some recent
arguments about the order in $N_c$ this parameter is, see \rcite{Perisraf}
and references therein. In the ENJL model it is obvious that this parameter
is of leading order in $1/N_c$. In the purely
fermionic picture it is obtained from the graphs shown in Fig.
\tref{figure4}.
\begin{figure}[htb]
\setlength{\unitlength}{1mm}
\begin{picture}(100.00,50.00)(0.,20.)
\put(40.00,20.25){\vector(1,0){19.25}}
\put(58.50,20.25){\vector(1,0){20.00}}
\put(78.75,20.25){\line(1,0){14.50}}
\put(66.50,55.50){\circle{14.00}}
\put(66.50,27.50){\circle{14.00}}
\put(66.50,41.50){\circle{14.00}}
\put(66.50,20.00){\circle*{2.50}}
\put(65.50,48.25){\circle*{2.50}}
\put(66.50,34.50){\circle*{2.50}}
\put(65.00,61.50){X}
\end{picture}
\caption{The set of diagrams summed to obtain $g_A(Q^2)$.
X is the insertion of the pion field and the other lines are fermions.}
\rlabel{figure4}
\end{figure}
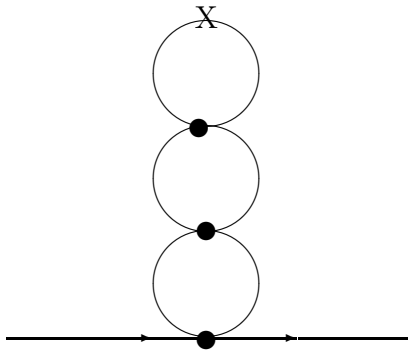
In general this parameter depends on the off-shellness of the pion but
in the low-energy approximation it becomes a constant.
In the language described above the parameter $g_A$ appears due
to the mixing of the pseudo-scalar fields and $L_\mu-R_\mu$.

In general the quark-meson models include kinetic terms for the mesons as
well. These are in the present approach of course assumed to be produced
from the integration over the quarks.

\section{Regularization independence}
\rlabel{regularization}

The general method we will use to argue independence of the regularization
procedure is the heat-kernel method. A review of this method can be
found in \rcite{Ball}. There exists various versions of the heat kernel
method. The version we use here is the most naive one. More careful
definitions also exist, see \rcite{Ball} and references therein.

The underlying problem is that, as can be seen in Sect. \tref{spontaneous},
the chiral symmetry is spontaneously broken by the quadratic divergence.
In a regulator that does not have the quadratic divergence, like
dimensional regularization, one always works in the phase where
chiral symmetry is explicitly realized in the spectrum. In the ENJL model
this means that we treat it as being in a phase with weakly interacting
massive quarks. The reason is that the logarithmic divergence in \rref{gap}
has a negative sign so the vacuum energy from the logarithmic term is positive.
To avoid this we have chosen a variation on the proper time
regularization.
Most regulators that preserve the presence of quadratic divergences do break
the underlying chiral symmetry explicitly. The Ward identities have then to
be used to determine the coefficients of the symmetry-breaking counterterms
that have to be added to obtain chirally symmetric results. In general this is
a very cumbersome method and we will use some simplified versions of it.

In general we will consider several options.
We can treat the heat kernel
regularized by a specific regularization scheme. The one used here is
the proper time heat kernel expansion. This is the scheme used to obtain
the low-energy expansion of Sect. \tref{p4analysis}.
We can then be more general in the heat-kernel expansion and leave the
coefficients of the terms in the heat-kernel expansion completely free.
This way we test a combination of the symmetry structure and the general
couplings of the mesonic fields to the quarks only. It is rather
surprising that in this case there are still several nontrivial results
left. These type of results are in fact the major improvement of the
methods used here as compared to the more traditional ones\rcite{NJL2}.

Since we would also like to go beyond the few first terms in the low-energy
expansion it is necessary to either go to very high orders in the explicit heat
kernel expansion or go to an alternative method where we directly regulate
the Feynman diagrams. Here there are also several options. In \rcite{BRZ}
it was shown how a regularization via dispersion relations and determining
the subtraction constants from the heat kernel expansion can be used
in this case. To go beyond two-point functions this method becomes very
cumbersome as well and there a simpler method\rcite{BP2,bernard} was used.
The essence of the method is to expand all one-loop diagrams of
the constituent quarks into the basic integrals by removing all dependencies
on the loop-momentum in the numerator via algebraic methods.
All combinations that involve only Lorentz structures without $g_{\mu\nu}$
are correctly reproduced this way. The Ward identities are then used to
determine the Lorentz structures involving $g_{\mu\nu}$. For the two-point
functions this procedure agrees with the dispersion relation technique
and for 3 and higher point functions it agrees with the results from
the heat-kernel expansion. The latter has been checked explicitly
for the first few terms by comparing results from the full expansion
with those from the heat kernel \rcite{BP2}.

Let us now show the last procedure on the simplest example.
We look at the one-loop contribution to the two-point function
\be
\Pi^V_{ij} = i \int d^4x e^{i q\cdot x}
\langle0|T\left(V^{ij}_\mu(x)V^{kl}(0)\right)|0\rangle
\ee
with $V^{ij}_\mu = \overline{q}_ii \gamma_\mu q_j$.
The relevant Feynman diagram is shown in Fig. \tref{figure5}.
\begin{figure}[htb]
\setlength{\unitlength}{1mm}
\begin{picture}(100.00,30.00)(30,20)
\thicklines
\put(97.50,35.00){\oval(15.00,10.00)}
\put(105.00,35.00){\circle*{2.00}}
\put(95.50,40.00){\vector(1,0){5.00}}
\put(99.00,30.00){\vector(-1,0){3.00}}
\put(90.00,35.00){\circle*{2.00}}
\end{picture}
\caption{The one-loop fermion diagram. The lines are constituent quarks.
The dots are insertions of the external currents.}
\rlabel{figure5}
\end{figure}
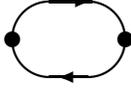
We will here for simplicity only quote the equal mass case.
The resulting feynman integral expression after doing the Dirac algebra in
four dimensions is proportional to
\be
\rlabel{feynint}
\int\frac{\dis d^4x}{(2\pi)^4}
\frac{\dis (- r^2 -q\cdot r +  m^2)g_{\mu\nu}
+ q_\mu r_\nu + r_\mu q_\nu + 2 r_\mu r_\nu}
{(r^2 - m^2)((q+r)^2 -m^2)}\,.
\ee
This integral has to be proportional to
\be
\rlabel{V1}
q_\mu q_\nu - q^2 g_{\mu\nu}
\ee
from the Ward identities.
Naively cutting of the integral in \rref{feynint} leads to
a piece of the form \rref{V1} but there is an extra term proportional
to $g_{\mu\nu}$. This term should be absent and has to be removed via the
Ward identities.

We now remove the $r^2$ via $r^2=r^2-m^2 + m^2$ and
$q.r$ via $2q.r=((q+r)^2-m^2)-q^2-(r^2-m^2)$.
This removes from the numerator a large fraction of the dependence on $r$.
As the next step we combine the two numerators using a Feynman parameter
\be
\frac{1}{(r^2 - m^2)((q+r)^2 -m^2)}
= \int_0^1 dx \frac{1}{\left( (r + xq)^2 - (m^2 -x(1-x)q^2)\right)^2}\,.
\ee
Then we perform in those integrals a shift to $p = r+xq$.
The integral with an odd number of $p$'s in the numerator vanishes.
Those with two powers are of the form $p_\mu p_\nu$ and are after
integration proportional to $g_{\mu\nu}$.
This procedure leads to the correct answer for the $q_\mu q_\nu$ term but needs
to have terms subtracted in the $g_{\mu\nu}$ piece. This is done by requiring
the full integral to be proportional to \rref{V1}.

The final result is then proportional to
\be
\left(q_\mu q_\nu - q^2 g_{\mu\nu}\right)
\int_0^1 dx \int d^4 p \frac{x(1-x)}{\left(p^2-(m^2-x(1-x)q^2)\right)^2}\,.
\ee
The integral we now regularize via
\be
\frac{1}{X^n} = \frac{1}{(n-1)!}\int_{1/\Lambda_\chi^2}^\infty d\tau
\tau^{n-1}\exp\left(-\tau X\right)\;.
\ee
Rotating the $p$ integral to Euclidean space finally leads to an  answer
proportional to
\be
\int_0^1 dx \Gamma\left(0,\frac{m^2-x(1-x)q^2}{\Lambda_\chi^2}\right)\,,
\ee
with
\be
\Gamma(n,\epsilon) = \int_\epsilon^\infty d\tau \tau^{n-1} e^{-\tau}\,.
\ee

This procedure can be easily generalized to the case with different masses
and higher than two-point functions. The requirement of being proportional
to \rref{V1} is then replaced by using the appropriate Ward identities.

The equivalent results to leaving the coefficients of the heat-kernel expansion
free, is to find out which identities exist between the different
one-loop Green functions and then to leave
only the ones not related to others as completely free functions.
In this case, similar to the low energy expansion, we are actually testing
a whole class of models where the one-loop expressions are left completely
free. A prominent example is the possible inclusion of extra low-energy
gluonic effects as described earlier.

\section{The anomaly}
\rlabel{anomaly}

There have been claims,
\rcite{RA} and references therein, that the
Extended Nambu--Jona-Lasinio model does not
reproduce the correct QCD
anomalous Ward identities. The correct result for the decay $\pi^0\to\gamma
\gamma$ was found but there were deviations from the anomalous Ward
identity prediction for the $\gamma\pi^0\pi^+\pi^-$ vertex.
Here we review the solution of Ref. \rcite{BP1} to this problem.
A similar problem was encountered in constructing anomalous effective
Lagrangians using full Vector Meson Dominance (VMD)\rcite{Brihaye}.
The same solution
also works in this case and it provides a simpler way to deal with the
Ward identities than the subtraction method used in ref. \rcite{Brihaye}.
The point of view taken here is that the ENJL model is looked upon
as a low energy approximation to QCD by only keeping the leading terms
in $1/\Lambda_\chi^2$. We know that the anomaly is a short-distance phenomenon
that is not suppressed by the cut-off so these terms can be subtracted
consistently to reproduce the correct anomalous Ward identities.
The procedure here restores the correct terms. The lowest order
terms thus become independent of the cut-off, but the higher
order contributions (like ${\cal O}(p^6)$) in the anomalous sector
will still depend on the cut-off $\Lambda_\chi$.

We will first
point out the underlying cause of the problem. This followed from the way
the four quark vertices in \rcite{RA} were treated. This is essentially
equivalent to requiring VMD.
The definition of the
abnormal intrinsic parity part of the effective action for effective theories
has already quite a history. After Fujikawa derived the anomalous Ward
identities\rcite{Bardeen} from the change in the measure in the functional
integral \rcite{Fujikawa}, Bardeen and Zumino clarified the relation between
the various forms of the anomaly found using this method \rcite{BardeenZumino}.
This paper also clarified the relation between the covariant and the
noncovariant (or consistent) forms of the anomalous current.
Leutwyler then showed how these different forms are visible in the
definition of the determinant of the Dirac operator\rcite{Leutwyler}. He also
discussed the relation of the anomalous current to this determinant.
At the same time Manohar and Moore
showed how the Wess-Zumino term\rcite{WZW}
can be derived from a change of variables
in the functional integral in a constituent chiral quark model
and how this can be used
to relate different anomalously inequivalent effective theories\rcite{Manohar}.

What we will show here is that the terms that violate the
anomaly generated by the procedure used in \rcite{RA} can be subtracted
consistently. We describe how the problem with the anomaly arises in the
standard treatment of the ENJL model. Then we illustrate a simpler way
to obtain the offending terms. This way will then show that these terms
can be subtracted in a consistent fashion. We also show that our prescription
does not influence the chirally covariant part of the effective
action.

Similar problems with the anomaly occur when one tries to formulate
quark-meson effective Lagrangians which include vector and axial-vector
meson couplings to the quarks. There the problem can be solved in a similar
way by subtracting terms that contain only (axial)vector mesons and external
fields. The same basic problem also occurs when trying to implement Vector
Meson Dominance for the anomalous terms. We show how it is related to the
problem in the ENJL model and can hence be solved similarly. Finally
we explicitly state what our prescription corresponds to.

In \rref{NJLpath} the measure ${\cal D}q{\cal D}\overline{q}$
has to be defined
in a way which reproduces the correct anomalous Ward identities. This means
that the cut-off procedure should be defined with a Dirac operator that
involves the external left and right handed vector fields.

The standard way to analyze the generating functional \rref{NJLpath} is
to introduce
a set of auxiliary variables
as described in section \tref{othermodels} to obtain an action bilinear in
fermion fields.
We will concentrate here on the vector--axial-vector
part since it is that one that
may generate the problems with the anomaly. The
scalar-pseudoscalar part is already treated in ref. \rcite{Manohar}.

Formally the Lagrangian in the exponential can be rewritten in terms of
the full fields $l^\prime_{\mu} = l_\mu + L_\mu$,
$r^\prime_{\mu} = r_\mu + R_\mu$ and $s^\prime,p^\prime$. The latter are
defined by $s^\prime - ip^\prime\gamma_5 = s - ip \gamma_5 + M\gamma_L
+M^\dagger \gamma_R$. We then have that
\be
\rlabel{fVMD}
\Gamma(l,r,s,p,L,R,M) = \Gamma(l^\prime,r^\prime,s^\prime,p^\prime)\ .
\ee
with $\Gamma(l,r,s,p)$ defined by
\be
\exp i\Gamma(l,r,s,p) = \int {\cal D}q {\cal D}\bar{q} \exp\left\{i\int d^4x
\left(\bar{q} \gamma^\mu i D_\mu q -
\bar{q}\left(s-ip\gamma_5\right)q\right)\right\}\ .
\ee

We can then integrate out the fermions
to obtain the effective generating functional as a function of the external
fields and the auxiliary fields. There is one caveat here and that is
precisely the cause of the problem observed in \rcite{RA}. The measure that
corresponds to the standard procedure is then defined by a Dirac operator that
is a function of $l^\prime_\mu, r^\prime_\mu$ rather than a function of
$l_\mu$ and $r_\mu$.

Let us show in a compact fashion how this problem occurs.
For simplicity we temporarily neglect the scalar-pseudoscalar part.
The effective action $\Gamma(l^\prime,r^\prime,s,p)$ can be related
simply to $\Gamma(l,r,s,p)$ by introducing the fields
\be
l^t_\mu = l_\mu + t L_\mu\qquad{\rm and}\qquad r^t_\mu = r_\mu + t R_\mu\ .
\ee
These fields transform under the chiral symmetry group in the same way as
$l_\mu,r_\mu$. We can now describe the effective action after integrating
the fermions as
\begin{eqnarray}
\rlabel{llprime}
\Gamma(l^\prime,r^\prime,s,p)
 & = &
\Gamma(l,r,s,p) + \int_0^1 dt
\frac{d}{dt}\Gamma(l^t,r^t,s,p)\nonumber\\
&=& \Gamma(l,r,s,p) + \int_0^1 dt \,
\tr \left( L_\mu \frac{\delta}{\delta l^t_\mu}
\Gamma(l^t,r^t,s,p) +
 R_\mu \frac{\delta}{\delta r^t_\mu}\Gamma(l^t,r^t,s,p) \right)\ .
\end{eqnarray}
The last two terms in \rref{llprime} correspond to the left and right handed
current. This current consists out of two pieces, a
non-anomalous and an anomalous part.
The part that is non-anomalous causes no problem and one can use
the standard heat kernel methods as used in Refs. \rcite{NJL2,BBR},
section \tref{p4analysis}, to obtain
information about the generating functional \rref{NJLpath}.

The anomalous part of the current can also be written as the sum of a local
chirally covariant part and a local polynomial of ${\cal O}(p^3)$
in $l^t,r^t$\rcite{BardeenZumino}.
If we now insist that at the first step,
where we integrate out the fermions, we should have the global chiral
symmetry exact (this corresponds to choosing the left-right form of the
anomalous current)
this local polynomial contains two pieces. One is a function of
$l^t$ and its derivatives only and the other one is a function of
$r^t$ and its derivatives. This globally
invariant form is precisely the form that a ``naive''
application of the heat kernel method would give\rcite{Leutwyler}.

The anomalous left and right currents [of ${\cal O}(p^3)$] in
eq. \rref{llprime}
have the following form in the left-right symmetric scheme
\ba
\rlabel{current}
\left.\frac{\delta\Gamma(l,r,s,p)}{\delta l^{\mu}}\right|_{\rm{an}}
 & \equiv & J^L_\mu + j^L_\mu \, ; \nonumber \\
\left.\frac{\delta\Gamma(l,r,s,p)}{\delta r^{\mu}}\right|_{\rm{an}}
 & \equiv & J^R_\mu + j^R_\mu \, ; \nonumber \\
J^{L\mu} & = & \frac{N_c}{48 \pi^2}
\varepsilon^{\mu \nu \alpha \beta} \left[i
{\cal L}_\nu {\cal L}_\alpha {\cal L}_\beta +
\left\{l_{\nu \alpha} + \frac{1}{2}U^\dagger
r_{\nu \alpha} U, {\cal L}_\beta \right\} \right] \, ; \nonumber \\
j^{L\mu} & = & \frac{N_c}{48 \pi^2}
\varepsilon^{\mu \nu \alpha \beta} \left[i
l_\nu l_\alpha l_\beta +
\left\{l_{\nu \alpha}, l_\beta\right\} \right]\, ;
\nonumber\\
l(r)_{\mu\nu} & = & \partial_\mu l(r)_\nu -\partial_\nu l(r)_\mu - i
[ l(r)_\mu , l(r)_\nu]\ ,
\nonumber\\
{\cal L}_\mu & = & i U^\dagger \left(\partial_\mu U - i r_\mu U + i U l_\mu
\right)\ .
\ea
The matrix $U$ is the ``phase'' of $M$. $U = \xi^2$ with $M = \xi H \xi$.
Here $H$ is hermitian and $\xi$ is unitary. The currents $J^R_\mu$ and
$j^R_\mu$ can be obtained from $J^L_\mu$ and $j^L_\mu$ by a parity
transformation.

Since in \rref{llprime} the part $\Gamma(l,r,s,p)$ already saturates
the inhomogeneous part of the
anomalous Ward identities the remainder should be
locally chirally invariant. The
parts that are not locally invariant in the last two terms of eq.
\rref{llprime}
should thus be subtracted. As can be seen from \rref{current} these terms
are a local function of $l,L$ and their derivatives (plus the right handed
counterpart). The change in the
definition of the measure involves only the fields
$l,L,r$ and $R$ so the local terms that can be added to the effective action
to obtain the correct Ward identities should only be functions of these
and their derivatives. The preceding discussion shows that the terms that
spoil the anomalous Ward identities are precisely of this type.

As a consistency check we will show that the contribution of the
local chirally covariant part of the anomalous current
to the resulting effective action can not be changed
by adding globally invariant counterterms that are functions of $l,L$ and
their derivatives only. The full list of terms that could contribute is
(an overall factor of $\varepsilon^{\mu\nu\alpha\beta}$ is understood).
\begin{eqnarray}
\tr (L_\mu L_\nu L_\alpha L_\beta) \,,&&
\tr (D_\mu L_\nu L_\alpha L_\beta)\,,
\nonumber\\
\tr (l_{\mu\nu}L_\alpha L_\beta )\,,&&
\tr (l_{\mu\nu}D_\alpha L_\beta)
\end{eqnarray}
and their right handed counterparts.
All others are related to these via partial integrations.
The first term vanishes because of the cyclicity of the trace. The second one
is a total derivative. The third one is forbidden by CP invariance and
the last one vanishes because of the Bianchi identities for $l_{\mu\nu}$.

This is just proving that the standard procedure of adding counterterms
and determining their finite parts by making the final effective
action satisfy the (anomalous) Ward identities also works here
giving an unambiguous answer.

We have used the left-right symmetric form of the anomaly.
But it is obvious from the discussion above, that by following an
analogous procedure to the one given here in any scheme of
regularization of the chiral anomaly one obtains the same result
since the difference between the anomalous current in two of these schemes
is a set of local polynomials that can only depend on $l_\mu'$ and
$r_\mu'$, ref. \rcite{Leutwyler}. A scheme of particular interest
is that where the vector symmetry is explicitly conserved.
In order to obtain this form of the Wess-Zumino action one
has to add a set of local polynomials that only depend on
$l'_\mu$ and $r'_\mu$
to the left-right symmetric one.
These are given explicitly in ref. \rcite{Bardeen}.

In the basis of fields we have been working until now the
effective action in the non-anomalous sector has generated a
quadratic form mixing the pseudoscalar field
and the axial-vector auxiliary field (roughly speaking $R_\mu-L_\mu$).
It is of common practice to change to a basis where this quadratic form
is diagonal (e.g. see \rcite{BBR}).
Afterwards the vector and axial-vector degrees of freedom can be removed
by using their equations of motion to obtain an effective action for the
pseudoscalars only.
In this way one also
introduces the axial coupling, the so-called $g_A$, which in
the chiral constituent quark model \rcite{GM}
corresponds to the axial vector coupling of constituent quarks to
pseudoscalar mesons. In our effective action, this change of basis
can only generate local chiral invariant terms that therefore
cannot modify the Wess-Zumino effective action \rcite{WZW}
and the standard predictions at order $p^4$ for $\pi^0\to
\gamma\gamma$ and $\gamma\pi^+\pi^-\pi^0$ will be satisfied. There will of
course be changes at higher orders due to the chiral
local invariant terms.
Thus, the value of $g_A$ is not constrained by the chiral anomaly
which is a low-energy theorem of QCD contrary to the conclusion
of ref. \rcite{RA} and in agreement with the results of ref.
\rcite{Manohar}.

These changes due to higher orders are very similar to the description using
the hidden symmetry approach\rcite{Fujiwara} (see also \rcite{Bijnens})
and the gauged Yang-Mills approach as given in
ref. \rcite{Wakamatsu}. This prescription is also
precisely the prescription that was used in ref. \rcite{Ximo} to
construct the lowest order anomalous effective chiral Lagrangian
involving vector and axial-vector fields and
obtain predictions for the ``anomalous'' decays of these particles
within the ENJL model.

We would like to add one small remark about ref. \rcite{RA}. In this reference
only the $p^2$ equations of motion
for $L_\mu,R_\mu$ were used. In principle there is also
a contribution from the $p^4$ part proportional to
$\varepsilon^{\mu\nu\alpha\beta}$ when substituted into the $p^2$ term of the
effective action. This contribution does however cancel between the
$p^2$ term and the ``mass'' term for the auxiliary fields.

Our discussion was in the framework of the ENJL model. The root of the problem
was the relation \rref{fVMD}. As mentioned before similar problems occur
in effective quark-meson models with explicit spin-1 mesons couplings to
the quarks and in the old approaches that require full vector meson
dominance (VMD). The basic requirement of VMD is that vector mesons couple
like the external fields. If we describe the physical
vector mesons by fields $L_\mu$
and $R_\mu$,
 this requirement can be cast in the form ($\phi$ stands for all
the other fields involved)
\ba
\rlabel{VMD2}
\left.\frac{\delta^n\Gamma(l,r,L,R,\phi)}{\delta^m L_\mu\delta^{n-m}R_\nu}
\right|_{L=R=0}
&\equiv&
\left.\frac{\delta^n\Gamma(l,r,L,R,\phi)}{\delta^m l_\mu\delta^{n-m}r_\nu}
\right|_{L=R=0}\ ; \nonumber\\*[0.5cm]
{\rm for} \,\, 0 \leq m \leq n \ .
\ea
Using the Taylor expansion
of $\Gamma(l,r,L,R,\phi)$ in $l,r,L$ and $R$ and applying
eq. \rref{VMD2} it can be shown that the action then only depends
on $l^\prime=l+L$ and $r^\prime=r+R$, i.e.
\be
 \Gamma(l,r,L,R,\phi) = \Gamma(l+L,r+R,0,0,\phi)\ .
\ee
This will lead to precisely the same type of problems as seen in the ENJL
model since this is the same relation as eq. \rref{fVMD}. Here again it
can be remedied by adding local polynomials in $l,L,r$ and $R$ precisely
as was done before.

Now, what does our prescription mean in practice?
It means that vector and axial-vector fields are consistently introduced
in the low-energy effective Lagrangian by requiring a slightly
modified VMD relation
\ba
\rlabel{VMD3}
\left.\frac{\delta^n\Gamma(l,r,L,R,\phi)}{\delta^m L_\mu\delta^{n-m}R_\nu}
\right|_{L=R=0}
&\equiv&
\left(\left.
\frac{\delta^n\Gamma(l,r,L,R,\phi)}{\delta^m l_\mu\delta^{n-m}r_\nu}
\right|_{L=R=0}\right)_{\begin{array}{l}{\rm local}\\* {\rm covariant}
\end{array}}
\ ;
\nonumber\\
{\rm for} \,\, 0 \leq m \leq n & {\rm with} \;\; m + n \geq 1&
\ea
\noindent
instead of the usual VMD requirement in eq. \rref{VMD2}. This is equivalent
to use the standard heat kernel expansion technique (for a review see
\rcite{Ball}) for the non-anomalous part,i.e.
no Levi-Civita symbol, and for the chiral orders larger than $p^4$
in the anomalous part, i.e. terms with a Levi-Civita symbol.
 For the ${\cal O}(p^4)$ part of the anomalous action
one has just the usual Wess-Zumino term and for the chiral
orders smaller than $p^4$ in the anomalous part one has
\be
\int^1_0 dt \ \tr\left( L^\mu J^L_\mu(l^t,r^t,U)
+ R^\mu J^R_\mu(l^t,r^t,U) \right)\ .
\ee
Here the anomalous currents $J^{L,R}_\mu$ are
those defined in eq. \rref{current}.

In the present work we have been implicitly using a
representation similar to the
so-called vector model (model II in ref. \rcite{37})
to represent vector and axial-vector fields as the
most natural way within the ENJL model we are working with.
However, it is straightforward to work out the analogous
prescription to eq. \rref{VMD3} for any other suitable
representation of vector and axial-vector fields
(tensor, gauge fields, $\cdots$)\rcite{37}
to implement
VMD in both the anomalous and the non-anomalous sectors
of the effective action.

\section{Analysis to order $p^4$}
\rlabel{p4analysis}

\subsection{The Mean Field Approximation}

We shall first discuss a particular case of
 $\Gamma_{eff}(H,\xi,L_{\mu},R_{\mu};v,a,s,p)$ as
defined in eq.~(\tref{61}). It is the case corresponding to the mean field
 approximation, where
\be\rlabel{73}
H(x) = <H> = M_Q \hbox{\bf 1},
\qquad{\rm and}\qquad L_{\mu} = R_{\mu} = 0.
\ee
The effective action
$\Gamma_{eff}(M_Q,\xi,0,0;v,a,s,p)$ coincides
 then with the one calculated in Ref.~\rcite{ERT},
except that the regularization  of the UV behaviour
 is different.
In Ref.~\rcite{ERT}, the regularization which is used is the $\zeta$
function regularization.
The results, to a first approximation where low-frequency gluonic
 terms are ignored, are
as follows:
\be\rlabel{75}
f_0^2 = {N_c\over 16\pi^2} 4\,M_Q^2 \Gamma(0,{M_Q^2\over \Lambda_{\chi}^2})
\ee
 and
\be\rlabel{76}
f_0^2 B_0 = - <\bar {\psi} \psi> =
{N_c \over  16 \pi^2} 4\,M_Q^3 \Gamma(-1,{M_Q^2\over \Lambda_{\chi}^2})
\ee
for the lowest $O(p^2)$ couplings of the low-energy effective
 Lagrangian in (\tref{22}).

For the $O(p^4)$ couplings which exist in the chiral limit we find
\be\rlabel{77}
L_2=2L_1={N_c\over 16\pi^2}{1\over 12}\Gamma(2,{M_Q^2\over \Lambda_{\chi}^2}),
\ee
\be\rlabel{78}
L_3={N_c\over 16\pi^2}{1\over 6}\left[\Gamma(1,{M_Q^2\over \Lambda_{\chi}^2})
- 2\Gamma(2,{M_Q^2\over \Lambda_{\chi}^2})\right]
\ee
for the four derivative terms; and
\be\rlabel{79}
L_9={N_c\over 16\pi^2}{1\over 3}\Gamma(1,{M_Q^2\over \Lambda_{\chi}^2}),
\ee
\be\rlabel{80}
L_{10}=-{N_c\over 16\pi^2}{1\over 6}\Gamma(1,{M_Q^2\over \Lambda_{\chi}^2})
\ee
for the two couplings involving external fields. If one lets
$ M_Q^2 /    \Lambda_{\chi}^2  \to 0$, then $\Gamma(n,0)=\Gamma(n)=(n-1)$! for
 $n\geq 1$, and these
results coincide with those previously obtained in Refs.~\rcite{8a},
\rcite{8b},
\rcite{10b}, \rcite{ERT} and \rcite{50} to \rcite{50c}.

When terms proportional to the quark mass matrix ${\cal M}$ are kept, there
 appear four new
$L_i$ couplings (see eq.~(\tref{27})). With
\be\rlabel{81}
\rho = {M_Q \over \vert  B_0 \vert} = {M_Q f_0^2 \over \vert <\bar{\psi} \psi>
 \vert},
\ee
the results we find for these new couplings are
\be\rlabel{82}
L_4=0
\ee
\be\rlabel{83}
L_5={N_c\over 16\pi^2} {\rho \over 2}
\left[\Gamma(0,{M_Q^2\over \Lambda_{\chi}^2})-\Gamma(1,{M_Q^2\over
 \Lambda_{\chi}^2})\right]
\ee
\be\rlabel{84}
L_6=0
\ee
\be\rlabel{85}
L_7={N_c \over 16\pi^2}{1 \over 12}\left[-\rho \Gamma(0,{M_Q^2\over
 \Lambda_{\chi}^2})
+{1 \over 6}\Gamma(1,{M_Q^2\over \Lambda_{\chi}^2})\right]
\ee
\be\rlabel{86}
L_8=-{N_c \over 16\pi^2} {1 \over 24}\left[6\rho(\rho - 1)\Gamma(0,{M_Q^2\over
 \Lambda_{\chi}^2})
+\Gamma(1,{M_Q^2\over \Lambda_{\chi}^2}) \right].
\ee
 If we identify $\Gamma(0, M_Q^2 / \Lambda_{\chi}^2 )
\equiv \hbox{log}(\mu^2 / M_Q^2)$, and take the limit
$\Gamma(n\geq 1, M_Q^2/\Lambda_{\chi}^2 \to 0)$, these results coincide
 then with those obtained in
Ref.~\rcite{ERT}.
(Notice that $\rho$ is twice the parameter $x$ of Ref.~\rcite{ERT}.)

The fact that $L_4=L_6=0$ and $L_2=2L_1$  is more general than the model
 calculations we are
discussing. As first noticed by Gasser and Leutwyler \rcite{GL}, these are
 properties of the large-$N_c$
 limit. The contribution we find for $L_7$ is in fact non-leading in the
 $1/N_c$ expansion. The
above result is
entirely due to the use of the lowest-order equations of motion
 (see the erratum to Ref.~\rcite{ERT}).
In the presence of the $U_A(1)$ anomaly, $L_7$ picks up a
 contribution from the
$\eta^{\prime}$ pole and becomes $O(N_c^2)$, \rcite{GL}
\footnote{This counting is somewhat misleading since it first relies
on $1/N_c$ to be small to have $m_{\eta'}^2$ of order $1/N_c$ and
then expands in $1/m_{\eta'}^2$, or $1/N_c$ large.}.

Finally, we shall also give the results for the $H_1$ and $H_2$ coupling
 constants of terms which
only involve external fields:
\be\rlabel{87}
H_1=-{N_c \over16\pi^2}{1 \over 12}\left[2\Gamma(0,{M_Q^2\over
 \Lambda_{\chi}^2})
-\Gamma(1,{M_Q^2\over \Lambda_{\chi}^2})\right]
\ee
\be\rlabel{88}
H_2={N_c \over16\pi^2}{1 \over 12}\left[6\rho^2\Gamma(-1,{M_Q^2\over
 \Lambda_{\chi}^2})
-6\rho(\rho + 1)\Gamma(0,{M_Q^2\over \Lambda_{\chi}^2}) +
\Gamma(1,{M_Q^2\over \Lambda_{\chi}^2})\right]
\ee

\subsection{Beyond the Mean Field Approximation}

In full generality,
\be\rlabel{89}
H(x) = M_Q {\LARGE\bf 1} + \sigma(x)\;,
\ee
and the effective action
$\Gamma_{eff}(H,\xi,L_{\mu},R_{\mu};v,a,s,p)$
has a non-trivial dependence on the auxiliary field variables $\sigma(x)$,
$L_{\mu}(x)$ and $R_{\mu}(x)$. It is convenient to trade the auxiliary left and
 right
vector field variables $L_{\mu}$ and $R_{\mu}$,
which were introduced in eq.
 (\tref{57}), by the new
vector fields
\be\rlabel{90}
W_{\mu}^{(\pm)} = \xi L_{\mu} \xi^{\dagger} \pm \xi^{\dagger} R_{\mu} \xi.
\ee
{}From the transformation properties
in eqs.~(\tref{31}) and (\tref{55}),  it follows that
$W_{\mu}^{\pm}$ transform homogeneously, i.e.
\be\rlabel{91}
W_{\mu}^{(\pm)} \to h(\Phi,g) W_{\mu}^{(\pm)} h^{\dagger}(\Phi,g).
\ee
We also find it convenient to rewrite the effective action in eq.
 (\tref{61}) in a basis of
constituent chiral quark fields
\be\rlabel{92}
Q = Q_L + Q_R\,\,\,\hbox{and}\,\,\,\bar Q = \bar Q_L + \bar Q_R\;,
\ee
where
\be\rlabel{93}
Q_L = \xi q_L\,,\,\bar Q_L = \bar q_L \xi^{\dagger}\,;\quad
\,Q_R = \xi^{\dagger} q_R \,,\,\bar Q_R = \bar q_R \xi,
\ee
which under the chiral group $G$, transform like
\be\rlabel{94}
Q \to h(\Phi,g) Q \,\,\,\hbox{and}\,\,\,\ \bar Q \to \bar Q
h(\Phi,g)^{\dagger}.
\ee
In this basis, the linear terms (in the auxiliary field variables) in
 the r.h.s. of eq.~(\tref{61}) become
\be\rlabel{96}
\bar Q \left( - H + {1\over 2} \gamma^{\mu}W_{\mu}^{(+)} -
{1\over 2} \gamma^{\mu}\gamma_5W_{\mu}^{(-)} \right) Q.
\ee

At this stage, it is worth pointing  out a formal symmetry which is useful to
 check explicit
calculations. We can redefine the external vector-field sources via
\be\rlabel{102}
l_{\mu} \to l_{\mu}^{\prime}=l_{\mu} + L_{\mu}
\ee
\be\rlabel{103}
r_{\mu} \to r_{\mu}^{\prime}=r_{\mu} + R_{\mu}
\ee
and
\be\rlabel{104}
{\cal M} \to {\cal M}^{\prime}(x) = {\cal M} + \xi \sigma(x) \xi.
\ee
The Dirac operator ${\cal D}_E$, when reexpressed  in terms of the
``primed'' external fields, is formally the
same Dirac operator as
 the one corresponding
to the ``mean   field approximation.''  In practice, it
means that once we have evaluated the formal effective action
\be\rlabel{107}
\hbox{exp } \Gamma_{eff}({\cal A}_{\mu},M) =
\int {\cal D}\bar{Q} {\cal D}Q \hbox{exp} \int d^4 x \bar{Q}{\cal D}_E Q =
 \hbox{det} {\cal D}_E
\ee
we can easily get the new terms
involving the new auxiliary fields
 $L_{\mu}$, $R_{\mu}$
and $\sigma$ by doing the appropriate shifts. The formal evaluation of
$\Gamma_{eff}({\cal A}_{\mu},M)$ to $O(p^4)$
in the chiral expansion has been
 made by several authors (see Refs.~\rcite{50} to \rcite{50c}).

\subsection{The constant $g_A$ and resonance masses}

When computing the effective action $\Gamma_{eff}({\cal A}_{\mu},M)$ in
 eq.
 (\tref{107}), there appears
a mixing term proportional to $\tr \xi_{\mu} {W^{(-)}}^{\mu}$.
More precisely,
 one finds a
quadratic form in $\xi_{\mu}$ and $W^{(-)}_{\mu}$ (in Minkowski space-time):
\be\rlabel{108}
\Gamma = \alpha <{W^{(-)}}_{\mu}W^{(-){\mu}}> + \beta
<\xi_{\mu}W^{(-){\mu}}> + \gamma <\xi_{\mu}\xi^{\mu}>
\ee
with
\be\rlabel{109}
\alpha = {N_c \over 16\pi^2}\left({1\over 4}{\Lambda_{\chi}^2 \over G_V} +
M_Q^2\Gamma(0,{M_Q^2 \over \Lambda_{\chi}^2}) \right)
\ee
and
\be\rlabel{110}
\beta = -2{N_c \over 16\pi^2}\Gamma(0,{M_Q^2 \over \Lambda_{\chi}^2}) M_Q^2
\,\,\,\hbox{and}\,\,\, \gamma = -{1\over 2} \beta \;.
\ee
The field redefinition
\be\rlabel{111}
{W^{(-)}}_{\mu} \to {\hat W^{(-)}}_{\mu} + (1-g_A)\xi_{\mu} \;,
\ee
with
\be\rlabel{112}
g_A = 1+{\beta \over 2\alpha}
\ee
diagonalizes the quadratic form. There is a very interesting physical
 effect due to this
diagonalization, which is that it redefines the coupling of the constituent
 chiral quarks to the
pseudoscalars. Indeed, the covariant derivative in eq.~(\tref{107}) becomes
\be\rlabel{113}
{\cal D}_E = \gamma_\mu \nabla_{\mu}=
\gamma_\mu\left(\partial_{\mu} + iG_{\mu} + \Gamma_{\mu} - {i
\over 2} \gamma_5  (g_A \xi_{\mu} - \hat W_{\mu}^{(-)}) - {i \over 2}
 W_{\mu}^{(+)}\right).
\ee
$g_A$ can be identified with
 the $g_A$ coupling constant of the constituent
chiral quark model of Manohar and Georgi \rcite{GM}.

In the calculation of $\Gamma_{eff}({\cal A}_{\mu},M)$
we also encounter kinetic-like terms
for the fields $\hat W^{(-)}_{\mu}$ and $W^{(+)}_{\mu}$.
Comparison with the standard vector and axial-vector kinetic terms
requires
a scale redefinition of the fields $W^{(+)}_{\mu}$
and $\hat W^{(-)}_{\mu}$ to obtain the correct
kinetic couplings, i.e.
\be\rlabel{116}
V_{\mu} = \lambda_V W^{(+)}_{\mu}\;,
\,\,\,\,\,\,\,\,
A_{\mu} = \lambda_A \hat W^{(-)}_{\mu},
\ee
with
\be\rlabel{117}
\lambda_V^2 = {N_c \over 16\pi^2} {1 \over 3}
\Gamma(0,{M_Q^2 \over \Lambda_{\chi}^2})
\ee
and
\be\rlabel{118}
\lambda_A^2 ={N_c \over 16\pi^2} {1 \over 3}
\left[\Gamma(0,{M_Q^2 \over \Lambda_{\chi}^2})-\Gamma(1,{M_Q^2 \over
 \Lambda_{\chi}^2})\right].
\ee
These $V_\mu$ and $A_\mu$ fields are the ones that transform in the standard
CCWZ way \rcite{CCWZ} for the vector and axial-vector fields.

This scale redefinition gives rise to mass terms (in Minkowski
 space-time)
\be\rlabel{119}
{1 \over 2} M_V^2\tr(V_{\mu}V^{\mu})+ {1 \over 2}M_A^2\tr(A_{\mu}A^{\mu})
\ee
with
\be\rlabel{120}
M_V^2 = {2\alpha +\beta \over \lambda_V^2}
\,\,\,\,\,\,\hbox{and}\,\,\,\,\,\,
M_A^2 = {2\alpha \over \lambda_A^2}.
\ee

The same comparison between the calculated kinetic and mass terms in the scalar
 sector, with the
standard scalar Lagrangian in eq.~(\tref{46}), requires the scale redefinition
\be\rlabel{121}
S(x) = \lambda_S \sigma(x),
\ee
with
\be\rlabel{122}
\lambda_S^2 = {N_c \over 16\pi^2} {2 \over 3} \left[3\Gamma(0,{M_Q^2 \over
 \Lambda_{\chi}^2})
- 2\Gamma(1,{M_Q^2 \over \Lambda_{\chi}^2}) \right].
\ee
The scalar mass is then
\be\rlabel{123}
M_S^2 = {N_c \over 16\pi^2} {8M_Q^2  \over \lambda_S^2} \Gamma(0,{M_Q^2 \over
 \Lambda_{\chi}^2}).
\ee

\subsection{The couplings of the ${\cal L}^R_{eff}$ Lagrangian}

The Lagrangian in question is the one that
we have written in section \tref{hadronic}, in
eqs.~(\tref{46}), (\tref{47}) and (\tref{48}), based on
chiral-symmetry requirements alone. These requirements  did not fix,
however,
the masses and the interaction couplings with the pseudoscalar fields and
external fields.
The results for the masses which we now find in the extended
Nambu--Jona-Lasinio model are given by
eqs.~(\tref{120}) and (\tref{123}) in the previous subsection.
These are the results in the limit where
low-frequency gluonic interactions in
${\cal L}^{\Lambda_{\chi}}_{QCD}$ in eq.~(\tref{LENJL}) are neglected, i.e.
the results
 corresponding to the first
alternative scenario we discussed in the introduction. For the other coupling
 constants, and also
in the limit where low-frequency gluonic interactions are neglected,
the results  are:
\be\rlabel{124}
{1\over 4}f_{\pi}^2 = {N_c\over 16\pi^2} M_Q^2 g_A\Gamma(0,{M_Q^2 \over
 \Lambda_{\chi}^2}),
\ee
instead of the mean
field approximation result in eq.~(\tref{75}):\footnote{This implicitly
changes the value of $B_0$ via eq. \protect{(\tref{76})}.}
\be\rlabel{125}
f_V = \sqrt 2 \lambda_V\,\,\,\,\, , \,\,\,\,\,f_A = \sqrt 2 g_A \lambda_A,
\ee
\be\rlabel{126}
g_V = {N_c \over 16\pi^2} {1 \over \lambda_V} {\sqrt 2 \over 6}
\left[(1-g_A^2) \Gamma(0,{M_Q^2 \over \Lambda_{\chi}^2}) +
2 g_A^2 \Gamma(1,{M_Q^2 \over \Lambda_{\chi}^2}) \right]
\ee
 for the vector and axial-vector coupling constants in (\tref{47}) and
 (\tref{48}); and
\be\rlabel{127}
c_m = {N_c \over 16\pi^2} {M_Q \over \lambda_S} \rho
\left[\Gamma(-1,{M_Q^2 \over \Lambda_{\chi}^2}) - 2 \Gamma(0,{M_Q^2 \over
 \Lambda_{\chi}^2}) \right],
\ee
\be\rlabel{128}
c_d = {N_c \over 16\pi^2} {M_Q \over \lambda_S} 2 g_A^2\left[\Gamma(0,{M_Q^2
 \over \Lambda_{\chi}^2})
-\Gamma(1,{M_Q^2 \over \Lambda_{\chi}^2})\right]
\ee
 for the scalar coupling constants in (\tref{46}).

There are a series of interesting relations between these results:
\be\rlabel{129}
M_V^2 = {3 \over 2} {\Lambda_{\chi}^2 \over G_V(\Lambda_{\chi}^2)}
{1 \over \Gamma(0,{M_Q^2 \over \Lambda_{\chi}^2})},
\ee
\be\rlabel{130}
M_A^2\left\{1- {\Gamma(1,{M_Q^2 \over \Lambda_{\chi}^2}) \over
\Gamma(0,{M_Q^2 \over \Lambda_{\chi}^2})}\right\} = M_V^2 + 6 M_Q^2,
\ee
\be\rlabel{131}
g_A = 1+{\beta \over 2\alpha} = {f_V^2 M_V^2 \over f_A^2 M_A^2}g_A^2,
\ee
 with the two solutions
\be\rlabel{132}
g_A = 0 \,\,\,\hbox{and}\,\,\, g_A = {f_A^2 M_A^2 \over f_V^2 M_V^2};
\ee
 and
\be\rlabel{133}
f_V^2 M_V^2 = f_A^2 M_A^2 + f_{\pi}^2.
\ee
 The last relation is the first Weinberg sum rule \rcite{40}.
Using this sum rule and the second solution for $g_A$, we also have
\be\rlabel{134}
g_A = 1 - {f_{\pi}^2 \over f_V^2 M_V^2}.
\ee
Therefore $g_A<1$. The
 two relations
in eqs.~(\tref{133}) and (\tref{134}) remain valid in the presence of gluonic
 interactions, i.e.  the
gluonic corrections do modify the explicit form of the calculation we have made
 of $f_{\pi}$, $f_V$,
$M_V$ and $g_A$, but they do it in such a way that eqs.~(\tref{133}) and
 (\tref{134}) remain unchanged.

\subsection{The coupling constants $L_i$'s, $H_1$ and $H_2$ beyond the mean
 field approximation}

These coupling constants are now modified because we no longer have
 $g_A=1$.
 With the short-hand
notation
\be\rlabel{135}
x = {M_Q^2 \over \Lambda_{\chi}^2},
\ee
 the analytic expressions we find from the quark-loop
integration are the following:
\be\rlabel{136}
 L_2 = 2\,L_1 = {N_c \over 16\pi^2}{1\over 24}\left[(1-g_A^2)^2\Gamma(0,x)
+ 4g_A^2 (1-g_A^2) \Gamma(1,x)+ 2g_A^4 \Gamma(2,x)\right],
\ee
\be\rlabel{137}
 \tilde L_3 = {N_c \over 16\pi^2}{1\over 24}\left[-3(1-g_A^2)^2\Gamma(0,x)
+ 4\left(g_A^4 - 3g_A^2 (1-g_A^2)\right)\Gamma(1,x) -
 8g_A^4\Gamma(2,x)\right],\ee
\be\rlabel{138}
L_4 =0,\ee
\be\rlabel{139}
\tilde L_5 = {N_c\over 16\pi^2} {\rho \over 2}
 g_A^2\left[\Gamma(0,x)-\Gamma(1,x)\right],\ee
\be\rlabel{140}
 L_6 = 0,\ee
\be\rlabel{141}
L_7 = O(N_c^2), \ee
\be\rlabel{142}
\tilde L_8 = -{N_c \over 16\pi^2} {1\over 24}\left[ 6 \rho (\rho - g_A)
\Gamma(0,x) + g_A^2\Gamma(1,x) \right],\ee
\be\rlabel{143}
L_9 = {N_c \over 16\pi^2}{1\over 6}\left[(1-g_A^2)\Gamma(0,x) +2
 g_A^2\Gamma(1,x)\right],\ee
\be\rlabel{144}
L_{10} = -{N_c \over16\pi^2}{1\over6}\left[(1-g_A^2)\Gamma(0,x) +
 g_A^2\Gamma(1,x)\right],\ee
\be\rlabel{145}
H_1 = -{N_c \over16\pi^2}{1\over12}\left[(1+g_A^2)\Gamma(0,x) -
 g_A^2\Gamma(1,x)\right],\ee
\be\rlabel{146}
\tilde H_2 = {N_c \over16\pi^2}{1 \over 12}\left[6 \rho^2 \Gamma(-1,x) -
6\rho (\rho + g_A)\Gamma(0,x) +g_A^2\Gamma(1,x)\right].\ee

 Three of the $L_i$ couplings ($i=$ 3, 5 and 8) as well as $H_2$
receive explicit contributions from the integration of scalar
fields. This is why we write $L_i = \tilde L_i + L_i^S$, $i=$~3,~5,~8;
$H_2 = \tilde H_2 + H_2^S$ with $\tilde L_i$, $\tilde H_2$ the contribution
from the quark-loop and $L_i^S$, $H_2^S$ that from the scalar field.
The results for $L_1$, $L_2$ and $\tilde L_3$ agree with those
of Ref.~\rcite{58},
where these couplings were obtained by integrating out the
constituent quark fields in the model of Manohar and Georgi \rcite{GM}. At
the level where possible gluonic corrections are neglected, the two
calculations are formally equivalent.
The results for $L_4$ to $L_{10}$ agree with those of
Ref.~\rcite{5}.

We note that between these results for the $L_i$'s, $H_1$ and
the results for couplings and masses of the  vector and axial-vector
Lagrangians, which we obtained before,
there are the following interesting relations:

\be\rlabel{147}
L_9 = {1 \over 2} f_V g_V,
\ee

\be\rlabel{148}
L_{10} = -{1 \over 4} (f_V^2 - f_A^2)\,\,\,\hbox{and}\,\,\,
2\,H_1 = -{1 \over 4} (f_V^2 + f_A^2).
\ee

 As we shall see in the next subsection,
these relations, like those in eqs.~(\tref{133}) and (\tref{134}), are also
valid
 in the presence of
gluonic interactions. The alerted reader will recognize that these relations
are
 precisely the QCD
short-distance constraints which,
as discussed in Ref.~\rcite{37}, are required to
 remove the ambiguities
in the context of chiral perturbation theory to $O(p^4)$ when vector and
 axial-vector degrees of freedom are integrated out.
They are the relations which follow from demanding
consistency  between the low-energy
effective action of vector and axial-vector
mesons and the QCD short-distance behaviour of two-point and
three-point functions. It is rather remarkable that the simple ENJL
model we have been discussing
incorporates these constraints automatically.

There is a further constraint that was
also invoked in Ref.~\rcite{37}. It has to
do with the asymptotic
behaviour of the elastic meson--meson scattering, which in
QCD is expected to  satisfy the Froissart bound \rcite{41}.
If that is the case, the authors of Ref.~\rcite{37} concluded that,
 besides the constraints already discussed, one also must have
\be\rlabel{149}
L_1 = {1 \over 8} g_V^2\,\,;\,\, L_2 = 2 L_1 \,\,;\,\, L_3 = - 6
 L_1.
\ee
 As already mentioned,
the second constraint is a property of QCD in the large $N_c$ limit.
The first and third constraints, however,
are highly non-trivial. We observe that,
to the extent that $O(N_c g_A^4)$ terms can be neglected,
these constraints
are then also satisfied in the ENJL model.

When the massive scalar field is integrated out \rcite{36},
there is a further
 contribution to the
constants $L_3$, $L_5$, $L_8$ and $H_2$ with the results:
\be\rlabel{152}
L_3^S = {c_d^2 \over 2 M_S^2} = {N_c \over 16\pi^2}{1 \over 4} g_A^4 {1 \over
 \Gamma(0,x)}
[\Gamma(0,x)-\Gamma(1,x)]^2,
\ee
\be\rlabel{153}
L_5^S = {c_m c_d \over M_S^2} = {N_c \over 16\pi^2}{1 \over 4} \rho g_A^2 {1
 \over \Gamma(0,x)}
[\Gamma(-1,x)-2\Gamma(0,x)][\Gamma(0,x)-\Gamma(1,x)],
\ee
\be\rlabel{154}
L_8^S = {c_m^2 \over 2 M_S^2} = {N_c \over 16\pi^2}{1 \over 16} \rho^2 {1 \over
 \Gamma(0,x)}
[\Gamma(-1,x)-2\Gamma(0,x)]^2,
\ee
\be\rlabel{155}
H_2^S = 2L_8^S.
\ee

This result for $L_3^S$ disagrees with the one found in Ref.
\rcite{5}. Also, contrary to what is found in Ref.~\rcite{5}, there is no
contribution from scalar exchange to $L_2$.

It is interesting to point out that $\tilde L_5$, $L_5^S$ and
$\tilde L_8$, $L_8^S$ each depend explicitly on the parameter $\rho$.
This dependence, however, disappears in the sums

\be \rlabel{155a}
L_5 = \tilde L_5 + L_5^S = {N_c \over 16 \pi^2} {1\over 4}g_A^3
\left[\Gamma(0,x)-\Gamma(1,x)\right]
\ee
 and
\be\rlabel{155b}
L_8 = \tilde L_8 + L_8^S = {1 \over 4} f_{\pi}^2 {g_A \over 16 M_Q^2} -
{N_c \over 16\pi^2} {1\over 24} g_A^4 \Gamma(1,x).
\ee
Similar simplified expressions for the other $L_i$ can be found in
\rcite{eduardo3}.

\subsection{Results in the presence of gluonic interactions}

The purpose of this section is to explore in more detail the second
 alternative,
which we described in the introduction, whereby the four-quark
operator terms
in eqs.~(\tref{LSP}) and (\tref{LVA}) are viewed as
the leading result of a first-step
renormalization \`a la Wilson,
once the quark and gluon degrees of freedom
have been integrated out down to a scale $\Lambda_{\chi}$. Within this
alternative, one is still left with a fermionic determinant,
 which has to be
evaluated  in the presence of gluonic interactions due to
fluctuations below
the $\Lambda_{\chi}$ scale. The net effect of these long-distance gluonic
interactions is to modify the various incomplete gamma functions
$\Gamma(n,x= M_Q^2 /\Lambda_{\chi}^2)$,
which modulate the calculation of
the fermionic determinant in the previous sections, into new
({\em a priori}
incalculable) constants. We examine first how many independent
unknown constants can appear at most.
Then, following the approach developed in Ref.~\rcite{ERT},
we shall proceed to an approximate calculation of the new
constants to order $\alpha_S N_c$.

\subsubsection{Book-keeping of ({\em a priori}) unknown constants}

The calculation of the effective action in the previous sections was
organized as a power series in proper time.

 In the presence of a
gluonic background, each term in the effective action,
which originates on a fixed power of the proper-time expansion of the heat
kernel, now becomes modulated by an infinite series in powers of
colour-singlet gauge-invariant combinations
of gluon field operators. Eventually,
we have to take the statistical gluonic average over each of these series.
In practice, each different average becomes an unknown constant. If we limit
ourselves to terms in the effective action to $O(p^4)$ at most, there can only
appear a finite number of these unknown constants. We can make their
book-keeping by tracing back all the possible different types of terms
that can appear.

In the presence of gluonic interactions, there then appear 10 unknown
constants:
$\gamma_{-1}$; $\gamma_{01}$, $\gamma_{02}$, $\gamma_{03}$; $\gamma_{11}$,
 $\gamma_{12}$,
$\gamma_{13}$, $\gamma_{14}$; $\gamma_{21}$, $\gamma_{22}$. To these, we have
to
 add the
original  $G_S$ and $G_V$ constants, as well as the scale $\Lambda_{\chi}$.
 However,
the unknown constant $(1+\gamma_{-1})$ in eq.~(\tref{170})
can be traded by an appropriate change of the scale $\Lambda_{\chi}$,
\be\rlabel{167}
\Gamma(-1,\tilde x)=\Gamma(-1,x)\{1+\gamma_{-1}\}\ ;\ \tilde
x={M^2_Q\over\tilde{\Lambda}^2_{\chi}}, \ee
and a renormalization of the constant $G_S$,
\be\rlabel{168}
G_S\to \tilde G_S={\tilde{\Lambda}^2_{\chi}\over\Lambda^2_{\chi}}G_S\ .
\ee

 Altogether, we then have 12 ({\em a priori} unknown) theoretical constants
and one scale $\Lambda_{\chi}$. They determine 18 non-trivial physical
couplings
(in the large-$N_c$ limit)
of the low-energy QCD effective Lagrangian:
 $<\bar{\psi} \psi>$, $f_{\pi}$, $L_1$, $L_3$, $L_5$, $L_8$,
$L_9$, $L_{10}$, $H_1$, $H_2$, $f_V$, $f_A$, $g_V$, $c_m$, $c_d$, $M_S$, $M_V$
and $M_A$.

In full generality, the results are:
\be\rlabel{170}
<\bar {\psi} \psi> =
-{N_c \over  16 \pi^2} 4\,M_Q^3 \Gamma_{-1}(1 + \gamma_{-1}),
\ee
\be\rlabel{171}
{1\over 4}f_{\pi}^2 = {N_c\over 16\pi^2} M_Q^2 g_A\Gamma_0(1 + \gamma_{01});
\ee
\vskip 1cm
\be\rlabel{172} L_2 = 2L_1 = {N_c \over 16\pi^2}{1\over 24}\times
\ee
\begin{displaymath}
\left[(1-g_A^2)^2\Gamma_0(1 + \gamma_{03})
+ 4g_A^2 (1-g_A^2) \Gamma_1(1 + {3 \over 2}\gamma_{12}-{1 \over 2}\gamma_{13})
+ 2g_A^4 \Gamma_2(1 + \gamma_{21})\right],
\end{displaymath}
\begin{displaymath}
 \tilde L_3 = {N_c \over 16\pi^2}{1\over 24}\left[-3(1-g_A^2)^2
\Gamma_0(1 + \gamma_{03}) + 4g_A^4 \Gamma_1(1 + \gamma_{13})\right.
\end{displaymath}
\be\rlabel{173}
\left.- 12 g_A^2 (1-g_A^2)\Gamma_1(1 + {3 \over 2}\gamma_{12}-{1 \over
 2}\gamma_{13})
- 8g_A^4\Gamma_2(1 + {1\over 2}(\gamma_{21}+\gamma_{22}))
\right] ,
\ee
\be\rlabel{174}
L^S_3={c^2_d\over 2M^2_S},
\ee
\be\rlabel{175}
L_5 = {N_c\over 16\pi^2} {1\over 4} g_A^3 {1 + \gamma_{01}\over 1 +
\gamma_{02}}
\left[\Gamma_0(1 + \gamma_{01})-\Gamma_1(1 + \gamma_{11})\right],
\ee
\be\rlabel{177}
L_8 = {N_c \over 16\pi^2} \left[{1\over 16}{1 + \gamma_{01}\over 1 +
 \gamma_{02}}
-{1\over 24}{\Gamma_1(1 + \gamma_{13})\over \Gamma_0(1 + \gamma_{01})}\right]
g_A^2\Gamma_0(1 + \gamma_{01}),
\ee
\be\rlabel{179}
L_9 = {N_c \over 16\pi^2}{1\over 6}\left[(1-g_A^2)
\Gamma_0(1 + \gamma_{03}) +
2 g_A^2\Gamma_1(1 + {3 \over 2}\gamma_{12}-{1 \over 2}\gamma_{13})\right],
\ee
\be\rlabel{180}
L_{10} = -{N_c \over16\pi^2}{1\over6}\left[(1-g_A^2)
\Gamma_0(1 + \gamma_{03})+
g_A^2\Gamma_1(1 + \gamma_{13})\right],
\ee
\be\rlabel{181} H_1 = -{N_c \over16\pi^2}{1\over12}\left[(1+g_A^2)
\Gamma_0(1 + \gamma_{03}) -
g_A^2\Gamma_1(1 + \gamma_{13})\right],
\ee
\be\rlabel{182}
\tilde H_2 = {N_c \over16\pi^2}{1 \over 12}\left[6 \rho^2
\Gamma_{-1}(1 + \gamma_{-1}) -
6\rho^2 \Gamma_0(1 + \gamma_{02}) -
6\rho g_A \Gamma_0(1 + \gamma_{01}) +
g_A^2\Gamma_1(1 + \gamma_{13})\right],
\ee
\be\rlabel{182bis}
H_2^S={c_m^2\over M^2_S};
\ee

\be\rlabel{183}
f_V = \sqrt 2 \lambda_V
\ee
and
\be\rlabel{184}
f_A = \sqrt 2 g_A \lambda_A,
\ee
 with
\be\rlabel{185}
\lambda_V^2 = {N_c \over 16\pi^2} {1 \over 3}\Gamma_0(1 + \gamma_{03})
\ee
 and
\be\rlabel{186}
\lambda_A^2 ={N_c \over 16\pi^2} {1 \over 3}
\left[\Gamma_0(1 + \gamma_{03})-
\Gamma_1(1 + \gamma_{13})\right].
\ee
\be\rlabel{187}
g_V = {N_c \over 16\pi^2} {1 \over \lambda_V} {\sqrt 2 \over 6}
\left[(1-g_A^2) \Gamma_0(1 + \gamma_{03}) +
2 g_A^2 \Gamma_1(1 + {3 \over 2}\gamma_{12}-{1 \over 2}\gamma_{13})\right],
\ee
\be\rlabel{188}
c_m = {N_c \over 16\pi^2} {M_Q \over \lambda_S} \rho
\left[\Gamma_{-1}(1 + \gamma_{-1}) -
2 \Gamma_0(1 + \gamma_{02})\right],
\ee
\be\rlabel{189}
c_d = {N_c \over 16\pi^2} {M_Q \over \lambda_S} 2 g_A^2\left[
\Gamma_0(1 + \gamma_{01})
-\Gamma_1(1 + \gamma_{11})\right],
\ee
 with
\be\rlabel{190}
\lambda_S^2 = {N_c \over 16\pi^2} {2 \over 3} \left[3
\Gamma_0(1 + \gamma_{01})
- 2\Gamma_1(1 + \gamma_{14}) \right];
\ee

\be\rlabel{191}
M_S^2 = {N_c \over 16\pi^2} {8M_Q^2  \over \lambda_S^2}
\Gamma_0(1 + \gamma_{02}),
\ee
\be\rlabel{192}
M_V^2 = {3 \over 2} {\Lambda_{\chi}^2 \over G_V(\Lambda_{\chi}^2)}
{1 \over \Gamma(0,x)(1 + \gamma_{03})},
\ee
\be\rlabel{193}
M_A^2\left\{1- {\Gamma(1,x)(1 + \gamma_{13}) \over
\Gamma(0,x)(1 + \gamma_{03})}\right\} = M_V^2 +
6 M_Q^2{1 + \gamma_{01} \over 1 + \gamma_{03}}.
\ee

There exist relations among the above physical couplings
which are independent
of the unknown gluonic constants. They are clean tests of the basic
assumption that the low-energy effective action of QCD follows from an ENJL
Lagrangian of the type considered here. The relations are
\be\rlabel{194}
f_V^2M_V^2 - f_A^2M_A^2 = f_{\pi}^2\,\,\,
(\hbox{first Weinberg sum rule}),
\ee

\be\rlabel{195}
L_9={1\over 2}f_V g_V\ ,
\ee
\be\rlabel{196}
L_{10}=-{1\over 4}f_V^2 + {1\over 4}f_A^2\ ,
\ee
\be\rlabel{197}
2H_1=-{1\over 4}f_V^2 - {1\over 4}f_A^2\ ,
\ee
 and
\be\rlabel{198}
{H_2+2L_8\over 2L_5}={c_m\over c_d}\ .
\ee

The first four relations have already been discussed in the previous
subsection.
The combination of couplings in the
r.h.s. of eq.~(\tref{198}) is the one that appears in the context
of non-leptonic weak interactions, when one considers weak decays
such as $K\to \pi H$ (light Higgs) \rcite{52}.
In fact, from the low-energy theorem derived in \rcite{GL}
it follows that
\be\rlabel{199}
{H_2+2L_8\over 2L_5}={1\over 4}{{<0|\bar ss|0>\over <0|\bar uu|0>}-1\over
 f_K/f_{\pi}-1}\ .
\ee

 Experimentally
\be\rlabel{200}
f_K/f_{\pi}-1=0.22\pm 0.01\ .
\ee

 Unfortunately, the numerator in the r.h.s. of (\tref{199})
is poorly known. If
we vary the ratio
\be\rlabel{201}
{<\bar ss>\over <\bar uu>}-1\quad\hbox{ from }\ -0.1\ \hbox{ to }\ -0.2\ ,
\ee
 as suggested by the authors of ref. \rcite{52}, then
eq.~(\tref{198}) leads to the estimate
\be\rlabel{202}
c_m/c_d=-1.1\times 10^{-1}\ \hbox{ to }\ -2.3\times 10^{-1}\ .
\ee
 With this estimate incorporated in eq.~(\tref{53}),
we are led to the
 conclusion that
\be\rlabel{203}
|c_d|\simeq 34~MeV\ .
\ee
 In the version corresponding to the first alternative,
the results for $c_m$ and $c_d$ are those in
eqs.~(\tref{127}) and (\tref{128}). We observe that in this case $c_m/c_d$
comes out always positive for reasonable values of $M^2_Q/\Lambda^2_{\chi}$.
In fact from the gap-equation discussed in section \tref{spontaneous} it
is obvious that $\qvev$ increases with increasing current quark mass for
not too high current masses.

\subsubsection{Gluonic correction to $O(\alpha_S N_c)$}

\quad We can make an estimate of the ten constants
$\gamma_{-1};$
$\gamma_{01},\gamma_{02},\gamma_{03};$ $\gamma_{11},\gamma_{12},
\gamma_{13},\gamma_{14};$ $\gamma_{21}$ and $\gamma_{22}$,
by keeping only the leading  contribution,
which involves the gluon vacuum condensate $ <{\alpha_S
\over \pi} GG>/ M_Q^4$ as was done in Ref.~\rcite{ERT}. The relevant
dimensionless parameter is

\be\rlabel{204}
g = {\pi^2 \over 6N_c}{<{\alpha_S \over \pi} GG> \over M_Q^4}.
\ee

 Notice that in the large-$N_c$ limit,
$g$ is a parameter of $O(1)$.
One should also keep in mind that the gluon average in (\tref{204})
is the one corresponding to fluctuations below
the $\Lambda_{\chi}$ scale. The relation
of $g$ to the conventional
gluon condensate that appears in the QCD sum rules \rcite{QCDsum,42}
is rather unclear. We are forced to consider $g$ as a free parameter.
Up to order $O(\alpha_S N_c)$,
this is the only unknown quantity which appears,
and we can express all the $\gamma$'s in terms of $g$. We find:
\be\rlabel{205}
\gamma_{-1} = {\Gamma(1,x) \over \Gamma(-1,x)}\,2\,g;
\ee
\be\rlabel{206}
\pmatrix{\gamma_{01}\cr \gamma_{02}\cr \gamma_{03}\cr}
= {\Gamma(2,x) \over \Gamma(0,x)}\pmatrix{1\cr 2\cr -3/5\cr}g;
\ee
\be\rlabel{207}
\pmatrix{\gamma_{11}\cr \gamma_{12}\cr \gamma_{13}\cr \gamma_{14}\cr}
= {\Gamma(3,x) \over \Gamma(1,x)}\pmatrix{1\cr 1/5\cr 3/5\cr 9/5\cr}g;
\ee
\be\rlabel{208}
\pmatrix{\gamma_{21}\cr \gamma_{22}\cr}
= {\Gamma(4,x) \over \Gamma(2,x)}\pmatrix{0\cr 2/5\cr}g.
\ee

 Notice that the combination ${3 \over 2}\gamma_{12} -
{1 \over 2}\gamma_{13}$ entering some of the $L_i$'s coupling constants
is zero. This is the reason why
it was found, in Ref.~\rcite{ERT},
that in the limit $g_A\to 1$, $L_2$ and $L_9$
have no gluon correction of $O(\alpha_sN_c)$.

To this approximation, we have then reduced the theoretical
parameters to three unknown constants
$G_S$, $G_V$ and g, and the scale $\Lambda_{\chi}$.

\subsection{Discussion of numerical results}

In the ENJL model, we have three input parameters:
\be\rlabel{222}
G_S\,\,,\,\,G_V\,\,\hbox{and}\,\,\Lambda_{\chi}.
\ee
 The gap equation introduces a constituent chiral
quark mass parameter  $M_Q$, and the ratio
\be\rlabel{223}
x = {M_Q^2 \over \Lambda_{\chi}^2}
\ee
 is constrained to satisfy the equation
\be\rlabel{224}
{1\over G_S} = x \Gamma(-1,x)(1+\gamma_{-1}).
\ee
 Once $x$ is fixed, the constants $g_A$ and $G_V$ are related by
the equation
\be\rlabel{225}
g_A = {1 \over 1 + 4 G_V x \Gamma(0,x)(1+\gamma_{01})}.
\ee
 Therefore, we can trade $G_S$ and $G_V$ by
$x$ and $g_A$;
but we need an observable to fix the scale $\Lambda_{\chi}$. This
 is the scale
which determines the $\rho$ mass in eq.~(\tref{192}), i.e.
\be\rlabel{226}
\Lambda_{\chi}^2 = {2 \over 3} M_V^2 G_V \Gamma(0,{M_Q^2 \over
 \Lambda_{\chi}^2})(1+\gamma_{03}).
\ee

There are various ways one can proceed. We find it useful to fix as input
 variables the values of
$M_Q$, $\Lambda_{\chi}$ and $g_A$. Then we have predictions for
\be\rlabel{227}
f_{\pi}^2\,\,\, , \,\,\, <\bar{\psi} \psi>
\ee
\be\rlabel{228}
M_S\,\,\, , \,\,\, M_V \,\,\, \hbox{and} \,\,\, M_A
\ee
\be\rlabel{229}
f_V\,\, , \,\,g_V\,\, \hbox{and} \,\,f_A\,\,\,;\,\,\, \hbox{and}
\,\,\,c_m\,\,\,
 \hbox{and} \,\,\,c_d
\ee
 and the $O(p^4)$ couplings:
\be\rlabel{230}
L_i\,\,\,(i=1,2,\ldots,10)\,\,\, \hbox{and} \,\,\,H_1\,\,\, , \,\,\,H_2.
\ee
 In principle we can
also calculate any higher-$O(p^6)$\rcite{Scherer} coupling which
 may become of
interest. So far, we have fixed twenty-two parameters.
Eighteen of them are
 experimentally known.

In the first column of Table \tref{table2}
we have listed the experimental values of
the parameters which we consider.
In comparing with the predictions of the ENJL model,
it should be kept in mind that the relations (\tref{194}) to (\tref{196}).
are satisfied by the model while
\be
L_1 =\frac{1}{8}g_V^2;\quad L_2 = 2L_1 = \frac{1}{4}g_V^2;
\quad L_3 = -3L_2
\ee
only have numerically small corrections. These relations are rather well
satisfied by the experimental values and thus constitute a large part
of the numerical success of the model.

We have also used the predictions leading in $1/N_c$, so that we have
$ L_1 = L_2 /2$, $L_4 =  L_6 = 0$,
and we do not consider $L_7$ since this is
given mainly by the $\eta^\prime$ contribution \rcite{GL}.
In evaluating the  predictions given in Table \tref{table2},
we have used the full expressions for the incomplete gamma functions
and the numerical value of the $\gamma_{ij}$ in terms of $g$ given in
eqs.~(\tref{205}) to (\tref{208}).

The first column of errors in Table \tref{table2}
shows the experimental ones. The second column
gives the errors we have used for the fits. When no error is indicated in this
column, it means
that we never use the corresponding parameter for fitting. This is the
case for $<\overline{q}q>$, which is quadratically divergent in the
cut-off and which is not very well known experimentally. This is also the
case for $c_m$, which depends on $<\overline{q}q>$.
Fit 1 corresponds to a least-squares fit with
the maximal set of parameters
and requiring $g \ge 0$. Fit 2 corresponds to a fit
where only $f_\pi$ and the $L_i$ are used as input in the fit,while
fit 3 has the vector and scalar mass as additional input.
The next column, fit 4, is the one where we require $g_A = 1$, i.e.
we start with a model without the vector four-quark interaction. Here
there are no explicit
vector (axial) degrees of freedom, so those
have been dropped in this case.
This fit includes
all parameters except $M_V$, $M_A$, $f_V$, $g_V$
and $f_A$. Finally, fit 5 is the fit to all data, keeping the gluonic
parameter $g$ fixed at a value of 0.5. The main difference with fit 1 is
a decrease in the value of $M_Q$. The value of $\Lambda_\chi$
changes very little.
In addition the result with the constraint $G_S=4 G_V$,
\rref{GSGV4}, included is shown as fit 6.
Fit 7 is the result without gluonic corrections and $G_V = 0$ as suggested
by \rcite{Zakharov}.

The expected value for the parameter $g$, if we take typical values
from, e.g, QCD sum rules, is of $O(1)$. None of the fits here really
makes a qualitative difference between a $g$ of about $0.5$ to $0$.
Numerically we can thus not decide between the two alternatives mentioned
in the introduction. This can be easily seen by comparing fit 1 and fit 5,
or fit 4 and fit 7,
in Table \tref{table2}.

In all cases acceptable predictions for all relevant parameters are
possible. The scalar sector parameters tend all to be a bit on the low
side; but so is the constituent quark mass. The
predictions for the $L_i$'s are reasonably stable versus a variation of
the input parameters.
For $L_5$ and $L_8$,
this is a major improvement as compared with the predictions
of the mean field approximation \rcite{ERT}.
The typical variation with input parameters can be seen in table 2
of \rcite{BBR}.

\begin{table}[htb]
\begin{center}
\begin{tabular}{|c|c|c|c|c|c|c|c|c|c|c|}
\hline
    & exp. & exp.  & fit   & fit 1 & fit 2 & fit 3 & fit 4&fit 5&fit 6&fit 7\\
    & value& error & error &       &       &       &       & & &      \\
\hline
\hline
$f_\pi$ & 86(${}^{\dagger}$) &  $-$  & 10     &  89  &  86  & 86    &  87&83&
86&86\\
\hline
$\sqrt[3]{-<\overline{q}q>}$&235(${}^{\#}$)&15(${}^{\#}$)&$-$&281& 260 &  255
& 178 & 254 &210&170\\
\hline
$ 10^3 \cdot L_2$ & 1.2  &  0.4 &0.5  & 1.7   &  1.6 & 1.6 & 1.6 &1.7&1.5&1.6\\
$ 10^3 \cdot L_3$ &$-3.6$&  1.3 &1.3  & $-4.2$&$-4.1$&
$-4.4$&$-5.3$&$-4.7$&$-3.1$&$-3.0$\\
$ 10^3 \cdot L_5$ &  1.4 &0.5   & 0.5 & 1.6   &  1.5 &  1.1
   &1.7&1.6&2.1&1.9\\
$ 10^3 \cdot L_8$ &  0.9 & 0.3  & 0.5 & 0.8   &  0.8 &  0.7    &1.1&1.0
 &0.9&0.8\\
$ 10^3 \cdot L_9$ &  6.9 &  0.7 &0.7  & 7.1   &  6.7 &  6.6   &5.8&7.1
&5.7&5.2\\
$ 10^3 \cdot L_{10}$&$-5.5$& 0.7 & 0.7&$-5.9$ &$-5.5$&$-5.8$&
$-5.1$&$-6.6$&$-3.9$&$-2.6$\\
$ 10^3 \cdot H_1$ & $-$  &$-$&$-$  & $-4.7$ &$-4.4$&$-4.0$&$-2.4$&
$-4.6$&$-3.7$&$-2.6$\\
$ 10^3 \cdot H_2$ & $-$  &$-$&$-$  & $ 1.4$ &$ 1.2$&$ 1.2$  &
 $ 1.0$&$ 2.3$&$-0.2$&0.8\\
\hline
$ M_V $&768.3 &0.5& 100 &    811  &  830 & 831     &  $-$ & 802&1260&$-$\\
$ M_A $& 1260&30&300 &     1331  & 1376 & 1609    &  $-$  & 1610&2010&$-$\\
$ f_V $& 0.20&(*)& 0.02&   0.18  & 0.17 & 0.17    &  $-$  & 0.18&0.15&$-$\\
$ g_V $&0.090&(*)& 0.009&    0.081  & 0.079&  0.079  & $-$ & 0.080&0.076&$-$\\
$ f_A $&0.097&0.022(*)& 0.022&   0.083  & 0.080& 0.068   &
$-$&0.072&0.084&$-$\\
\hline
$ M_S $&983.3&2.6&200&            617  & 620  &  709    & 989 & 657&643&760\\
$ c_m $& $-$&$-$& $-$ &           20  &  18  &  20     & 24  & 25&16&6\\
$ c_d $&  34 &(*)&  10&          21  &  21  &  18     & 23   & 19&26&27\\
\hline
\hline
$x$  & & & &              0.052  & 0.063 &  0.057 &  0.089 & 0.035&0.1&0.2\\
$g_A$& & & &               0.61  &  0.62 &  0.62  &  1.0   & 0.66&0.79&1.0\\
$M_Q$& & & &                265  & 263   &  246   &  199   & 204&262&282 \\
$g$  & & & &                0.0  &  0.0  &  0.25  &  0.58  & 0.5&0.0&0.0 \\
\hline
\end{tabular}
\end{center}
\caption{Experimental values and predictions of the ENJL model
for the various low-energy parameters discussed in the text.
All dimensionful quantities are in MeV. The difference
between the predictions is explained in the text.
The numerical error in \protect{\rcite{BBR}} for $H_2$ has been
corrected. All masses are determined from the low-energy expansion,
not the pole position of the 2-point functions.}
\rlabel{table2}
\end{table}

\section{Two-point functions}
\rlabel{twopointsect}
\rlabel{xtwop}

This section is a discussion of the results in Refs. \rcite{BRZ,BP2}
about two-point functions. These two-point functions were
studied before in \rcite{Vogl} but there they were discussed as quark
form factors. What is new here is that the explicit dependence on
the regularization scheme has been put into two arbitrary functions,
namely, $\ovpi_V^{(0)}+\ovpi_V^{(1)}$ and $\ovpi^P_M$
(see this section below for definitions). This
also shows that these results are valid in a class of models where the
one-loop result can be expanded
in a heat-kernel expansion using the same basic quantities $E$ and $R_{\mu\nu}$
as used here. This includes the ENJL model with low-energy gluons described
by background expectation values.

The two-point functions are of course important quantities and have played
historically an important role in understanding the high-energy behaviour
of the strong interaction\rcite{40,Floratos}.
In addition some of the consequences for the
mesonic sector were also valid in the low-energy expansion of the ENJL
model as discussed in the previous section. Here we would like to
study the two-point functions directly in the ENJL model to all
orders in the current quark masses and momenta. This method was
developed for the chiral limit case in \rcite{BRZ} and then extended to
include nonzero quark masses in \rcite{BP2}. The discussion here follows the
 latter reference closely.

\subsection{Definition of the two-point functions}

We shall discuss two--point functions of the vector,
axial--vector, scalar and pseudoscalar quark currents
with the following definitions,
\begin{eqnarray}
\rlabel{x1}
V^{ij}_\mu (x)&\equiv& \bar q_i(x)\gamma_\mu
q_j(x)\, ,
\\
\rlabel{x2}
A^{ij}_\mu (x)&\equiv& \bar q_i(x)\gamma_\mu\gamma_5
q_j(x)\, ,
\\
\rlabel{x3}
S^{ij} (x)&\equiv& -\, \bar q_i(x) q_j(x)\, ,
\\
\rlabel{x4}
P^{ij} (x)&\equiv& \bar q_i(x)\, i\gamma_5 q_j(x) \, ,
{}~.
\end{eqnarray}
The indices $i,j$ are flavour indices and run over $u,d,s$. The two-point
functions themselves are defined as
\ba
\rlabel{x9}
{ \Pi}^V_{\mu\nu} (q)_{ijkl}&=& i\int {\rm d}^4x e^{iq\cdot x}
<0|T\left( V_\mu^{ij} (x)V_\nu^{kl}(0)\right) |0> \, ,
\\
\rlabel{x10}
{ \Pi}^A_{\mu\nu} (q)_{ijkl}&=& i\int {\rm d}^4x e^{iq\cdot x}
<0|T\left( A_\mu^{ij} (x)A_\nu^{kl}(0)\right) |0>\, ,
\\
\rlabel{x11}
{ \Pi}^S_{\mu} (q)_{ijkl}&=& i\int {\rm d}^4x e^{iq\cdot x}
<0|T\left( V_\mu^{ij} (x)S^{kl}(0)\right) |0>\, ,
\\
\rlabel{x12}
{     \Pi}^P_{\mu} (q)_{ijkl}&=& i\int {\rm d}^4x e^{iq\cdot x}
<0|T\left( A_\mu^{ij} (x)P^{kl}(0)\right) |0>\, ,
\\
\rlabel{x13}
{     \Pi}^S (q)_{ijkl}&=& i\int {\rm d}^4x e^{iq\cdot x}
<0|T\left( S^{ij} (x)S^{kl}(0)\right) |0>\, ,
\\
\rlabel{x14}
{     \Pi}^P (q)_{ijkl}&=& i\int {\rm d}^4x e^{iq\cdot x}
<0|T\left( P^{ij} (x)P^{kl}(0)\right) |0>\ .
\ea
In the leading order in the number of colours these are all
proportional to $\delta_{ijkl} \equiv
 \delta_{il}\delta_{jk}$, with $\delta_{il}$
the Kronecker delta. Using Lorentz-invariance these functions can then
be expressed as follows
\ba
\rlabel{x15}
{     \Pi^V_{\mu\nu} (q)_{ijkl}}&=& \left\{
(q_\mu q_\nu -q^2g_{\mu\nu}){ \Pi_V^{(1)}}(Q^2)_{ij} +
q_\mu q_\nu {\Pi_V^{(0)}}(Q^2)_{ij} \right\}
\delta_{ijkl} \, ,
\\
\rlabel{x16}
{\Pi^A_{\mu\nu} (q)_{ijkl}}&=& \left\{ (q_\mu q_\nu -q^2g_{\mu\nu}){
   \Pi_A^{(1)}}(Q^2)_{ij} + q_\mu q_\nu {\Pi_A^{(0)}}(Q^2)_{ij}
\right\} \delta_{ijkl} \, ,
\\
\rlabel{x17}
{     \Pi^S_{\mu} (q)_{ijkl}}&=& q_\mu \, {     \Pi^M_
{S}}(Q^2)_{ij} \delta_{ijkl} \, ,
\\
\rlabel{x18}
{     \Pi^P_{\mu} (q)_{ijkl}}&=& i q_\mu \,
{     \Pi^M_{P}}(Q^2)_{ij} \delta_{ijkl} \, ,
\\
\rlabel{x19}
{     \Pi^S(q)_{ijkl}}&=& {\Pi_{S}}(Q^2)_{ij}\delta_{ijkl} \, ,
\\
\rlabel{x20}
{     \Pi^P(q)_{ijkl}}&=& {     \Pi_
{P}}(Q^2)_{ij}\delta_{ijkl}\ .
\ea
Here $Q^2 = -q^2$. We shall discuss the Weinberg Sum Rules
and numerical results
for the two-point functions only in the Euclidean domain,
i.e. $Q^2$ positive.
Using Bose symmetry on the definitions of the two-point functions
it follows
that $\Pi^{(0)}_{V}(Q^2)_{ij}$,
$\Pi^{(1)}_{V}(Q^2)_{ij}$,
$\Pi^{(0)}_{A}(Q^2)_{ij}$,
$\Pi^{(1)}_{A}(Q^2)_{ij}$,
$\Pi_{S}(Q^2)_{ij}$ and
$\Pi_{M}(Q^2)_{ij}$ are all symmetric in the flavour indices $i$ and $j$.
The remaining ones need the Ward-identities to prove
their flavour structure. From the
identities in the appendix \tref{xAppB}
it follows that $\Pi^M_S(Q^2)_{ij}$
is also symmetric in $i,j$; while
$\Pi^M_S(Q^2)_{ij}$ is anti-symmetric.

\subsection{Lowest order results in Chiral
Perturbation Theory} From Chiral Perturbation
Theory to order $p^4$ in the expansion
we obtain the following
low energy results for the two-point functions. The orders mentioned behind
are the orders in Chiral Perturbation Theory that are neglected.
\ba
\rlabel{x21}
{    \Pi_V^{(1)}}(Q^2)_{ij} &=& -4(2H_1 + L_{10}) + {\cal O}(p^6) \, ,
\\
\rlabel{x22}
{    \Pi_V^{(0)}}(Q^2)_{ij} &=& {\cal O}(p^6)\, ,
\\
\rlabel{x23}
{\Pi_A^{(1)}}(Q^2)_{ij}&=& {2f_{ij}^2\over Q^2}-4(2H_1 - L_{10}) +
{\cal O}(p^6)\, ,
\\
\rlabel{x24}
{\Pi_A^{(0)}}(Q^2)_{ij}&=& 2f_{ij}^2 \left(\frac{1}{m_{ij}^2
+ Q^2} - \frac{1}{Q^2}\right) + {\cal O} (p^6)\, ,
\\
\rlabel{x25}
{    \Pi^M_S    }(Q^2)_{ij}&=& {\cal O}(p^6) \, ,
\\
\rlabel{x26}
{    \Pi^M_P    }(Q^2)_{ij}&=& {2B_0f_{ij}^2\over
m_{ij}^2+Q^2}+{\cal O}(p^6) \, ,
\\
\rlabel{x27}
{    \Pi_S    }(Q^2)_{ij}&=& 8B_0^2(2L_8 + H_2) + {\cal O}(p^6) \, ,
\\
\rlabel{x28}
{\Pi}_P (Q^2)_{ij}&=& {\dis 2B_0^2 f_{ij}^2 \over  m_{ij}^2 + Q^2}
+ 8B_0^2(-2L_8+H_2) + {\cal O} (p^6)\ .
\ea
With $m_{ij}$ the mass of the lightest pseudoscalar meson with flavour
structure $ij$.
These are obtained in the leading $1/N_c$ approximation so loop-effects are
not needed. Notice that these expressions
are valid to chiral order $p^4$. From a term of the
form ${\rm tr}\{D_\mu\chi D^\mu\chi^\dagger\}$ there
are contributions of order $(m_i-m_j)^2/Q^2$  to the vector two-point function
$\Pi^{(0)}_V(Q^2)_{ij}$
and of order $(m_i-m_j)$ to the mixed scalar vector function
$\Pi^M_S(Q^2)_{ij}$.

The functions ${\Pi_A^{(0)}}$, ${\Pi^M_P}$ and ${
\Pi_P}$
get their leading behaviour from the pseudoscalar Goldstone pole.
In addition $\Pi_A^{(1)}$ and $\Pi_A^{(0)}$ contain a kinematical
pole at $Q^2=0$.
The residue of the physical
pole is proportional to the decay constant $f_{ij}$
for the relevant meson,
(for the $\bar u d$ ones, $f_{ud}\simeq f_\pi \simeq 92.5$ MeV).
In $\chi$PT, the constant $B_0$ is related to the vacuum
expectation value in the chiral limit.
In the large $N_c$ limit and away from the chiral limit there are
corrections due to the terms proportional to combination of
${\cal O}(p^4)$ couplings $2L_8+H_2$ \rcite{GL}.
\be\rlabel{x29}
<0|:{\overline \Psi} \Psi :|0>_{\left|\Psi=u,d,s\right.}\equiv
-f_0^2 B_0\, \left ( 1+{\cal O}(p^4)\right)\ .
\ee
The vacuum expectation value here, $<0|: {\overline \Psi} \Psi : |0>$,
is the one used in $\chi$PT in the chiral limit and $f_0$ is the
pseudoscalar meson decay constant in the chiral limit.
The constants $L_8$, $L_{10}$, $H_1$ and $H_2$ are coupling constants
of the ${\cal O}(p^4)$ effective chiral Lagrangian in the notation of
Gasser and Leutwyler \rcite{GL}, section \tref{hadronic}.
The constants $L_8$ and $L_{10}$
are known from the comparison
between $\chi$PT and low energy hadron phenomenology.
At the scale of the $\rho$ meson mass they are
$L_8=(0.9\pm 0.3)\times 10^{-3}$  and
$L_{10}=(-5.5\pm 0.7)\times 10^{-3}$.
The high energy constants
 $H_1$ and $H_2$ correspond to couplings which
involve external source fields only and therefore can only
be extracted from experiment given a prescription.

\subsection{The method and Ward identities}

The method used here is identical to the one used in \rcite{BRZ}.
The full two-point functions are the sum of diagrams like those in
figure \tref{xFig2pt}a.
The one-loop two-point functions are those obtained by the graph in
figure \tref{xFig2pt}b.
\begin{figure}
\begin{center}
%
%
%
\thicklines
\setlength{\unitlength}{1mm}
\begin{picture}(140.00,50.00)
\put(97.50,35.00){\oval(15.00,10.00)}
\put(103.00,33.50){$\bigotimes$}
\put(17.50,35.00){\oval(15.00,10.00)}
\put(25.00,35.00){\circle*{2.00}}
\put(32.50,35.00){\oval(15.00,10.00)}
\put(40.00,35.00){\circle*{2.00}}
\put(47.50,35.00){\oval(15.00,10.00)}
\put(55.00,35.00){\circle*{2.00}}
\put(62.50,35.00){\oval(15.00,10.00)}
\put(08.00,33.50){$\bigotimes$}
\put(68.00,33.50){$\bigotimes$}
\put(88.00,33.50){$\bigotimes$}
\put(38.50,19.00){(a)}
\put(95.50,19.00){(b)}
\put(14.50,40.00){\vector(1,0){3.00}}
\put(29.50,40.00){\vector(1,0){3.50}}
\put(44.00,40.00){\vector(1,0){5.00}}
\put(60.50,40.00){\vector(1,0){3.00}}
\put(95.50,40.00){\vector(1,0){5.00}}
\put(99.00,30.00){\vector(-1,0){3.00}}
\put(64.00,30.00){\vector(-1,0){3.00}}
\put(49.50,30.00){\vector(-1,0){3.50}}
\put(34.00,30.00){\vector(-4,1){2.00}}
\put(18.00,30.00){\vector(-1,0){2.50}}
\end{picture}
\caption{The graphs contributing to the two point-functions
in the large $N_c$ limit.
a) The class of all strings of constituent quark loops.
The four-fermion vertices are either
${\cal L}^{\rm S,P}_{\rm NJL}$ or
${\cal L}^{\rm V,A}_{\rm NJL}$ in eqs. \protect{\rref{LSP}}
and \protect{\rref{LVA}}.
The crosses at both ends are the insertion of the external sources.
b) The one-loop case.}
\rlabel{xFig2pt}
\end{center}
\end{figure}
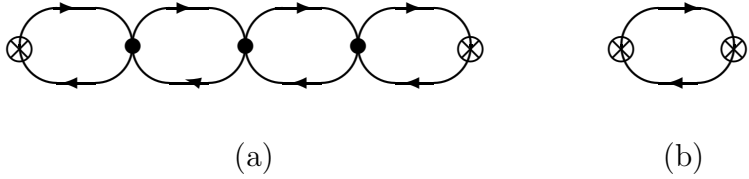
Using a recursion formula that relates the
$n$-loop graph to a product of the one-loop and the ($n$--1)-loop graph
and the relevant combination of kinematic factors and $G_V$ and $G_S$ the
whole class of graphs can be easily summed. Some care must be taken in the
case where different two-point functions can mix so a matrix
inversion is necessary (see ref. \rcite{BRZ}).

The two-point functions defined above satisfy the following
Ward identities.
(We suppress the argument $Q^2$ for brevity.)
\ba
\rlabel{xward1}
-Q^2{\Pi^{(0)}_V}_{ij} &=& \left( m_i - m_j \right){\Pi_S^M}_{ij} \, ,
\\
\rlabel{xward2}
-Q^2{\Pi_S^M}_{ij} &=& \left( m_i - m_j \right){\Pi_S}_{ij}
+\langle \overline{q}_i q_i\rangle - \langle \overline{q}_j q_j
\rangle \, ,
\\
\rlabel{xward3}
-Q^2{\Pi^{(0)}_A}_{ij} &=& \left( m_i + m_j \right){\Pi_P^M}_{ij} \, ,
\\
\rlabel{xward4}
-Q^2{\Pi_P^M}_{ij} &=& \left( m_i + m_j \right){\Pi_P}_{ij}
+\langle \overline{q}_i q_i\rangle + \langle \overline{q}_j q_j
\rangle \, .
\ea
These are derived in the appendix \tref{xAppB}. From these the
flavour symmetry of
the mixed two-point functions can be derived from the vector ones.

The one-loop expressions, which we shall
denote by $\overline{\Pi}$ and use
further the same conventions as given for the full ones above
are given in appendix \tref{xAppC}. They
satisfy the same identities but with the current quark masses $m_i$ replaced
by the constituent ones, $M_i$. In addition to these, there are two
more relations that follow in general if the one-loop part can be
described by a heat-kernel expansion in terms of the quantities $E$ and
$R_{\mu\nu}$ of appendix \tref{xAppB}.
These identities are (with the flavour subscript $ij$ and argument
suppressed)
\ba
\rlabel{xward5}
\overline{\Pi}^{(1)}_V +  \overline{\Pi}^{(0)}_V &=&
\overline{\Pi}^{(1)}_A +  \overline{\Pi}^{(0)}_A \, ,
\\ \rlabel{xward6}
\overline{\Pi}_S + Q^2 \overline{\Pi}^{(0)}_V &=&
\overline{\Pi}_P + Q^2 \overline{\Pi}^{(0)}_A \, .
\ea

Let us show in the most simple case\rcite{brijuni} how there are general
consequences of this approach. We will derive here the relation
between the scalar mass and the constituent quark mass
in the chiral limit. The set of diagrams
that contributes is drawn in Fig. \ref{xFig2pt}a. The series can be rewritten
as a geometric series and can be easily summed in terms of the one-loop
2-point function $\overline{\Pi}_S$. The full result for the scalar-scalar
two-point function (we only treat the case with equal masses here, see
\cite{BP2} for the general case) is:
\begin{equation}
\Pi_S = \frac{\overline{\Pi}_S}{1- g_S \overline{\Pi}_S}\ .
\end{equation}
The resummation has generated a pole that corresponds to a scalar particle.
Can we say more already at this level?

We can in fact. The Ward identities for the one loop functions become:
\begin{eqnarray}
\label{WI1}
\ovpi_S &=& \ovpi_P - q^2\ovpi^{(0)}_A  \ ,\\
\label{WI2}
\ovpi_P &=& \frac{q^4}{4M_Q^2}\ovpi^{(0)}_A - \frac{\qvev}{M_Q}\ .
\end{eqnarray}
(\ref{WI1}) is a consequence of using the heat kernel for the one-loop
functions and (\ref{WI2}) is a direct consequence of the symmetry.
Using these two relations we can rewrite
\begin{equation}
1-g_S\ovpi_S = 1 + \frac{g_S}{M_Q} +
(q^2-4M_Q^2)\frac{q^2\ovpi^{(0)}_A}{4M_Q^2}
\ .
\end{equation}
The first two terms vanish due to the gap equation so this two-point function
has a pole at twice the constituent mass.
Let us now derive the general cases.

\subsection{The transverse vector sector}
\rlabel{xvector1}

We use here the abbreviations \rref{gsgv}.
The full resummed transverse vector two-point function is then
\be
\rlabel{xvector}
{\Pi^{(1)}_V}_{ij} =
\frac{\overline{\Pi}^{(1)}_{Vij}}
{1 + g_V \overline{\Pi}^{(1)}_{Vij}}\ .
\ee
This can be simply written in a form resembling the
one in the complete VMD limit
with couplings $f_S$, $f_V$ and $M_V$ depending on
$Q^2$ and flavour and defined by
\ba
\rlabel{xtransvec}
\Pi^{(1)}_V(Q^2)_{ij} &=& \frac{2f_S^2(Q^2)_{ij}}{Q^2} +
\frac{2 f_V^2(Q^2)_{ij} M_V^2(Q^2)_{ij}}{M_V^2(Q^2)_{ij} + Q^2} \, ,
\\
2f_S^2(Q^2)_{ij}&=& \frac{-Q^2\ovpi^{(0)}_V(Q^2)_{ij}}
{1-g_V \ovpi^{(0)}_V(Q^2)_{ij}} \, ,
\\
\rlabel{xfvmv}
2 f_V^2(Q^2)_{ij} M_V^2(Q^2)_{ij} &=& \frac{N_c\Lambda_\chi^2}
{8\pi^2 G_V} \frac{1}{1-g_V \ovpi^{(0)}_V(Q^2)_{ij}} \, ,
\\
2 f_V^2(Q^2)_{ij} &=& {\overline{\Pi}^{(0+1)}_V(Q^2)_{ij}}\ .
\ea
Where we have used the fact that (see appendices
 \tref{xAppB} and \tref{xAppC})
$\ovpi^{(0+1)}_V \equiv \ovpi^{(0)}_V + \ovpi_V^{(1)}$
has no pole at $Q^2 = 0$.
There is a correction here (in $\ovpi^{(0)}_V$)
due to the mixing with the scalar sector, which is allowed
by  the presence of explicit breaking of the vector symmetry
(see the scalar mixed sector subsection \tref{xscalar0}).
 For the diagonal case, this is defined
as $m_i = m_j$ or $M_i = M_j$,
$\overline{\Pi}^{(0)}_V$ vanishes and the formulas above
simplify very much.

The pole mass of the vector corresponds to the pole in
this two point function
or to the solution of
${\rm Re}\left( Q^2 + M_V^2(Q^2)_{ij} \right) = 0$.
Alternatively, one can define the VMD values for the vector
parameters ($f_V$ and $M_V$)
as the best parameters of a linear fit of
the inverse of $\Pi^{(1)}_V(Q^2)_{ij}-2f^2_S(Q^2)_{ij}/Q^2$.
These definitions have the advantage
that they are also valid for the Euclidean region ($Q^2>0$)
where the vector cannot decay into two constituent
quarks. See sections
on numerical applications \tref{xnumbers} and
Vector-Meson-Dominance  \tref{VMD} for further comments.

\subsection{The transverse axial-vector sector}
\rlabel{xaxialvector1}

The transverse
axial-vector two-point function derivation is also identical to
the one in ref. \rcite{BRZ,BP2}.
\be
\rlabel{xaxvector}
{\Pi^{(1)}_A}_{ij} =
\frac{\overline{\Pi}^{(1)}_{Aij}}{1 + g_V \overline{\Pi}^{(1)}_{Aij}}\ .
\ee
Using the identity \rref{xward5}
it can be seen that this has a pole
at $Q^2$=0 because $\ovpi^{(0)}_{A}$ has it. As can be seen from
the explicit expression and is proved in general
in appendix \tref{xAppB}, the
combination $\overline{\Pi}^{(0)}_V + \overline{\Pi}^{(1)}_V$ is regular
at $Q^2$ going to zero.  This again allows us
to separate the pole at $Q^2=0$ in a simple fashion.
\ba \rlabel{x46}
 {\Pi}_A^{(1)}(Q^2)_{ij}
& =&  {2f_{ij}^2(Q^2)\over Q^2}
 + {2f_A^2(Q^2)_{ij} M_A^2(Q^2)_{ij}\over M_A^2(Q^2)_{ij}+Q^2}\ ,
\\ \rlabel{x104}
f_{ij}^2(Q^2) &=& g_A(Q^2)_{ij}\bar f_{ij}^2(Q^2) \ ,
\\ \rlabel{x105}
2\bar f_{ij}^2(Q^2)&=&-Q^2 \overline{\Pi}^{(0)}_A(Q^2)_{ij}\ ,
\\
\left(g_A(Q^2)_{ij}\right)^{-1} &=&
1- g_V \overline{\Pi}^{(0)}_A(Q^2)_{ij}\ ,
\\
\rlabel{xfa}
2 f_A^2(Q^2)_{ij} M_A^2(Q^2)_{ij} &=&
\frac{N_c\Lambda_\chi^2}{8\pi^2 G_V} g_A(Q^2)_{ij} \, ,
\\
2 f_A^2(Q^2)_{ij} &=& g_A^2(Q^2)_{ij}
\ovpi^{(0+1)}_V(Q^2)_{ij}\ .
\ea
There is a correction here (in $\ovpi^{(0)}_A$)
due to the mixing with the pseudo-scalar sector due to
the presence of both spontaneous and
explicit breaking of the axial-vector symmetry
(see the pseudo-scalar mixed sector subsection).
For further
 discussion of these expressions and the ones in the previous
section we refer to the subsection \tref{xWSRsec} on
Weinberg Sum Rules.

\subsection{The pseudo-scalar mixed sector}
\rlabel{xpseudo0}

The results obtained in \rcite{BRZ,BP2} are, with
the summed functions given in terms of the function
$\Delta_P(Q^2)$ and the one loop two-point functions
(with flavour subscripts $ij$ suppressed),
\ba
\rlabel{x131}
 {    \Pi}_A^{(0)}(Q^2)&=& {1\over \Delta_P(Q^2)} \left[
(1-g_S \overline{ \Pi}_P(Q^2))\overline{\Pi}_A^{(0)}(Q^2)
+ g_S{(\overline{\Pi}_P^M(Q^2))}^2
\right] \ ,
\\
\rlabel{x132}
{\Pi}_P^M(Q^2)&=&{1\over \Delta_P (Q^2)}
{\overline\Pi}_P^M(Q^2)\ ,
\\
\rlabel{x133}
 {    \Pi}_P(Q^2)&=& {1\over \Delta_P (Q^2)} \left[
(1-g_V\overline{    \Pi}_A^{(0)}(Q^2))\overline{\Pi}_P (Q^2)+
 g_V{(\overline{    \Pi}^M_P(Q^2))}^2 \right] \, ,\\
\rlabel{x134}
\Delta_P (Q^2)&=&\left( 1-g_V\overline{    \Pi}_A^{(0)}
(Q^2)\right)
 \left( 1 -g_S\overline{    \Pi
}_P(Q^2)\right)   - g_Sg_V{\left( \overline{    \Pi}^M_P
(Q^2)\right) }^2 .
\ea
Using the identities for the one-loop case it can be shown that the resummed
ones satisfy the Ward identities of appendix \tref{xAppB} with the
current quark masses. To show this it is also necessary to use the
Schwinger-Dyson equation for the constituent quark masses in
eq.\rref{gap}.

In order to rewrite this in terms of a nicer notation
we first express $\Delta_P(Q^2)_{ij}$
in a different form using the identities for the one-loop
two-point functions.
\ba
\Delta_P(Q^2)_{ij} &=& \frac{g_S \ovpi^M_P (Q^2)_{ij}}{M_i+M_j}
\left( m_{ij}^2(Q^2) + Q^2\right)
\\
\rlabel{xmpi}
{\rm with } \hspace*{0.2cm}
m_{ij}^2(Q^2) &\equiv& \frac{\left(m_i+m_j\right)}
{g_S g_A(Q^2) \ovpi^M_P(Q^2)_{ij}}\ .
\ea
Inserting the definition of $f_{ij}^2(Q^2)$ and
$1/g_S = -\langle \overline{q}_iq_i\rangle/\left(M_i-m_i\right)$ we
recover the Gell-Mann--Oakes--Renner (GMOR) relation for the pion mass
\rcite{Gell} when eq. \rref{xmpi} is expanded in powers of $m_i$.
For further discussion on corrections to the GMOR relation
in this model we refer to the section on numerical applications
\tref{xnumbers}.
Formula \rref{xmpi} gives
the expression for the pole due to the lightest pseudoscalar
mesons in the presence of explicit
chiral symmetry breaking.

This then allows us to
rewrite the full two-point functions in a very simple fashion:
\ba
\rlabel{xpipa0}
\Pi^{(0)}_A(Q^2)_{ij} &=& 2 f_{ij}^2(Q^2)
\left(\frac{1}{m_{ij}^2(Q^2)+Q^2}
 - \frac{1}{Q^2}\right) \, ,
\\ \rlabel{xpipm}
\Pi_P^M(Q^2)_{ij} &=& \frac{M_i + M_j}{g_S}\frac{1}
{m_{ij}^2(Q^2)+Q^2} \, ,
\\ \rlabel{xpip}
\Pi_P(Q^2)_{ij} &=& -\frac{1}{g_S} + \frac{\left(M_i+M_j\right)^2}
{2 f_{ij}^2(Q^2)} \frac{1}{g_S^2}\frac{1}
{m_{ij}^2(Q^2)+Q^2}\ .
\ea

Here we want to point out that the two-point functions $\Pi^M_P$
and
$\Pi_P$ suffer from the same ambiguity (via its dependence on $g_S$)
as the quark-antiquark one point-function (see discussion at the end
of section \tref{QCD}) when compared with the $\chi$PT results.

\subsection{The scalar mixed sector}
\rlabel{xscalar0}

This can be done in the same way as in the previous subsection with the
result (with flavour subscripts $ij$ suppressed)\rcite{BP2},
\ba
\rlabel{x62}
{\Pi}_V^{(0)}(Q^2)&=& {1\over \Delta_S (Q^2)} \left[
(1-g_S
\overline{\Pi}_S(Q^2))\overline{\Pi}_V^{(0)}(Q^2)
+ g_S{(\overline{\Pi}_S^M(Q^2))}^2
\right] \ ,
\\ \rlabel{x63}
{\Pi}_S^M(Q^2)&=&{1\over \Delta_S (Q^2)}  {\overline\Pi}_S^M(Q^2)\ ,
\\ \rlabel{x64}
 {\Pi}_S(Q^2)&=& {1\over \Delta_S (Q^2)} \left[
(1-g_V\overline{\Pi}_V^{(0)}(Q^2))\overline{\Pi}_S (Q^2) +
 g_V{(\overline{\Pi}^M_S(Q^2))}^2
\right] \ , \\ \rlabel{x65}
\Delta_S (Q^2)&=&\left( 1-g_V\overline{\Pi}_V^{(0)}(Q^2)\right)
 \left( 1 -g_S\overline{\Pi}_S(Q^2)\right)   - g_S g_V{\left(
\overline{\Pi}^M_S(Q^2)\right) }^2 .
\ea
To rewrite this in a simple fashion we would again like to expand
$\Delta_S$ in a simple pole like fashion. Using the identities
for the one-loop
two-point functions this can almost be done, we obtain
\ba
\rlabel{xdeltas}
\Delta_S(Q^2)_{ij}&=&\frac{g_S\ovpi_P^M(Q^2)_{ij}}{M_i+M_j}
\left(\left(M_i+M_j\right)^2 + g_A(Q^2)_{ij} m_{ij}^2(Q^2)
+ Q^2 \right)
\nonumber\\\hspace*{0.5cm} &+&
\ovpi^{(0)}_V(Q^2)_{ij}\left( Q^2
g_S - g_V \frac{m_i-m_j}{M_i-M_j}\right)\ .
\ea
It can be seen that in the diagonal case a simple
expression for the scalar meson pole can be found,
\be
\rlabel{xms}
\left.{M_S^2(-M_S^2)}\right|_{m_i=m_j} =
(M_i+M_j)^2 + g_A(-M_S^2)_{ii}  m_{ii}^2(-M_S^2)\ .
\ee
The expression for the scalar two-point function $\Pi_S (Q^2)$
is in this  case
\ba \rlabel{xpisca}
\left.{\Pi_S(Q^2)}\right|_{m_i=m_j}
&=& \left\{
-\frac{1}{g_S} + \frac{g_A(Q^2)_{ij} \left(M_i+M_j\right)^2}
{2 f_{ij}^2(Q^2)} \frac{1}{g_S^2}\frac{1}{M_S^2(Q^2)+Q^2}\right\}
_{m_i=m_j} \ .
\ea
So in the
diagonal case a simple relation between the scalar mass, the constituent
masses and the pseudoscalar mass remains valid to all orders in the masses.
In this case $\Pi_V^{(0)}=\Pi^M_S=0$.

For the off-diagonal case, i.e. $m_i \ne m_j$,
 the corresponding expressions
for $\Pi_V^{(0)}$, $\Pi_S^M$ and $\Pi_S$ can be obtained from eqs.
\rref{x62}-\rref{x65} and the explicit $\ovpi$ functions in appendix
\tref{xAppC}.
There is a small shift in the pole compared to eq. \rref{xms} for the case
$m_i\ne m_j$. From appendix \tref{xAppC}, in
eq. \rref{xpiv0}, it can be seen that
$\ovpi^{(0)}_V$ itself has a zero close to a value
 of $Q^2 = M_S^2$ of eq. \rref{xms}.  In addition
$\ovpi^{(0)}_V$ is suppressed by $\left(M_i-M_j\right)^2/Q^2$.
Therefore the value of the pole in the off-diagonal case is not too
far from that in eq. \rref{xms}.

Here we want to point out that (as in the mixed pseudoscalar sector)
the two-point functions $\Pi_V^{(0)}$, $\Pi^M_S$ and
$\Pi_S$ suffer from the same ambiguity (via its dependence on $g_S$)
as the quark-antiquark one point-function (see discussion at the end
of section \tref{QCD}) when compared with the $\chi$PT results.

\subsection{Weinberg Sum Rules}
\rlabel{xWSRsec}

The Weinberg Sum Rules are general restrictions on the short-distance behaviour
of various two-point functions \rcite{40}.
They were first discussed within QCD in ref.
\rcite{Floratos}. A low-energy model of QCD should have a behaviour at
intermediate energies that matches on reasonably well with the QCD behaviour.
The general behaviour should be ($\Pi_{LR} \equiv \Pi_V - \Pi_A$.)
\ba
\rlabel{xwsr1}
\lim_{Q^2\to\infty}\left(Q^2 \Pi^{(0+1)}_{LR}(Q^2)\right) = 0
&&{\rm First WSR} \,,
\\ \rlabel{xwsr2}
\lim_{Q^2\to\infty}\left(Q^4 \Pi^{(1)}_{LR}(Q^2)\right) = 0&&
{\rm Second WSR} \, ,
\\ \rlabel{xwsr3}
\lim_{Q^2\to\infty}\left(Q^4 \Pi^{(0)}_{LR}(Q^2)\right) = 0&&
{\rm Third WSR}\ .
\ea
Let us review first the QCD behaviour of these Sum Rules.
In  the large $N_c$ limit the three WSRs are
theorems of QCD in the chiral limit (i.e., ${\cal M} \to 0$).
The first WSR is still fulfilled in the large $N_c$ limit with
non-vanishing current quark masses. However the second and the
third ones are violated as follows \rcite{Pascual},
\ba
\lim_{Q^2\to\infty}\left(Q^4 \Pi^{(1)}_{LR}(Q^2)\right) &=&
- \lim_{Q^2\to\infty}\left(Q^4 \Pi^{(0)}_{LR}(Q^2)\right)
\nonumber \\ &=&
 2 \left( m_i\langle \bar q_j q_j  \rangle +
 m_j \langle \bar q_i q_i \rangle \right) \ .
\ea

As shown in \rcite{BRZ} the class of ENJL-like models does satisfy the
three WSRs in the chiral limit.
We shall now check how well this does in the case
of explicit breaking of chiral symmetry.

The high-energy behaviour  of the two-point functions
$\Pi^{(0,1)}_{V,A}$ needed for the three WSRs can be easily
obtained from the
expressions in sections \tref{xvector1},
\tref{xaxialvector1}, \tref{xpseudo0} and \tref{xscalar0}.
The first and second WSRs are satisfied in these ENJL-like models
even with non-vanishing and all different current quark-masses.
The  high energy behaviour ($Q^4$)
of these models is thus too strongly
suppressed for $\Pi^{(1)}_{LR}(Q^2)$ to reproduce the QCD behaviour
in the second WSR.
The third one is violated as in QCD and one has
\be
\lim_{Q^2\to\infty}\left( Q^4\Pi_{LR}^{(0)}(Q^2)\right)
 = \frac{\dis 2}{\dis g_S} \left(m_i M_j + m_j M_i\right) \ .
\ee

Let us now see what relations between low-energy hadronic
couplings do these Sum Rules imply for this ENJL cut-off model.
In the equal mass sector, $m_i = m_j \ne 0 $, one has
\ba
\rlabel{xwsrel}
f_V^2 M_V^2 = f_A^2 M_A^2 + f^2_\pi \, , \\
f_V^2 M_V^4 = f_A^2 M_A^4 \, .
\ea
Remember that in QCD one has in this case
\ba
\rlabel{xwsqcd}
f_V^2 M_V^2 = f_A^2 M_A^2 + f^2_\pi \, , \\
f_V^2 M_V^4 = f_A^2 M_A^4 + m^2_\pi f^2_\pi\, .
\ea
In the off-diagonal case, $m_i\ne m_j$, the situation becomes
a lot more complicated.  However, since the off-diagonal
part is suppressed by $(M_i-M_j)^2/Q^2$ one does not expect
qualitatively different results.

\subsection{Some numerical results}
\rlabel{xnumbers}
As can be seen from the explicit formulas the change with respect to
ref. \rcite{BRZ} is in most cases a (small)
shift in the two-point function mass pole positions.
Therefore we do not plot too many of the two-point functions.
As numerical input we use for $G_S$, $G_V$ and $\Lambda_\chi$ the values from
fit 1 in ref. \rcite{BBR}. These are $\Lambda_\chi = 1.160$ GeV and
$G_S = 1.216$. The value of $g_A(Q^2=0)$ there was 0.61. This is
$G_V = 1.263$. For the current quark masses we use the value of the quark
mass for ${\overline m}\equiv m_u=m_d$ that reproduces the physical
neutral pion and kaon masses. With
the other parameters as fixed above this is ${\overline m}=3.2$ MeV
and $m_s/{\overline m}=26$.

As an example
we have plotted the inverse of
the transverse vector two-point function in eq. \rref{xtransvec} in
figure \tref{xFigV1} for the values of $G_S$ and $\Lambda_\chi$
corresponding above mentioned.
\begin{figure}
\rotate[r]{\epsfysize=13.5cm\epsfxsize=8cm\epsfbox{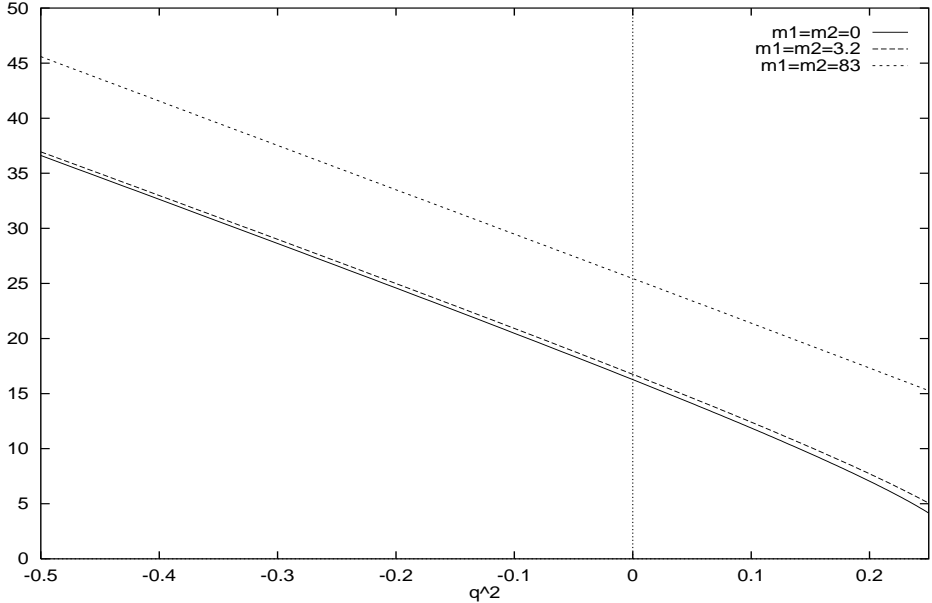}}
\caption{The inverse of the transverse vector
two-point function for equal
quark masses in the chiral limit, i.e. ${\cal M} \to 0$;
 for the $\rho$ meson, i.e. $m_1=m_2=3.2$ MeV
and for the $\phi$ meson, i.e. $m_1=m_2= 83$ MeV.
The units of $q^2$ are GeV$^2$}
\rlabel{xFigV1}
\end{figure}
The full curve is the result in the chiral
limit (${\cal M} \to 0$) and the
dashed is the result with $m_i=m_j=\overline m$ the value above.
The reason we have
plotted the inverse will become clear in section \tref{VMD}.
We also show the inverse for $m_i=m_j=m_s$ the value above
in the short-dashed curve.
To show the result for unequal quark masses we have plotted in
figure \tref{xFigV2} the transverse vector
two-point function itself for the chiral
limit  case and for the $\bar u s$ case with $m_s$ and
$\overline m$ above. Notice that the two-point function now has
a kinematical pole at $q^2=0$.
\begin{figure}
\rotate[r]{\epsfysize=13.5cm\epsfxsize=8cm\epsfbox{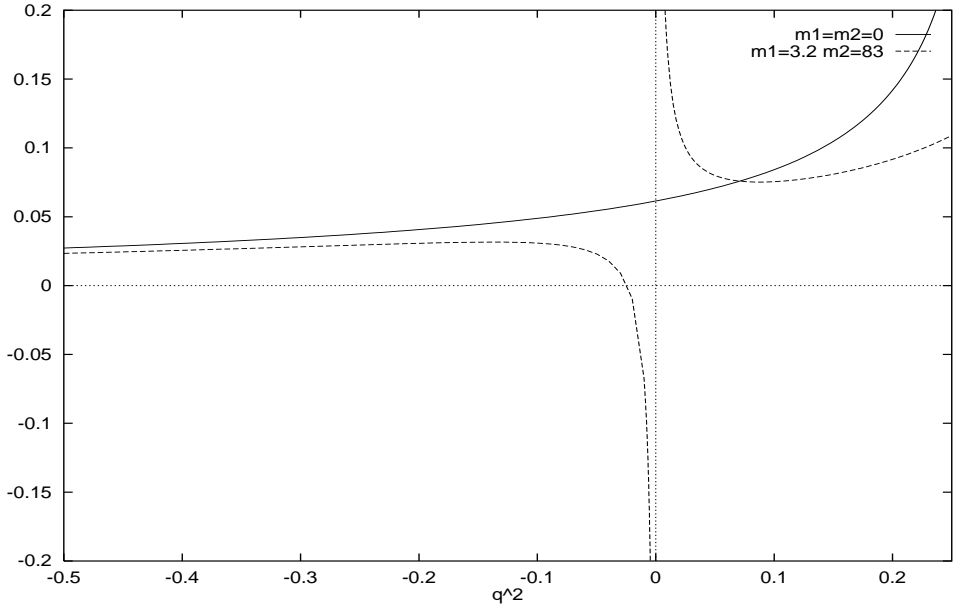}}
\caption{The transverse vector-two-point function for the chiral limit
and for unequal quark masses, $m_1 = \overline m$ and $m_2 = m_s$.
Note the kinematical pole at $q^2 = 0$. The units of $q^2$ are GeV$^2$.}
\rlabel{xFigV2}
\end{figure}

We have also plotted in figure
 \tref{xFigmpi} for the parameters quoted above the
dependence of the pion mass on $Q^2$.
Since $f_{ij}^2 m_{ij}^2$ is a constant,
see eq. \rref{xmpi} this is also the $Q^2$ dependence of the
inverse of the $f_{ij}$ decay constant squared.
\begin{figure}
\rotate[r]{\epsfysize=13.5cm\epsfxsize=8cm\epsfbox{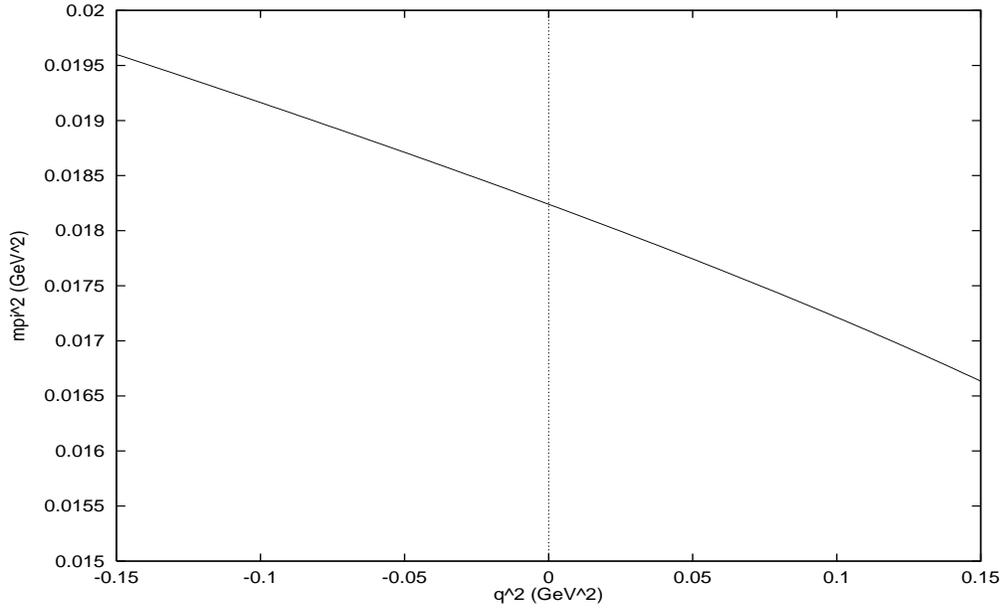}}
\caption{The running pseudoscalar mass squared, $m_{ij}^2(-q^2)$, as a
function of $q^2$ for $m_i = m_j = 3.2$ MeV.}
\rlabel{xFigmpi}
\end{figure}

Let us make some comments on the corrections we find to the GMOR
relation \rref{xmpi} in this model.
The corrections to the GMOR relation \rcite{Gell} can be
calculated here in an analogous expansion to the one in
$\chi$PT.
Then the GMOR relation can be written as follows
\rcite{GL} (for the diagonal flavour case, i.e. $i=j$)
\ba
2 m_i \langle {\overline q}_i q_i \rangle &=& - m_{ii}^2(-q^2)
f_{ii}^2(-q^2) \left(1 - 4 \frac{\dis m_{ii}^2(0)}
{\dis f_{ii}^2(0)} \left( 2 L_8 - H_2 \right)
+ {\cal O} (p^6)\, \right) \ . \nonumber \\
\ea
Here we have included all the chiral
corrections to the quark condensate, to the pion mass
and to the pion decay constant in their respective values.
Then the remaining is a correction to the GMOR relation.
We have also calculated this correction in this model
and it turns out to be
\ba
\rlabel{xh2}
1-4 \frac{\dis m_{ii}^2(0)}
{\dis f_{ii}^2(0)} \left( 2 L_8 - H_2 \right)
+ {\cal O} (p^6)
&=& A\left(1- \frac{\dis m_i}{\dis M_i}\right) \, .
\ea
Notice that the r.h.s. contains all the orders
in the $\chi$PT expansion in the large $N_c$ limit.
$A$ is the reducing factor discussed in eq. \rref{A2}.
Numerically, this correction is around 0.2 $\%$
for pions and 3 $\%$ for kaons and
disagrees with the one found in QCD Sum Rules, see
the book by Narison in \rcite{QCDsum} and
\rcite{BPR}.
 The expression in \rref{xh2} is the  one consistent with
the use of Ward identities to sum the infinite string
of constituent quark bubbles. Numerically we get
$2 L_8 - H_2 \simeq
0.2 \cdot 10^{-3}$ for the input parameters above.
  This agrees with the one found
at the one-loop level, Sect. \tref{p4analysis}.

\subsection{Inclusion of gluonic effects}

The inclusion of extra gluonic effects like in \rcite{ERT,BBR,BRZ}
can be done simply by replacing the one-loop functions of
appendix \tref{xAppC} by the ones including the effects of
$\langle \alpha_S G^2\rangle$. the expressions for the chiral case can be
found in \rcite{BRZ}, these can also be used for the case of equal but
nonzero current quark masses. In the general case the expressions needed
can be found in the QCD sum rules literature, see e.g. \rcite{QCDsum}.

The effects are in general rather small because the main behaviour
is produced by the resummation and not so much by the slow variation
with $Q^2$ of the one-loop functions. An example was shown
in Fig. 4 of \rcite{BRZ}.

\section{Some results on three-point functions and Meson Dominance}
\rlabel{threep}

In general the same procedure as used in the previous section can be extended
to three-point functions. In fact most of them were used in the calculation
of the $B_K$ factor\rcite{BP3,BP4}. Here we will discuss one example
extensively and a second one in a short form.
Some comments about meson dominance in the three-point functions are also
given. The discussion closely follows Ref. \rcite{BP2}.

\subsection{VPP with the use of the Ward identities}
\rlabel{vppsub}
In this subsection we calculate the
Vector Pseudoscalar Pseudoscalar (VPP) three-point
function to all orders in $\chi$PT
using the same type of methods as those used for the two-point
functions. The three-point function we calculate is the following
\be
\rlabel{VPP}
\Pi_\mu^{VPP}(p_1,p_2) \equiv
i^2 \int {\rm d}^4x \int {\rm d}^4y e^{i(p_1\cdot x + p_2 \cdot y)}
\langle 0 | T\left(V^{ij}_\mu(0)P^{kl}(x)P^{mn}(y)\right)|
0 \rangle \ .
\ee
Where $i,j,k,l,m$ and $n$ are flavour indices.
In the limit of large $N_c$ the flavour structure is limited because
of Zweig's rule (this flavour structure is general for any
three-point function of three quark currents),
\be
\Pi_\mu^{VPP}(p_1,p_2) \equiv
\Pi^{+}_\mu(p_1,p_2)_{ikm}  \delta_{il}\delta_{kn}\delta_{mj}
+\Pi^{-}_\mu(p_1,p_2)_{ikm} \delta_{in}\delta_{kj}\delta_{ml}\ .
\ee
Bose symmetry requires that
\be
\rlabel{pi+}
\Pi^{+}_\mu(p_1,p_2)_{ikm} = \Pi^{-}_\mu(p_2,p_1)_{imk}\ .
\ee
The three-point function $\Pi^{VPP}_\mu (p_1,p_2)$
 can then be simply calculated by only taking one
particular flavour combination.
Finally we can use Lorentz-invariance to rewrite
\be
\Pi^{+}_\mu(p_1,p_2)_{ikm} = p_{1\mu} \Pi^A_{ikm}(p_1^2,p_2^2,q^2)
+ p_{2\mu} \Pi^B_{ikm}(p_1^2,p_2^2,q^2)\ ,
\ee
where we have defined $q \equiv p_1 + p_2$.

We shall limit ourselves to the vector diagonal case, i.e.
 $m_i = m_j$.
In the vector off-diagonal case there will also
be non-trivial mixings with the
scalar-pseudoscalar-pseudoscalar three-point function.
Here a relatively
simple Ward identity for this three-point function can be derived from
$\partial^\mu V^{ij}_\mu = 0$ and the equal-time commutation relations.
It is
\be
\rlabel{WVPP}
q^\mu \Pi^{+}_\mu (p_1,p_2)_{ikm} =
-\Pi_P(-p_1^2)_{ki}+\Pi_P(-p_2^2)_{mk}\ .
\ee
So the Ward identity relates the three-point function to a combination of
two-point functions. This determines one of the two functions
$\Pi^A, \Pi^B$ in terms of the other. The Ward identity gives,
for instance, the following constraint (for $p_1^2=p_2^2$ and $i=m$)
\be
\rlabel{WAB}
\Pi^B_{iki}(p^2,p^2,q^2)= -\Pi^A_{iki}(p^2,p^2,q^2)\ .
\ee

 The type of graphs that need to be summed
are depicted in figure \ref{Fig3graf}.
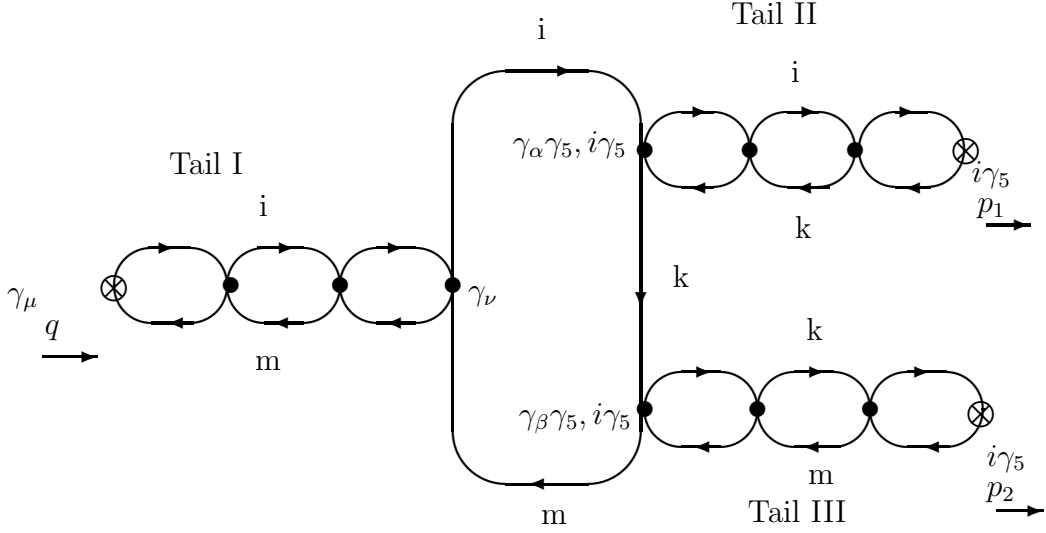
\begin{figure}
\begin{center}
%
%
%
\thicklines
\setlength{\unitlength}{1mm}
\begin{picture}(140.00,80.00)(10.,0.)
\put(133.00,18.50){\oval(15.00,10.00)}
\put(136.25,51.75){$\bigotimes$}
\put(118.00,18.50){\oval(15.00,10.00)}
\put(95.50,18.50){\circle*{2.00}}
\put(103.00,18.50){\oval(15.00,10.00)}
\put(136.00,13.50){\vector(-1,0){3.50}}
\put(121.50,13.50){\vector(-1,0){5.00}}
\put(105.00,13.50){\vector(-1,0){3.00}}
\put(101.50,23.50){\vector(1,0){3.00}}
\put(116.00,23.50){\vector(1,0){3.50}}
\put(131.50,23.50){\vector(1,0){2.00}}
\put(32.50,35.00){\oval(15.00,10.00)}
\put(47.50,35.00){\oval(15.00,10.00)}
\put(23.00,33.50){$\bigotimes$}
\put(62.50,35.00){\oval(15.00,10.00)}
\put(70.00,35.00){\circle*{2.00}}
\put(29.50,40.00){\vector(1,0){3.50}}
\put(44.00,40.00){\vector(1,0){5.00}}
\put(60.50,40.00){\vector(1,0){3.00}}
\put(64.00,30.00){\vector(-1,0){3.00}}
\put(131.00,53.00){\oval(14.00,10.00)}
\put(123.50,53.00){\circle*{2.00}}
\put(116.50,53.00){\oval(14.00,10.00)}
\put(109.50,53.00){\circle*{2.00}}
\put(102.50,53.00){\oval(14.00,10.00)}
\put(133.50,48.00){\vector(-1,0){3.50}}
\put(120.00,48.00){\vector(-1,0){4.50}}
\put(104.50,48.00){\vector(-1,0){3.00}}
\put(101.00,58.00){\vector(1,0){3.00}}
\put(115.00,58.00){\vector(1,0){3.50}}
\put(129.50,58.00){\vector(1,0){2.00}}
\put(110.50,18.50){\circle*{2.00}}
\put(96.00,18.50){\circle*{0.00}}
\put(95.50,53.00){\circle*{2.00}}
\put(82.50,36.00){\oval(25.00,55.00)}
\put(35.50,30.00){\vector(-1,0){3.50}}
\put(95.00,35.50){\vector(0,-1){3.50}}
\put(82.50,63.50){\vector(1,0){3.50}}
\put(83.50,8.50){\vector(-1,0){3.50}}
\put(49.50,30.00){\vector(-1,0){3.50}}
\put(125.50,18.50){\circle*{2.00}}
\put(138.25,16.75){$\bigotimes$}
\put(10.75,32.75){$\gamma_\mu$}
\put(72.00,33.00){$\gamma_\nu$}
\put(78.00,52.75){$\gamma_\alpha\gamma_5,i\gamma_5$}
\put(78.75,16.50){$\gamma_\beta\gamma_5,i\gamma_5$}
\put(139.00,48.75){$i\gamma_5$}
\put(141.00,11.00){$i\gamma_5$}
\put(141.00,43.00){\vector(1,0){5.75}}
\put(142.25,5.00){\vector(1,0){6.25}}
\put(139.75,44.75){$p_1$}
\put(141.00,7.00){$p_2$}
\put(107.00,69.75){Tail II}
\put(109.25,3.50){Tail III}
\put(32.25,49.75){Tail I}
\put(44.25,44.25){i}
\put(43.75,23.75){m}
\put(81.25,67.75){i}
\put(115.00,62.00){i}
\put(115.50,41.25){k}
\put(99.25,34.75){k}
\put(117.00,27.75){k}
\put(117.25,8.50){m}
\put(81.75,3.25){m}
\put(15.50,25.50){\vector(1,0){7.25}}
\put(15.75,28.50){$q$}
\put(55.00,35.00){\circle*{2.00}}
\put(40.50,35.00){\circle*{2.00}}
\end{picture}
\caption{The graphs that need to be summed in the large $N_c$ limit for
the Vector-Pseudoscalar-Pseudoscalar three-point function. See text for
explanation.}
\rlabel{Fig3graf}
\end{center}
\end{figure}
Each of the three tails  here
is  the diagram in figure  \tref{xFig2pt}a with the same explanation
as there.
We have there depicted one particular flavour combination.
This is the one that corresponds to the
function  $\Pi^{+}_\mu$ given above.
The $i,k,m$ written above the lines are the flavours of each line.

All graphs are formed by having the tails summed over 0, 1, 2,
$\cdots$, $\infty$
loops connected by four-fermion couplings. These then couple to
the one-loop
three-point function (or vertex) $\ovpi^{+}_\mu$,
with various possibilities for the insertion in the
three-point vertex.
These possibilities for the $\gamma$-matrices are written
in figure \ref{Fig3graf} inside the main loop.

In this figure the left-hand side depicts the insertion of
 the current $V^{ij}_\mu (0)$ and
Tail I is the connection to this current. On the end connecting to
the one-loop three-point function it is only nonzero for another
vector insertion since in the diagonal case we consider, the mixed
vector--scalar two-point function vanishes.
Its expression is given by
\be
g_{\mu\nu} + \frac{-8\pi^2 G_V}{N_c\Lambda^2_\chi}
\Pi^V_{\mu\nu}(-q^2)_{mi}\ .
\ee
Here the first term comes from where the external current directly connects
to the one-loop three-point function and the second term is with the
two-point function in between. The sum of both is
\be
\frac{g_{\mu\nu}M_V^2(-q^2)_{mi}-q_\mu q_\nu}
{M_V^2(-q^2)_{mi}-q^2}.
\ee

A similar discussion can be done for Tail II and Tail III.
First we have the
insertion of the current $P^{kl}(x)$
at the external end. On the end connecting
to the one-loop three-point function we can have $i\gamma_5$ or an
axial-vector insertion since the mixed axial-vector--pseudoscalar
two-point function is nonzero. The $i\gamma_5$ insertion tail is :
\ba
\rlabel{ga5}
1 + \frac{4\pi^2 G_S}{N_c\Lambda^2_\chi}\Pi_P(-p_1^2)_{ki}
\nonumber \\ \hspace*{1cm}
= \frac{\left(M_k+M_i\right)^2}{2 g_S f_{ki}^2(-p_1^2)
\left(m_{ki}^2(-p_1^2) - p_1^2\right)}\ .
\ea
For the connection with the axial-vector insertion it is instead
\ba
\rlabel{axiv}
\frac{8\pi^2 G_V}{N_c\Lambda^2_\chi}
i p_1^\alpha \Pi^M_P(-p_1^2)_{ki} \nonumber
\\ \hspace*{1cm} =
\frac{ip_1^\alpha}{2 f_V^2 M_V^2}
\frac{\left(M_k+M_i\right)}{g_S
\left(m_{ki}^2(-p_1^2) - p_1^2\right)}\ .
\ea
The combination $f_V^2 \, M_V^2$ is here flavour and $p_1^2$ independent,
it is the combination in eq. \rref{xfvmv} with
$\ovpi^{(0)}_V(Q^2)_{ij}=0$ since we are in the diagonal flavour case.
The way both these types of insertions
can appear due to the tail are how within this formulation the mixing
of pseudoscalar and axial-vector degrees comes about. These will
be described by factors of $g_A^2$ (see below).
Tail III is identical to Tail II with the substitutions $p_1\to p_2$ and
$i,k\ \to\ k,m$.

The full expression for $\Pi^+_\mu$ is
\be
\renewcommand{\arraystretch}{1.5}
\begin{array}{l}
\rlabel{fvpp} \dis
\Pi^{+\mu}(p_1,p_2) = \left\{
 g^{\mu\nu} + \frac{-8\pi^2 G_V}{N_c\Lambda^2_\chi}
\Pi^{V\mu\nu}(-q^2)_{mi} \right\} \nonumber \\ \dis
\hspace*{1cm}
\times \left\{ \ovpi^+_\nu(p_1,p_2) \left(
1 + \frac{4\pi^2 G_S}{N_c\Lambda^2_\chi} \Pi_P(-p_1^2)_{ki} \right)
\left( 1 + \frac{4\pi^2 G_S}{N_c\Lambda^2_\chi} \Pi_P(-p_2^2)_{mk}
\right) \right. \nonumber \\ \dis \hspace*{1cm}
+ \ovpi^{VPA}_{\nu\beta}(p_1,p_2)
\left(1 + \frac{4\pi^2 G_S}{N_c\Lambda^2_\chi} \Pi_P(-p_1^2)_{ki}
\right) \left( \frac{8\pi^2 G_V}{N_c\Lambda^2_\chi}
i p_2^\beta \Pi^M_P(-p_2^2)_{mk} \right) \nonumber \\ \dis
\hspace*{1cm}
+  \ovpi^{VAP}_{ \nu\alpha}(p_1,p_2)
\left(\frac{8\pi^2 G_V}{N_c\Lambda^2_\chi}
i p_1^\alpha  \Pi^M_P(-p_1^2)_{ki} \right)
\left(1 + \frac{4\pi^2 G_S}{N_c\Lambda^2_\chi}
\Pi_P(-p_2^2)_{mk} \right)  \nonumber \\ \dis
\hspace*{1cm} + \left. \ovpi^{VAA}_{\nu\alpha\beta}(p_1,p_2)
\left( \frac{8\pi^2 G_V}{N_c\Lambda^2_\chi}
i p_1^\alpha \Pi^M_P(-p_1^2)_{ki} \right)
\left( \frac{8\pi^2 G_V}{N_c\Lambda^2_\chi}
i p_2^\beta \Pi^M_P(-p_2^2)_{mk} \right) \right\} \, .
\nonumber \\
\end{array}
\renewcommand{\arraystretch}{1.5}
\ee
Where the one-loop three-point functions $\ovpi^{VPA}_{\mu\nu}$,
$\ovpi^{VAP}_{\mu\nu}$ and $\ovpi^{VAA}_{\mu\nu\alpha}$
are the one fermion-loop result for
\ba
\rlabel{VPA}
\Pi_{\mu\nu}^{VPA}(p_1,p_2) &\equiv &
i^2 \int {\rm d}^4x \int {\rm d}^4y e^{i(p_1\cdot x + p_2 \cdot y)}
\langle 0 | T\left(V^{im}_\mu(0)P^{ki}(x)A^{mk}_\nu (y)\right)|
0 \rangle \, , \nonumber \\ \\
\rlabel{VAP}
\Pi_{\mu\nu}^{VAP}(p_1,p_2) &\equiv &
i^2 \int {\rm d}^4x \int {\rm d}^4y e^{i(p_1\cdot x + p_2 \cdot y)}
\langle 0 | T\left(V^{im}_\mu(0)A^{ki}_\nu(x)P^{mk}(y)\right)|
0 \rangle \, , \nonumber \\ \\
\rlabel{VAA}
\Pi_{\mu\nu\alpha}^{VAA}(p_1,p_2) &\equiv &
i^2 \int {\rm d}^4x \int {\rm d}^4y e^{i(p_1\cdot x + p_2 \cdot y)}
\langle 0 | T\left(V^{im}_\mu(0)A^{ki}_\nu(x)A^{mk}_\alpha
 (y)\right)| 0 \rangle \, . \nonumber \\
\ea

To obtain the full expression in eq. \rref{fvpp}
it now remains to calculate these
VPP, VAP, VPA and VAA one-loop three-point functions (or vertices).
 The axial-vector
ones always come multiplied with the relevant momentum. So we
always have the scalar products
$p_1\cdot A^{ki}(x)$ and $p_2\cdot A^{mk} (y)$.
 That means that using the
Ward identities we can relate the VAA, VAP, VPA to
the VPP one plus possibly two-point function
terms resulting from equal time commutators.

These Ward identities
are (remember we assume $M_i = M_j$ here).
\ba
i p_1^\nu \ovpi^{VAA}_{\mu \nu \alpha} (p_1,p_2) &=&
- \left(M_k + M_i\right) \ovpi^{VPA}_{\mu \alpha}(p_1,p_2)
\nonumber\\
&& + i \, \ovpi^V_{\mu\alpha}(-q)_{mi} - i \,
 \ovpi^A_{\mu\alpha}(-p_2)_{mk}\ ;
\\
i p_1^\nu \ovpi^{VAP}_{\mu \nu} (p_1,p_2) &=&
- \left(M_k + M_i\right) \ovpi^+_\mu (p_1,p_2) \nonumber\\
&&  + i \, \ovpi_{P \mu}(-p_2)_{mk}\ .
\ea
The other needed ones can be derived from this using Bose-symmetry.
Notice that there is no contribution here from the flavour chiral
anomaly.

We can now use these identities to obtain the final result for the three-point
function we want. The terms which after the use of the one-loop identities
above are proportional to VPP can be combined into a simple form
using $g_A(p^2)$. The result is (we have $M_i = M_j$ and $j=m$
in this flavour configuration).
\be
\renewcommand{\arraystretch}{1.5}
\begin{array}{l} \dis
\rlabel{VPPres}
\Pi^{+ \mu}(p_1,p_2)\,=\nonumber \\ \dis
\hspace*{3cm} \left(\frac{\left(M_i+M_k\right)^4}
{4 g_S^2 f_{ki}^2(-p_1^2)f_{mk}^2(-p_2^2)} \right)
\left(\frac{g^{\mu\nu} M_V^2(-q^2)_{mi} - q^\mu q^\nu}
{M_V^2(-q^2)_{mi} - q^2}\right) \nonumber \\ \dis
\times \, \frac{1}{\left( m_{ki}^2(-p_1^2)-p_1^2 \right)
\left(m_{mk}^2(-p_2^2)-p_2^2 \right)}
\Bigg\{g_A(-p_1^2)_{ki} g_A(-p_2^2)_{mk} \ovpi^+_\nu (p_1,p_2)
 \nonumber \\ \dis
\hspace*{1cm} + \frac{\left(1-g_A(-p_1^2)_{ki}\right)
\left(1-g_A(-p_2^2)_{mk}\right)}{\left(M_i+M_k\right)^2} \left\{
\left(p_2\cdot q \right) p_{1\nu} - \left( p_1\cdot q \right)
p_{2\nu}\right\} \ovpi^{(1)}_V(-q^2)_{mi}
\nonumber\\ \dis \hspace*{1cm}
- \frac{g_A(-p_1^2)_{ki}\left(1-g_A(-p_2^2)_{mk}
\right)}{M_i+M_k}p_{1\nu} \ovpi_P^M(-p_1^2)_{ki} \nonumber
\\ \dis \hspace*{1cm}
+ \frac{g_A(-p_2^2)_{mk}\left(1-g_A(-p_1^2)_{ki}
\right)}{M_i+M_k}p_{2\nu}
\ovpi_P^M(-p_2^2)_{mk} \Bigg\}\ . \nonumber \\
\end{array}
\renewcommand{\arraystretch}{1}
\ee
This result satisfies the Ward identity \rref{WVPP}
if the one-loop function $\ovpi^+_\mu$ one
satisfies the same one with the one-loop functions. This provides
a rather non-trivial check on the result \rref{VPPres}.

It now only remains to calculate the one-loop form factor
$\ovpi^+_\mu(p_1,p_2)$. We give its expression in
appendix \ref{xAppD}. At this point we can see in  eq. \rref{VPPres}
how far regularization ambiguities affect the result. We first have to
define the two-point functions. Here all ambiguities are restricted to two
bare functions (see section \tref{xtwop} for details).
This three-point function adds one more in general, the
three-propagator function $I_3(p_1^2,p_2^2,q^2)$ (see explicit expression
in appendix \ref{xAppD}). Of course, this one-loop form factor
$\ovpi^+_\mu(p_1,p_2)$, satisfies all the identities eqs. \rref{VPP}
to \rref{WAB} as well. We refer to section \tref{threevpp}
for the definition of the physical vector form factor
after reducing this $VPP$ three-point function. We shall also discuss
there  the VMD limit in this form factor and give some numerics.

The same three-point function can be calculated in Chiral Perturbation
Theory. The result is
\be
\Pi^{+\mu}(p_1,p_2) =
\frac{2 B_0^2 f_{mi}^2}{(m_{ki}^2-p_1^2)(m_{mk}^2-p_2^2)}
\left(p_2-p_1\right)^\mu
\left(1+\frac{2 L_9}{f_{mi}^2}q^2 + {\cal O}(p^6) \right)\ .
\ee
 Pulling out the pion poles (see section \tref{threevpp}
for technical details) and taking the low-energy limit
 and the value of $L_9$ in this class of models our
full result in eq. \rref{VPPres}
reduces to this, providing one more non-trivial check.

\subsection{PVV with a discussion about its Ward identity}
\rlabel{pi0gg}

In this subsection we calculate the Pseudoscalar Vector Vector
(PVV) three-point function to all orders in $\chi$PT
with the same method as the one used before.
\be
\rlabel{PVV}
\Pi_{\mu\nu}^{PVV}(p_1,p_2) \equiv
i^2 \int {\rm d}^4x \int {\rm d}^4y e^{i(p_1\cdot x + p_2 \cdot y)}
\langle 0 | T\left(P^{ij}(0)V^{kl}_\mu(x)V^{mn}_\nu(y)\right)|
0 \rangle \ .
\ee
The pseudoscalar can
couple with at the one-loop end, both an axial-vector and
pseudoscalar two-point function. These have the same form as equations
\rref{ga5} and \rref{axiv} in the previous section with $p_1\to q$.
Summing up the three tails the total result is then\rcite{BP2}
\be
\rlabel{wardan}
\renewcommand{\arraystretch}{1.5}
\begin{array}{l}
\Pi^{+\mu\nu}(p_1,p_2) = \nonumber \\ \dis
 \left( g^{\mu\alpha} +
\frac{-8\pi^2 G_V}{N_c\Lambda^2_\chi} \Pi^{V \mu\alpha}(-p_1^2)
\right) \left( g^{\nu\beta} +
\frac{-8\pi^2 G_V}{N_c\Lambda^2_\chi} \Pi^{V \nu\beta}(-p_2^2)
\right) \nonumber \\ \dis
\times \left[ \ovpi^+_{\alpha\beta}(p_1,p_2) \left\{ 1 +
\frac{\dis 4 \pi^2 G_S}{\dis N_c \Lambda_\chi^2}
\Pi_P(-q) \right\}+\ovpi^{AVV}_{\rho \alpha \beta}
(p_1,p_2) \left( \frac{\dis 8 \pi^2 G_V}{\dis N_c
\Lambda_\chi^2} i q^\rho \Pi^M_P (-q)
 \right)\right] \nonumber \\
\end{array}
\renewcommand{\arraystretch}{1}
\ee
with $\ovpi^{AVV}_{\alpha\mu\nu}$ the one-loop result for
the following three-point function
\ba
\rlabel{AVV}
\Pi_{\rho\mu\nu}^{AVV}(p_1,p_2) &\equiv &
i^2 \int {\rm d}^4x \int {\rm d}^4y e^{i(p_1\cdot x + p_2 \cdot y)}
\langle 0 | T\left(A^{im}_\rho(0)V^{ki}_\mu(x)V^{mk}_\nu (y)\right)|
0 \rangle \, . \nonumber \\
\ea

The main new part here is that at the
one-loop level we now have to include the anomalous part of
the Ward identities.
There has been in fact quite some confusion whether this can be done
consistently, see Sect. \tref{anomaly}. Here we want to apply
that method to the PVV three-point function to all orders in external momenta
and quark masses.
The prescription is essentially
to use the anomalous QCD Ward identities  for the axial current
consistently. We shall use the scheme where vector currents are
conserved \rcite{Bardeen}. When we use the one-loop anomalous
Ward identity
to reduce the right-hand side of the pseudoscalar to
a part with only pseudoscalar couplings to the one-loop
vertex, we obtain a local chiral invariant result plus
an extra part where the tail couples directly to the
external vector sources $v^{kl}_\mu(x) v^{mn}_\nu(y)$.
This extra part is
of order $p^4$ and is the subtraction the anomalous
Ward identity
imposes to obtain the correct QCD flavour anomaly.

The full result in terms of the one-loop $\ovpi^+_{\mu\nu}$
three-point function is given by
\ba
\rlabel{PVVr1}
\Pi^+_{\mu\nu}(p_1,p_2) &=& \ovpi^+_{\mu\nu}(p_1,p_2)
\left(\frac{\dis
M_V^2(-p_1^2)_{ii} M_V^2(-p_2^2)_{ii}}{\dis \left(
M_V^2(-p_1^2)_{ii} - p_1^2\right) \left( M_V^2(-p_2^2)_{ii}
- p_2^2 \right)} \right) \nonumber \\ &\times&
\left\{ 1 + \frac{\dis 4 \pi^2 G_S}{\dis N_c \Lambda_\chi^2}
\Pi_P(-q) - \frac{\dis 8 \pi^2 G_V}{\dis N_c \Lambda_\chi^2}
2 M_i \Pi^M_P (-q) \right\} \nonumber \\
&+& \ovpi^+_{\mu\nu}(p_1,p_2)\Bigg|_{p_1^2=p_2^2=q^2=0}
\frac{\dis 8 \pi^2 G_V}{\dis N_c \Lambda_\chi^2}
2 M_i \Pi^M_P (-q)  \, .
\ea
Where the one-constituent quark loop function $\ovpi^+_{\mu\nu}$
is given by
\ba
\rlabel{onam}
\ovpi^+_{\mu\nu} (p_1,p_2)&=&
\frac{\dis N_c}{\dis 16 \pi^2}
\varepsilon_{\mu\nu\beta\rho} p_1^\beta p_2^\rho
\, F(p_1^2,p_2^2,q^2) \frac{\dis 2}{\dis M_i} \nonumber \\
{\rm with} \hspace*{1.5cm}
F(p_1^2,p_2^2,q^2)&=&1+I_3(p_1^2,p_2^2,q^2)-I_3(0,0,0)
\ea
where the form factor $I_3(p_1^2,p_2^2,q^2)$ is  the one given in
appendix \ref{xAppD}
and which appeared before in the study of the
VPP three-point function  in section \rref{vppsub}.
This form factor coincides with  the one
found in the context of constituent quark-models (see for instance
\rcite{Ametller}) when the cut-off $\Lambda_\chi$ is sent to
$\infty$. Here, this is a physical scale of
the order of the spontaneous symmetry breaking scale and
therefore we have to keep it finite.
The anomalous Ward
identities are telling us that terms
which are of chiral counting different to
${\cal O}(p^4)$ have to be local chiral invariant \rcite{BP1}
but they do not fix the regularization for those terms.
We therefore use  here
consistently the same regularization for them
as in the non-anomalous sector. At ${\cal O}(p^4)$
the chiral anomaly also uniquely fixes the
 one-loop constituent chiral quark anomalous form factor
 to be the one in eq. \rref{onam}
when $p_1^2=p_2^2=q^2=0$ \rcite{AB}.

Here we have used the anomalous Ward identity in
eq. \rref{wardan}.
A naive use of the two-point functions and Ward identities
would have led only
to the first term  in the sum in eq. \rref{PVVr1}.
The second term is the result of
enforcing the validity of the QCD flavour anomaly.
Substituting the results on the two-point functions in section
\ref{xtwop} we  can write down the following explicit expression
\be
\begin{array}{l} \dis
\rlabel{PVVres}
\Pi^+_{\mu\nu}(p_1,p_2) = \frac{\dis N_c}{\dis 16 \pi^2}
 \varepsilon_{\mu\nu\beta\rho} p_1^\beta p_2^\rho
\left( \frac{\dis 4 M_i}{\dis g_S f_{ii}^2(-q^2)
 \left(m_{ii}^2(-q^2)-q^2\right)} \right)
\nonumber \\ \dis
\left\{ 1 - g_A(-q^2)_{ii} \left[ 1 -
 F(p_1^2,p_2^2,q^2)\frac{\dis
M_V^2(-p_1^2)_{ii} M_V^2(-p_2^2)_{ii}}{\dis \left(
M_V^2(-p_1^2)_{ii} - p_1^2\right) \left( M_V^2(-p_2^2)_{ii}
- p_2^2 \right)} \right] \right\}  \ . \nonumber \\
\end{array}
\ee
We refer to section \tref{threepvv}
for the definition of the physical anomalous
$\pi^0\gamma^*\gamma^*$
 form factor after reducing this $PVV$ three-point function.
We shall also discuss there on the VMD limit in this process and give some
numerics.

\subsection{Meson-Dominance}
\rlabel{VMD}

We already saw that in the low energy limit we had a lot of relations that
were equivalent to various meson dominance relations. Here we discuss the
extension of those relations to the all-order case.

\subsubsection{Two-point functions}

Here we shall discuss the vector case, the axial-vector
case is similar. The transverse vector two-point function
in eq. \rref{xtransvec} reduces in the diagonal case, $m_i=m_j$
(the off-diagonal case can be done analogously) to the following
simple expression
\ba
\rlabel{twovec}
\Pi^{(1)}_V(-q^2) &=&
2 f_V^2(-q^2)\,
\frac{M_V^2(-q^2)}{M_V^2(-q^2)-q^2}  \,
\\ {\rm with} \hspace*{0.5cm} \rlabel{fvmv2}
2 f_V^2(-q^2) M_V^2(-q^2) &=& \frac{N_c\Lambda_\chi^2}
{8\pi^2 G_V} \, \\ {\rm and} \hspace*{2cm}
2 f_V^2(-q^2) &=& {\overline{\Pi}^{(1)}_V(-q^2)}\, .
\ea
In the complete VMD limit this two-point function
has the same form but with $f_V$ and $M_V$ constants. Let
us see how complete VMD works in this model. For that, we shall
study the inverse of $\Pi^{(1)}_V(-q^2)$, which in the complete
VMD limit is a straight line. This function was plotted
in section \tref{xnumbers} in figure \tref{xFigV1}. There we can see
that $\Pi^{(1)}_V(-q^2)$ in this model is very near to reproducing
the complete VMD linear form. Moreover, we can perform a linear
fit to the inverse of $\Pi^{(1)}_V(-q^2)$ to obtain the best VMD
values for the $f_V$ and $M_V$ parameters. These parameters
are in this way meaningfully defined in the Euclidean region
$-q^2>0$ where the model is far from  the two constituent
quark threshold. Doing this type of fit for the values of
the input parameters $\Lambda_\chi$, $G_V$, $G_S$ discussed
in section \tref{xnumbers} leads
to $M_V \simeq 0.644$ GeV for the vector mass in the chiral limit
(remember that we are always in the large $N_c$ limit)
and $f_V \simeq 0.17$ for the decay constant.
For current quark masses values discussed also in section
\tref{xnumbers},  we obtain for the $\rho$ meson
flavour configuration $M_\rho \simeq 0.655$ GeV and
$f_\rho \simeq 0.17$ and  for the $\phi$ meson
one $M_\phi \simeq 0.790$ GeV
and $f_\phi \simeq 0.14$. We see thus that the $\rho$ mass is
very close in the large $N_c$ limit,
to the one in the chiral limit, $M_V$.
Notice that these values for $M_V$ are far away from those quoted in
ref. \rcite{BBR}. The underlying reason is that
 in ref. \rcite{BBR} $f_V$ and $M_V$ were determined
directly from the  Lagrangian at ${\cal O} (p^2)$ in the ENJL expansion,
identifying them with their values at $q^2=0$.
What we find here is that even though the
two-point function in eq. \rref{twovec} has the correct $q^2 \to
0$ limit behaviour it does have, with the choice of vector fields
to represent vector particles in ref. \rcite{BBR}, substantial
contributions from higher order terms (mainly of ${\cal O} (p^4)$
in the ENJL expansion). A physical
vector field that would include these contributions can in
principle be defined as is shown by the fact that the inverse
of $\Pi^{(1)}_V(-q^2)$ is a rather straight line. What has happened
is that
\ba
\Pi^{(1)}_V(-q^2) &\simeq& \frac{\dis \left(2 f_V^2 M_V^2
\right)_{q^2=0}}{\dis M_V^2(0)-q^2 \left(1+\lambda+
{\cal O}(q^2/\Lambda_\chi^2)\right)} \, .
\ea
The vector meson mass derived in \rcite{BBR} was $M_V(0)$
while the slope of the physical two-point function
(for $|q^2|/\Lambda_\chi^2<<1$ that is where this
ENJL cut-off model makes sense) corresponds to rather
$M_V \sim M_V(0)/\sqrt{1+\lambda}$. We find from the calculation
that indeed $\lambda$ is of order 1 ($\lambda \simeq 0.7$),
explaining the difference in the slope from the ${\cal O}
(p^2)$ ENJL calculation in ref. \rcite{BBR}
of the two-point function to the ${\cal O}(p^4)$ one.

We can also see from eqs. \rref{x46}, \rref{xpipa0}-\rref{xpip}
and \rref{xpisca} that the forms of these two-point functions
are very similar to the corresponding ones
 in the meson dominance limit
but with couplings varying with $q^2$. The identification
of the corresponding physical values will involve analogous
procedures to the one described above for the transverse
vector two-point function one.

\subsubsection{VPP three-point function and the KSRF relation}
\rlabel{threevpp}

In this subsection we discuss how the result for the three-point
function $\Pi^{VPP}_\mu(p_1,p_2)$ obtained in section
\tref{vppsub} can be used to determine the physical
pion electromagnetic form factor in this model.
We shall discuss the $VPP$ three-point function flavour
structure corresponding to the
three-point function $\Pi^+_{\mu}(p_1,p_2)$
in eq. \rref{VPPres} for $m \equiv m_i=m_j=m_k$
and $p^2 \equiv p_1^2=p_2^2$ for definiteness.

Since this $\Pi^+_{\mu}(p_1,p_2)$
is a Green's function we first have to reduce
the external legs to properly normalized pion fields.
The vector leg acts here as an external source and is properly
reduced without bringing in any factor.
For this, we first look at the pseudoscalar two-point function
in eq. \rref{xpip} obtained using the same external fields
and parametrize it around the pole as
\be \rlabel{pip2}
\Pi_P(-p^2)=-\frac{\dis 1}{\dis g_S} +
\frac{\dis Z_\pi}{\dis p^2- m_\pi^2} \left( 1 +
{\cal O}(m_\pi^2/\Lambda_\chi^2) \right) \, .
\ee
The reducing factor $Z_\pi$ is
\ba
\rlabel{zpi}
Z_\pi &\equiv&-\frac{\dis \left( M_i +M_j \right)^2}{\dis 2 f_\pi^2
(-m_\pi^2) g_S^2} \, \frac{\dis 1}{\dis A^2}\, \hspace*{1.5cm}
{\rm with}  \nonumber \\ \rlabel{A2}
A^2&=& 1 - \left. \frac{\dis \partial m_{ij}^2(-p^2)}
{\dis \partial p^2} \right|_{p^2=m_\pi^2}
\, \nonumber \\
&=& 1+ \frac{\dis g_A^2(-m_\pi^2)}{\dis 2 f_\pi^2(-m_\pi^2)}
\left[\overline f^2_\pi(0) - \overline f^2_\pi(-m_\pi^2)
+ 2 m_\pi^2 I_3(m_\pi^2,m_\pi^2,0) \right] \nonumber \\
\ea
where $I_3(p_1^2,p_2^2,q^2)$ defined in appendix \tref{xAppD}
and $f_\pi^2(-q^2)$ and  ${\overline f}_\pi^2(-q^2)$
in eqs. \rref{x104}-\rref{x105}.
The quantity $A$ is very close to one and exactly one in the chiral limit.
Each pion leg brings a factor
$Z_\pi^{1/2}$ after reducing the Green's function
to the physical amplitude.
Rewriting the pseudoscalar two-point function in the form
in eq. \rref{pip2} gives that $m_\pi^2$ is the
solution of $ m_\pi^2= m_{ij}^2(-m_\pi^2)$.

Reducing the $VPP$ three-point function $\Pi^+_{\mu}(p_1,p_2)$
in eq. \rref{VPPres} we find
that it can be written as follows\footnote{To obtain the $\gamma^*\pi^+\pi^-$
three-point function from this $\Pi^+_\mu$ is necessary to
multiply it by the electric charge of the pion.}
(we shall suppress the flavour indices
which are always $ii$)
\be
\renewcommand{\arraystretch}{1.5}
\begin{array}{l}
\Pi^{+\mu}(p_1,p_2) = \frac{\dis Z_\pi}{\dis (p^2-m_\pi^2)^2}
\, F_{VPP}(p^2,q^2) \, (p_2-p_1)^\mu \,
\nonumber \\
\end{array}
\renewcommand{\arraystretch}{1}
\ee
which defines  the electromagnetic pion form factor
(or in general the pseudoscalar vector form factor)
$F_{VPP}(p^2,q^2)$  in this model. (The
general pion form factor, i.e different quark masses and
$p_1^2 \ne p_2^2$ can be obtained similarly from
\rref{VPPres}.) This form factor in the ENJL model
is  expected to be a good approximation
at intermediate and low-energy energies,
within the validity of the ENJL
model we are working with, i.e. for $|q^2| << \Lambda_\chi^2$.
The explicit expression for this form factor\footnote
{This form factor
was also calculated in ref. \rcite{bernard}. With
the appropriate changes of notation it agrees
with the one found there.}  is
\ba
\rlabel{fvpp2}
F_{VPP}(m_\pi^2,q^2) &=& \frac{\dis 1}{\dis 2 A^2 f_\pi^2
(-m_\pi^2)} \, \frac{\dis M_V^2(-q^2)}{\dis M_V^2(q^2)-q^2}
\left\{ 2 f_\pi^2(-m_\pi^2) \right. \nonumber \\
&-&  q^2 (1-g_A(-m_\pi^2))^2
f_V^2(-q^2)  +  \frac{\dis 2 g_A^2(-m_\pi^2)}{\dis
q^2 -4m_\pi^2} \nonumber \\ &\times& \left.
\left[(q^2-2 m_\pi^2)(\overline f^2_\pi(-q^2)
- \overline f^2_\pi(-m_\pi^2)) - 4 m_\pi^4 I_3(m_\pi^2,
m_\pi^2,q^2) \right] \right\} \, . \nonumber \\
\ea
Notice that this form factor has  no pole at $q^2=4m_\pi^2$.
The value of $A^2$ in eq. \rref{A2} is precisely the one
that ensures that $F_{VPP}(m_\pi^2,0)=1$ in the large $N_c$
limit as is required by the electromagnetic gauge invariance.
This must be so since we have imposed the Ward identities
to obtain this form factor.
In figure \ref{figVPP} we have plotted the inverse of this
form factor for the parameters quoted in section \tref{xnumbers}
in the chiral case ($\overline m=0$)
and in the case corresponding to the
physical pion mass ($\overline m=3.2$ MeV).
\begin{figure}
\rotate[r]{\epsfysize=13.5cm\epsfxsize=8cm\epsfbox{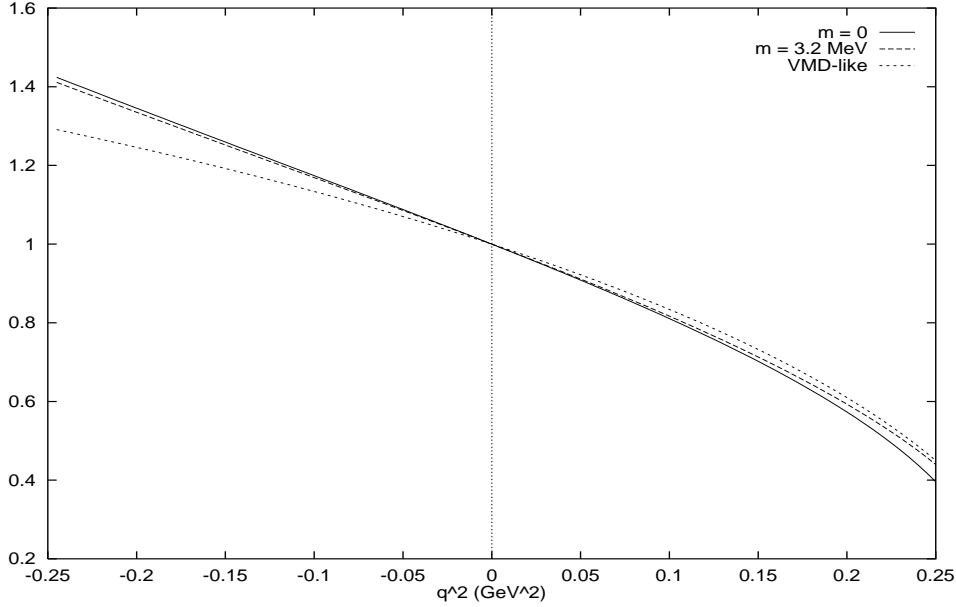}}
\caption{The inverse of the
vector form factor of the pion of eq. \protect{\rref{fvpp2}}.
For the chiral limit and with all current quark masses
equal to 3.2 MeV. Also plotted is the VMD approximation
$M_V^2(-q^2)/(M_V^2(-q^2)-q^2)$ for the latter case.}
\rlabel{figVPP}
\end{figure}
As can be seen from the picture, it is a rather straight line so
the complete VMD result for this form factor, i.e.,
\ba
\rlabel{vppvmd}
F_{VPP}^{VMD}(m_\pi^2,q^2) =
\frac{\dis M_\rho^2}{\dis M_\rho^2 - q^2}
\ea
with constant vector mass $M_\rho$ works rather well.
The slope of the linear fit of the inverse of the form factor
in eq. \rref{fvpp2} to this VMD form
gives a vector mass which is
$M_\rho \simeq 0.77$ GeV. This mass is
very close to the physical value and rather
different from the one found for the transverse vector
two-point function in the VMD limit $M_\rho \simeq
0.655$ GeV in the large $N_c$ limit.
This explains why using the physical
$\rho$ meson mass and the VMD dominance works so well
but it also shows that this $M_\rho$ ``mass'' in eq.
\rref{vppvmd} has not, in principle, to be the
same as the mass of the vector meson described
by the transverse two-point vector function.

The same three-point function $VPP$ also
contains implicitly
the $\rho \to \pi \pi$ coupling constant $g_V$.
(See Sect. \tref{hadronic} for its definition. Notice that is
 different from the symbol defined in \rref{gsgv}.)
Again, to obtain
the physical $\rho \to \pi \pi$ amplitude
we should first reduce the vector leg that now corresponds
to the $\rho$ particle, (remember that the pion legs have
been already reduced). This will bring a factor which
is similar to the factor
$1+\lambda$ discussed in the previous subsection.
We shall, as before, first determine the reducing vector
factor from the vector two-point function in eq.
\rref{twovec}. The reducing factor $Z_\rho$ is
\ba
\rlabel{zrho}
Z_\rho &\equiv& - \left(\dis 2 f_V^2 M_V^2\right)
\left(1- \left. \frac{\dis \partial M^2_V(-q^2)}{\dis
\partial q^2}\right|_{q^2=M_\rho^2} \right)^{-1}
\nonumber \\
&\equiv& - \frac{\dis \dis 2 f_V^2 M_V^2}{\dis B^2}  \, .
\ea
In this equation the combination $2f_V^2M_V^2$ is the one given
in eq. \rref{fvmv2} and is independent of $q^2$.
The vector mass $M_\rho$ is again given by the solution to
$M_\rho^2= M_V^2(-M_\rho^2)$.

One also can rewrite down the electromagnetic
pion form factor showing explicitly the coupling constant
of the $\rho$ meson to pions, $g_V$, as follows
\ba
\rlabel{defvgv}
F_{VPP}= 1 + f_V g_V \,
\frac{\dis q^2}{\dis f_\pi^2} \frac{\dis M_\rho^2}{\dis
M_\rho^2 -q^2} \,  .
\ea
Then, in the complete VMD limit one has $f_V g_V = f_\pi^2/M_\rho^2$.
In this ENJL model this relation is equivalent
to $g_V = (1-g_A) f_V$, i.e. one has complete
VMD and the KSRF relation \rcite{KSRF}
$2 g_V = f_V$ satisfied for $g_A=1/2$.

One can see in the eq. \rref{defvgv}, that
reducing the $\rho$ vector leg brings in a factor $B^2$ in the
numerator and another factor $B^2$ in the denominator with
the net result that $f_V(-q^2)g_V(-q^2)$ is not affected
by  reducing of the vector leg as much as  happens to
$f_V^2(-q^2)$ in eq. \rref{twovec}.

We now define an off-shell coupling $g_V(-q^2)$ by
Eqs. \rref{defvgv} and \rref{fvpp2}.
When expanded in $q^2$ and with $m_\pi^2=0$
one gets $f_V(0) g_V(0) = 2 L_9$, where $L_9$
 is the one found in ENJL in ref. \rcite{BBR}, Sect. \tref{p4analysis}.
As discussed there, at $q^2=0$
one has the KSRF \rcite{KSRF} relation,
i.e. $f_V(0) = 2 g_V(0)$  approximately
satisfied.
The definition above is the off-shell
equivalent to the KSRF relation in this model. For
$g_A=0$ the vector mass vanishes and the $\rho$ meson
couples as an SU(3)$_V$  gauge boson, in fact in this limit
one recovers the results of the Hidden Gauge Symmetry
model \rcite{Bando} for the non-anomalous sector.
In particular, when $g_A =0$ we have that the reducing
factor $B$ is 1
as corresponds to external gauge sources. In this limit
($g_A=0$),
one still has the KSRF relation analytically satisfied
off-shell, i.e. $f_V(-q^2)= 2 g_V(-q^2)$ for all
$q^2$.

In the limit $g_A \to 1$ one obtains
the constituent quark model result.
Let us see how $g_V(-q^2)$ works numerically
compared with $f_V(-q^2)$ for a definite value of $g_A$.
In figure \ref{fvgvfig}
we plot $f_V(-q^2)/2$ and $g_V(-q^2)$ for the values of parameters
discussed in section \tref{xnumbers}. These values
correspond to $g_A(0)=0.61$.
\begin{figure}
\rotate[r]{\epsfysize=13.5cm\epsfxsize=8cm\epsfbox{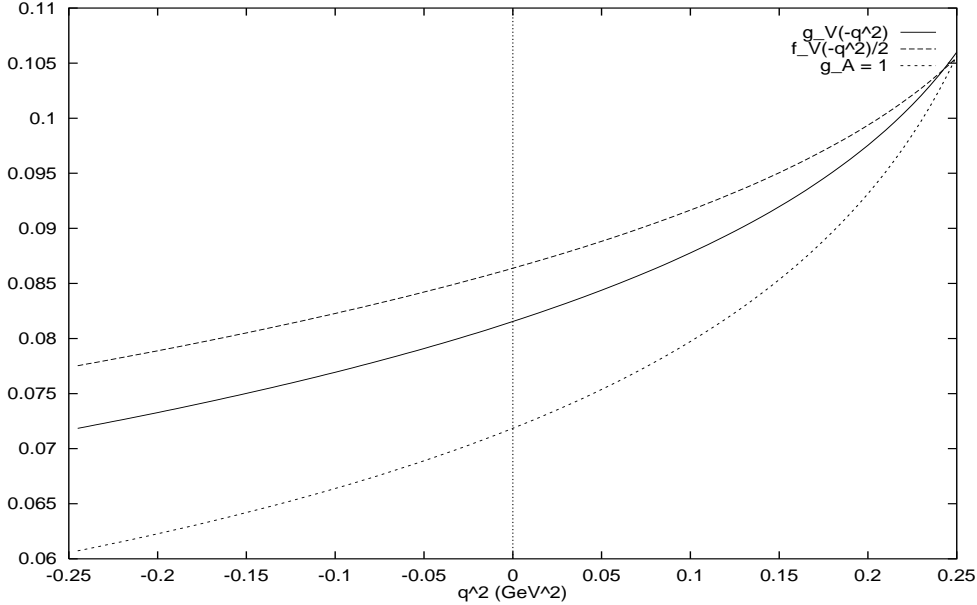}}
\caption{The generalized KSRF relation. We plot $g_V(-q^2)$ for
$g_A =0.61$ (solid line); $g_A \to 1$ (short-dashed line)
and $f_V(-q^2)/2$ (dashed line).
 The difference between the curves gives the violation
of the KSRF relation. See text for further comments.}
\rlabel{fvgvfig}
\end{figure}
The form factor $g_V(-q^2)$ is somewhat dependent
on $q^2$ with $ (2.1\sim 2.2) \,
g_V(-q^2) \simeq f_V(-q^2)$ in the Euclidean region.
In this figure we also plot the case $g_A \to 1$
where the same features can be seen. The form factor
$g_V(-q^2)$ for any value of
$g_A$ will be between the line $f_V/2$ (i.e., the $g_A=0$
limit) and the line for $g_A=1$, therefore the KSRF relation
is approximately satisfied off-shell for any value of $g_A$.

\subsubsection{PVV three-point function}
\rlabel{threepvv}

In this subsection we want to study the $\pi^0\gamma^*\gamma^*$
anomalous form factor. For that we shall reduce
the $PVV$ Green's function in eq. \rref{PVVres} calculated in
section \tref{pi0gg} to the physical amplitude following the
same procedure that in the previous section (for details
see there). Now, we have to reduce one pion leg, this will
bring in a factor $\sqrt Z_\pi$ and two external vector sources
legs which are properly reduced without bringing any factor.
Then the $PVV$ three-point function in eq. \rref{PVVres}
\footnote{To obtain the $\pi^0\gamma^*\gamma^*$  three-point
function from this $\Pi^+_{\mu\nu}$ is necessary to multiply it by
a factor $\sqrt 2$ coming from the $\pi^0$ flavour structure
and a factor $e^2/3$ from the quarks electric charge.}
can be rewritten as follows
\ba
\Pi^+_{\mu\nu}(p_1,p_2) &=& \frac{\dis \sqrt Z_\pi}{\dis
q^2-m_\pi^2} \,
\frac{\dis N_c}{\dis 16\pi^2} i \varepsilon_{\mu\nu\beta\rho}
\, p^\beta_1 p^\rho_2 \,
\frac{\dis 2 \sqrt 2 }{\dis f_\pi(-m_\pi^2)}
\nonumber \\
&\hspace*{1.5cm} \times&  F_{PVV}(q^2,p_1^2,p_2^2) \,
\ea
where $F_{PVV}$ is the $\pi^0 \to \gamma^* \gamma^*$
form factor in this model.
Notice that the reducing factor $A$ in eq.
\rref{A2} goes to one in the chiral limit preserving,
in that way, the chiral anomaly condition $F_{PVV}(0,0,0)=1$.
We plot the inverse of
this form factor for the case $p_2^2=0$ in figure
\ref{pvvfig}.
\begin{figure}
\rotate[r]{\epsfysize=13.5cm\epsfxsize=8cm\epsfbox{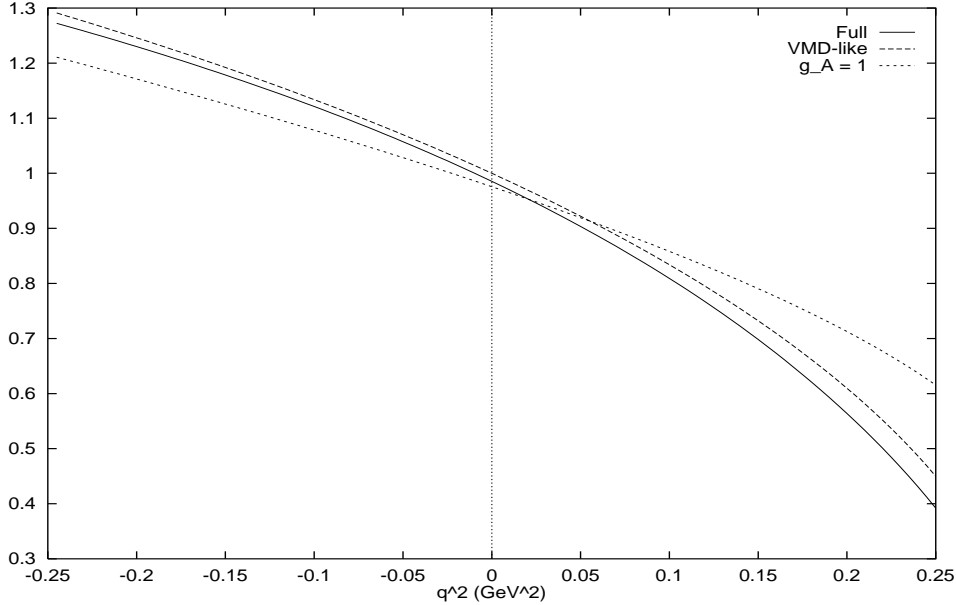}}
\caption{The
inverse of the $\pi^0\gamma^*\gamma$ form factor for one photon on-shell and
one off-shell as a function of the photon mass squared, $q^2$. Notice the
linearity in the Euclidean region. Plotted are the full result,
$M_V^2(-q^2)/(M_V^2(-q^2)-q^2)$(VMD-like) and the ENJL model without
vector and axial-vector mesons ($g_A = 1$).}
\rlabel{pvvfig}
\end{figure}
Notice that there $F_{PVV}(m_\pi^2,0,0) \neq 1$
and the difference comes from the reducing factor $A$
and is of chiral counting ${\cal O}(p^6)$.
We can expand this form factor for small $p_1^2,p_2^2$
and pion mass \footnote{For the $\pi^0$ decay we are on the pole
and hence $q^2=m_\pi^2$} as follows
\ba
F_{PVV} (m_\pi^2,p_1^2,p_2^2)
 &=& 1 + \rho \, (p_1^2+p_2^2)  + \rho' \,
m_\pi^2 + {\cal O} (q^4) \, ,
\ea
this expansion defines the slopes $\rho$ and $\rho'$
which in this model are
\ba
\rho &=& g_A(0) \left(\frac{\dis 1}
{\dis M_V^2(0)} + \frac{\dis \Gamma(2,M^2/\Lambda_\chi^2)}
{\dis 12 M^2} \right) \, ,
\nonumber \\ {\rm and} \hspace*{1cm}
\rho' &=& g_A(0) \left(\frac{\dis \Gamma(2,M^2/\Lambda_\chi^2)}
{\dis 12M^2}-\frac{\dis \Gamma(1,M^2/\Lambda_\chi^2)}
{\dis 12 \Gamma(0,M^2/\Lambda_\chi^2) M^2}
\right)\, .
\ea
Where the second term in $\rho'$ comes from the reducing
factor $A$ defined.
The constituent quark mass $M$ here
is the one corresponding to the current quark mass value
$\overline m = 3.2$ MeV
used in the numerical applications section \tref{xnumbers}.
Using $M_V^2(0)=6 M^2 g_A(0) / (1-g_A(0))$ \rcite{BBR}
we can write down them as
\ba
\rlabel{slope}
\rho &=& \frac{\dis 1}
{\dis 12 M^2}  \left(2 -
\left(2-\Gamma(2,M^2/\Lambda_\chi^2)\right) g_A(0)\right)
\, \nonumber \\
{\rm and} \hspace*{1cm} && \nonumber \\
\rho' &=& \frac{\dis g_A(0)}
{\dis 12 M^2} \left(\Gamma(2,M^2/\Lambda_\chi^2)
- \frac{\dis \Gamma(1,M^2/\Lambda_\chi^2)}
{\dis \Gamma(0,M^2/\Lambda_\chi^2)} \right) \, \nonumber \\
\ea
which interpolate between the constituent quark-model
result $g_A(0)=1$ and the gauge vector meson result $g_A(0)=0$.

With the input parameters we have been using
(see numerical application section \tref{xnumbers})
we get
\ba
\rho=(0.86+0.67)=1.53 \, {\rm GeV}^{-2} \,
\nonumber \\ {\rm and} \hspace*{1cm}
\rho'=(0.67-0.27)=0.40 \, {\rm GeV}^{-2} \, .
\ea
Where for $\rho$ the first number between brackets
is the vector meson exchange contribution and the second
is the constituent quark contribution (up to $g_A(0)$). We see that
both contributions are very similar  giving some
kind of complementarity between both approaches and
explaining the relative success of both when
used to describe this slope. For $\rho'$ they are the constituent quark
contribution and the one coming from the
pion leg reducing factor $1/A$. (Notice the cancellation
there.)
Experimentally \rcite{CELLO}
\ba
\rho=(1.8 \pm 0.14) \, {\rm GeV}^{-2} \, .
\ea
Taking into account that the $1/N_c$
corrections from $\chi$PT loops are estimated \rcite{Hans}
to be twice the experimental error
we consider the result as good. The cancellation in $\rho'$ is also welcome
since otherwise the $SU(3)$ breaking corrections in the $\eta$ decay would have
been much too large.

Let us compare this full result in eq. \rref{slope}
with the one obtained in ref. \rcite{Ximo} in this same
model assuming complete VMD in the chiral limit.
There, the same prescription
to include the QCD chiral anomaly that here \rcite{BP1}
was used at the one-loop level with the result
\ba
\rlabel{slope0}
\rho &=& \frac{\dis 1}{\dis 12 M^2} \,
\frac{\dis 1-g_A^2(0)}{\dis g_A(0)} \, .
\ea
Of course, this complete VMD result vanishes when $g_A=1$
where vector mesons decouple.

\section{Hadronic Matrix elements}
\rlabel{matrixelem}

\subsection{The $\pi^+-\pi^0$ Mass Difference}

In this section we will discuss the general philosophy behind  the $1/N_c$
method of calculating nonleptonic matrix elements. A good review where
also the references to the original papers can be found are the lecturers
by G\'erard\rcite{Gerard}. The application of this method to the
$\pi^+-\pi^0$ mass difference can be found in Ref.\rcite{BBG}
and the calculation within the QCD effective action model and the ENJL
model is in Refs.\rcite{BR,BRZ}.

We look at this quantity because it is the simplest nonleptonic
matrix elements in several respects. There is no factorizable
contribution because the photon is spin 1 and the pion spin 0. It
involves only pions so we expect the limit where the current quark masses
vanish to be a good approximation and (unlike $B_K$) it doesn't vanish and is
well defined in this limit. The latter remark has one very useful consequence.
Using PCAC it can be shown\rcite{Das} that this matrix element can be related
to a vacuum matrix element. So the mass difference becomes a
vacuum matrix element
of the photon propagator integrated over all momenta in the presence
of the strong interactions.
Schematically, the matrix element $\langle\pi^+|J^2|\pi^+\rangle$
can be rewritten in terms of
$\langle 0|J^2|0\rangle$. The precise expression in terms of
the hadronic two-point functions is given by\rcite{Das,BBG,BR}
\be
\label{ydass}
m_{\pi^+}^2 - m_{\pi^0}^2 =
-\frac{3\alpha_{em}}{8\pi f_\pi^2}
\int^\infty_0 dQ^2\ Q^2\left(
\Pi^{(1)}_V(Q^2)-\Pi^{(1)}_A(Q^2)\right)\ .
\ee
Eq. \rref{ydass}
involves an integral over all distance scales. The underlying idea
is now to split this integral into two parts,
$
\int_0^\infty = \int_0^{\mu^2} + \int_{\mu^2}^\infty
$,
and then to evaluate both pieces separately.

The long distance part in $1/N_c$ can be calculated in models since in $1/N_c$
the only quantities needed are the couplings of currents to hadrons and not
of full four-quark operators to hadrons.
The essence of the $1/N_c$ method is to do the short-distance part using the
operator expansion and then use $1/N_c$ to evaluate the matrix element.
Here this corresponds to using as the
difference of 2-point
functions:
\be
\left(
\Pi^{(1)}_V(Q^2)-\Pi^{(1)}_A(Q^2)\right)
=
\frac{-1}{Q^6}{ 8\pi \alpha_S(Q^2)} \left(\qvev\right)^2\ .
\ee
This leads to\rcite{BBG}
\be
\left.\Delta m_\pi^2\right|_{\rm SD} = \frac{3\alpha\alpha_S}{f_\pi^2 \mu^2}
\left(\qvev\right)^2\ .
\ee
One can then still do a renormalization group improvement of this\rcite{BRZ}.

\begin{figure}
\hspace{1.5cm}\epsfxsize=12cm\epsfbox{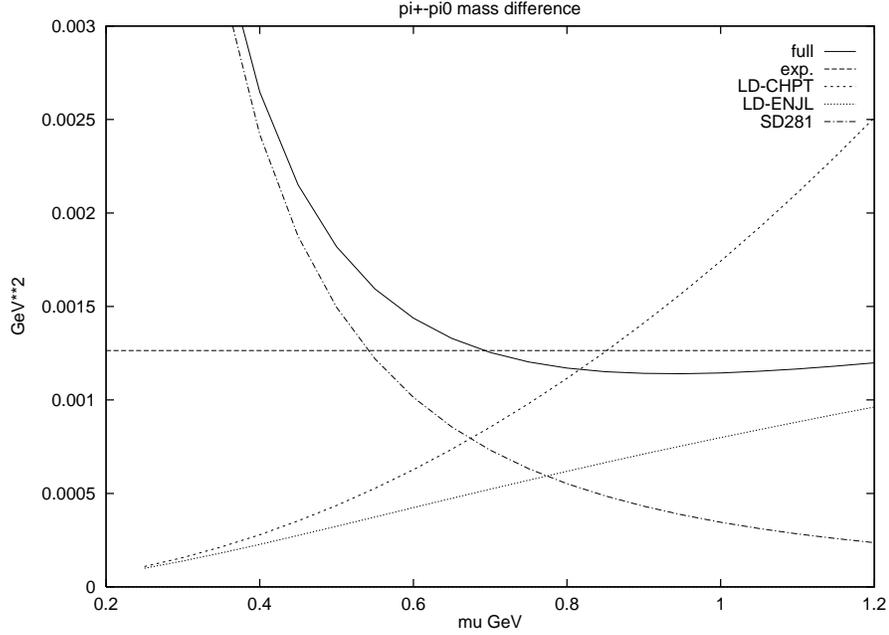}
\caption{The results for $m_{\pi^+}^2-m_{\pi^0}^2$:
long-distance result with only the pion (LD-CHPT);
ENJL long-distance (LD-ENJL); experimental
value (exp.); the short-distance result (SD281)
and the sum of short- and long-distance ENJL (full).}
\label{yfigure14}
\end{figure}

The long distance part of the integral requires more care. There are several
approaches.
\begin{enumerate}
\item One can take the measured spectral functions and use these to evaluate
the two-point functions needed in the integral. The most recent evaluation
of this is in Ref.\rcite{Don2}.
\item The two-point functions can be approximated by including the
$\rho$, $\pi$ and $a_1$ contribution. This was done neglecting the QCD part
in the original paper\rcite{Das} and more recently in \rcite{BBG}.
\item We can take only the $\pi$ contribution\rcite{BBG}.
This is most like the original
$1/N_c$ method(Ref.\rcite{Gerard} and references therein). This leads to
$\Delta m_\pi^2 = 3\alpha\mu^2/(4\pi)$.
\item One can use the QCD effective action approach\rcite{BR}.
\item The ENJL model can be used\rcite{BRZ}.
\end{enumerate}
All of these approaches give a good result for the mass difference.
In cases 1,2 and 5 a good matching was also obtained. This means that we
can vary $\mu$, the split between the short- and long-distance part of the
integral, over a reasonable interval without changing the result. In
figure \tref{yfigure14} the long-distance result with only the pion
is shown and the ENJL long-distance result. Also shown is the experimental
value, the short-distance result and the sum of short- and long-distance
for the ENJL case. The value of $\qvev$ used is the one given by the
ENJL model.

At this point I would like to remark that for this quantity in the QCD
effective
action approach one only obtains a gauge invariant result if the pion
is explicitly taken as propagating (see\rcite{BR}). This shows that in
this model the pion degree of freedom has to be added by hand. The gauge
dependence then cancels between a two- and a three-loop diagram.

\subsection{$B_K$}

In this section the extension to weak nonleptonic matrix elements of
the methods in the previous section is discussed on the example
of $B_K$. Here again the pure $1/N_c$ method\rcite{Gerard,BBG2},
the QCD effective action model\rcite{AP1} and the ENJL model\rcite{BP3,BP4}.
An overview of theoretical situation a few years ago can be found in
Ref.\rcite{Buras1}. The main alternatives to the present method are lattice
calculations\rcite{lattice,sharpe}
and 2 and 3-point QCD sum rules\rcite{QCD1}.

The short-distance integration here is done using the renormalization group.
This sums the possible large logarithms involving $\log(m_W^2/\mu^2)$.
The problem then reduces to the study of
\be
\rlabel{ydefbk}
\langle \overline K^0  | {\cal O}_{\Delta S=2} (x)
| K^0  \rangle \equiv \frac{\dis 4}{\dis 3}
B_K (\mu) f^2_K m_K^2
\ee
with
the $\Delta S=2$ operator
$
{\cal O}_{\Delta S=2} (x)  \equiv L^{sd}_\mu(x) L_{sd}^\mu (x)
$;
$2L^{sd}_\mu(x) = \overline s (x) \gamma_\mu
\left( 1-\gamma_5\right) d (x)$ and
summation over colours is understood.
Eq. \rref{ydefbk} is also the definition of the $B_K$ parameter.
The different approximations give
\begin{enumerate}
\item Vacuum Insertion : $B_K(\mu) = 1$.
\item Leading in $1/N_c$: $B_K(\mu) = 3/4$.
\item The standard $1/N_c$ result: $B_K(\mu) = 3/4(1-2\mu^2/(16\pi^2f_\pi^2))$.
\item $1/N_c$ with inclusion of vector mesons\rcite{Gerard}:
\be
 B_K(\mu) = \frac{3}{4}\left(1+\frac{1}{16\pi^2f_\pi^2}
\left(
-\frac{7}{8}\mu^2-\frac{3}{4}\frac{m_V^2\mu^2}{\mu^2+m_V^2}
-\frac{3}{8}m_V^2\log\frac{\mu^2+m_V^2}{m_V^2}\right)\right)\ .
\ee
\item The QCD effective action result\rcite{AP1}:
\be
B_K(\mu)=\frac{3}{4}\left(1+\frac{1}{N_c}\left(
1-\frac{N_c}{32\pi^2 f_\pi^4}\langle\frac{\alpha_S}{\pi}G^2\rangle+\ldots
\right)\right)\ .
\ee
\end{enumerate}
We would also like to study the effects of off-shellness. Therefore we do not
directly study the matrix element in Eq. \rref{ydefbk} but the Green function
\be
\rlabel{ytwopoint}
G_F\,\Pids(q^2)  \equiv
 i^2 \int d^4 x \, e^{iq\cdot x}
\langle 0 | T \left( P^{ds}(0)P^{ds}(x) \Gamma_{\Delta S =2}
\right)| 0 \rangle
\ee
in the presence of strong interactions. We use the ENJL model
for scales
below or around the spontaneous symmetry breaking scale. Here
$G_F$ is the Fermi coupling constant, we use
$P^{ds}(x) = \overline d (x)i\gamma_5 s (x)$, with summation over colour
understood and
$
\rlabel{yoperator}
\Gamma_{\Delta S=2} = - G_F\,
\int d^4 y \, {\cal O}_{\Delta S = 2} (y)
$.
 The reason to calculate this two-point function rather than
directly the matrix element is that we can now perform the calculation
fully in the Euclidean region so we do not have the problem
of imaginary scalar products. This also allows us in principle to
obtain an estimate of off-shell effects in the matrix elements. This
will be important in later work to assess the uncertainty when trying
to extrapolate from $K\to\pi$ decays to $K\to 2\pi$.
This quantity is also very similar to what is used in
the lattice and QCD sum rule calculations of $B_K$.

The $\Delta S = 2$ operator can be
rewritten as
\be
\rlabel{yoperator3}
\Gamma_{\Delta S=2} = - G_F \, \int \frac{d^4 r}{(2\pi)^4}
\int d^4 x_1 \int d^4 x_2 \,
e^{-i r \cdot(x_2 - x_1)} L^{sd}_\mu(x_1) L_{sd}^\mu(x_2) .
\ee
This allows us to consider this operator as
being produced at the $M_W$ scale
by the exchange of a heavy $X$ $\Delta S = 2$ boson.
We will work in the Euclidean domain
where all momenta squared are negative.
The integral in the modulus of the momentum $r$ in
\rref{yoperator3} is then split into two parts,
$
\int_0^{M_W} d |r| = \int_0^\mu d |r| + \int_\mu^{M_W} d |r|
$.
In principle one should then evaluate both parts separately as was done
for the $\pi^+-\pi^0$ mass difference
in the above quoted references. Here we will do
the upper part of the integral using the renormalization group.
This results in the integral being of the same form but multiplied
with the Wilson coefficient $C(\mu)$,
\be
\rlabel{yoperator2}
\Gamma_{\Delta S=2} = - G_F \, C(\mu) \int_0^\mu \frac{d^4 r}{(2\pi)^4}
\int d^4 x_1 \int d^4 x_2 \, e^{-ir\cdot(x_2 - x_1)}
L^{sd}_\mu(x_1) L_{sd}^\mu (x_2)\ .
\ee

This can now be studied using the $1/N_c$ expansion.We can first do this within
a chiral expansion leading to the result
in the chiral limit (see Ref.\rcite{BP3} for details):
\be
\rlabel{ybknlo}
B_K(\mu)_{CHPT} =
\frac{3}{4}\left( 1 -\frac{1}{16\pi^2 f_0^2}
\left[2 \mu^2 + \frac{\dis q^2}{\dis 2}\right] \right)\ .
\ee
The correction is negative. It disagrees somewhat with the result obtained
in \rcite{BBG2}
because there no attempt at identifying the cut-off
across different diagrams was made.
Since we work at leading level in $1/N_c$
in the NLO CHPT corrections we have included the relevant
singlet ($\eta_1$) component as well using nonet symmetry.
The correction in \rref{ybknlo} has precisely
the right behaviour to cancel partly $C(\mu)$ which increases with
increasing $\mu$.

The same calculation can now be performed for the ENJL model. Here the major
complication is the number of different diagrams that has to be evaluated.
An example of one of the classes is shown in figure \tref{yafigure4}.
\begin{figure}[htb]
\hspace{3.5cm}\epsfxsize=12cm\epsfbox{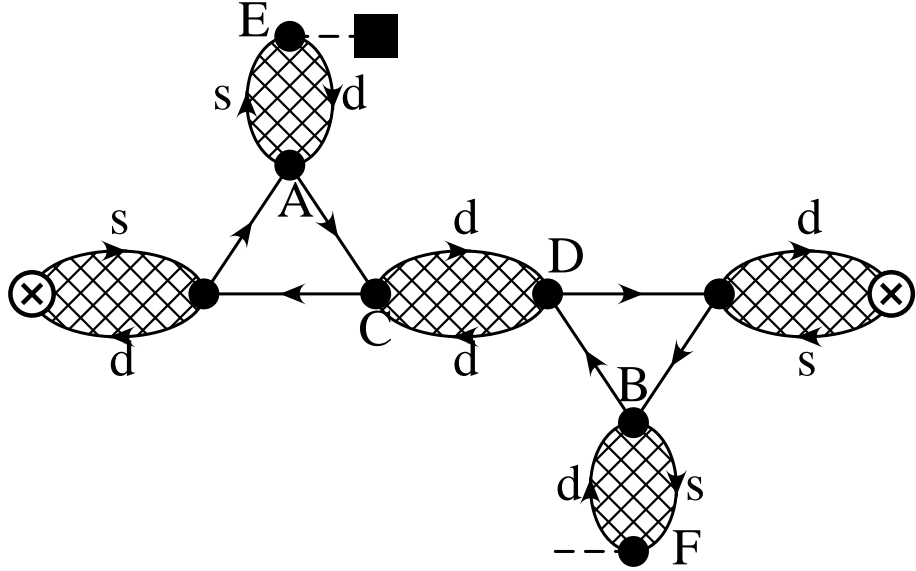}
\caption{A leading $1/N_c$ contribution to
the nonfactorizable part of $\Pids(q^2)$
in the NJL model. The crosshatched areas are the full two-point functions
as discussed in subsection 3.1. Point E and F are connected via
$\Gamma_{\Delta S=2}$.}
\rlabel{yafigure4}
\end{figure}

We now evaluate all contributions numerically to the two-point function
of Eq. \rref{ytwopoint}. The results for several input values are
in table \tref{ytable2}.
\begin{table}[hbt]
\caption{Results for $B_K$ and $\hat B_K$ in the ENJL model.}
\begin{center}
\begin{tabular}{c|cc|ccc|cc}
$\mu$ (GeV) & $B_K^\chi(\mu)$ & $\hat B{}_K^\chi$ & $B_K^m(\mu)$ &
$B_K^a(\mu)$ &
 $\hat B_K^m$ & $B_K^{\rm eq}(\mu)$ & $\hat B_K^{\rm eq}$ \\
\hline
0.3  & 0.68 &0.50 & 0.74&   0.50&0.55  &0.74 &0.55 \\
0.5  & 0.59 &0.59 & 0.71&$-$0.44&0.71  &0.72 &0.72 \\
0.7  & 0.53 &0.58 & 0.69&$-$2   &0.75  &0.68 &0.75 \\
0.9  & 0.48 &0.55 & 0.66&$-$3   &0.76  &0.65 &0.75 \\
1.1  & 0.45 &0.54 & 0.64&$-$4   &0.76  &0.64 &0.76
\end{tabular}
\end{center}
\rlabel{ytable2}
\end{table}
We have studied three cases, namely, the chiral case, $m_d = m_s =0$,
the case with SU(3) symmetry breaking $m_s = 83 ~{\rm MeV} \ne m_d
= 3~{\rm MeV}$  and the case with $m_s = m_d = 43~{\rm MeV}$.
The other parameters are
$G_S = 1.216$, $\Lambda_\chi = 1.16~{\rm GeV}$ and $G_V = 0$.
The latter simplifies the calculation by about an order of magnitude.
Preliminary results for the $G_V\ne 0$ case have the same qualitative
conclusion\rcite{BP4} but typically somewhat lower values of $B_K$ and
less good matching.

The procedure we have followed to analyze the numerical results
is the following.  We fit the ratio between the correction and the
leading $1/N_c$
result for a fixed scale $\mu$ to $a/q^2 + b + c q^2$ which always gives
a very good fit ($a$, $b$ and $c$ are $\mu$ dependent).
 Once we have this fit we can extrapolate our
$B_K$ form factor (remember that we have calculated it for
Euclidean $q^2$) to the physical $B_K$, i.e. to
$q^2=m_K^2\approx 0,0.13~{\rm GeV}^2$ (chiral,other cases).

Let us first treat the chiral or massless quarks case.
Here a nontrivial check on the
results is that the diagrams have a behaviour which sums to $1/q^2$,
i.e. $a$ should be zero.
The individual contributions do not have this behaviour.
$b$
is the relevant contribution to $B_K$ since $m_K^2|_\chi=0$.
The first three
columns in table \tref{ytable2} are $\mu$, $B_K^\chi(\mu)$ and
$\hat B_K^\chi = B_K(\mu)\alpha_S(\mu)^{a_+}$ with
$a_+ = -2/9$ and $\Lambda^{(3)}_{\overline{MS}} = 250$ MeV. The hatted
quantity is the scale independent quantity. Good matching is obtained if
this value is stable within a range of $\mu$.

In the second case ($m_s \ne m_d$),
because the chiral symmetry is broken, there is
a possibility for contributions to $B_K(q^2)$ that are not
proportional to $q^2$, i.e. $a\ne 0$.
In fact a CHPT calculation predicts
precisely the presence of this type of terms\rcite{BP4}. For
small values of $q^2$ the part due to $a$ dominates even though it is only a
small correction when extrapolating to the physical $B_K^m$ at $q^2=m_K^2$.
This can be found in column 4.
The fifth column is the form factor $B_K^a$ for $q^2=-0.001$
 GeV$^2$ where the correction due to the $a$ term is sizeable.
Notice the difference
between these two columns.
This same feature should be visible in the lattice
calculations as soon as they are done with different
quark masses. The invariant $\hat B_K^m$ for this case is in column 6.

In the the last case, i.e. $m_d=m_s$, which is similar to the present
lattice QCD calculations,  the fit gives $a=0$ to a good
precision and  the value of $B_K^{\rm eq}$ extracted is rather
independent of $q^2$. The invariant $\hat B_K^{\rm eq}$ in this case is in
column 8.

In view of the results of \rcite{BBR,NJL2,BP2} we expect to get
a good prediction for the effects of non-zero and different current quark
masses.
We see those and find a significant
change due to both:
\be
\hat B_K^m(m_K^2\approx 0.13~{\rm GeV}^2) \approx 1.35
 \, \hat B_K^\chi(m_K^2= 0)
\ee
for scales $\mu\approx (0.7\sim 1.1) ~{\rm GeV}$.
For the extrapolation to the kaon pole the
difference between the masses has a much smaller effect than the fact that
they were non-zero. In order to compute $B_K$ in the general case a careful
extrapolation to the poles was needed. The final correction to the $B_K$
parameter compared to its leading value of $3/4$ turns out to be rather
small.

\section{Conclusions}

In this Physics Reports an overview of the results and the methods of
references \rcite{BBR,BRZ,Ximo,BP1,BP2,BP3} was given. There have also been
several short versions given in talks and
lectures\rcite{brijuni,dafne,eduardo2,eduardo3}.
The main conclusion is that it is possible to use constituent chiral quark
models and in particular the extended Nambu-Jona-Lasinio models to obtain
results in the hadronic sector. The method underlying most of the
results reviewed here allows for a clean separation of cut-off effects and
consequences of the general structure of the model.
This was especially clear in the case of the low-energy expansion,
section \tref{p4analysis} and the two-point functions, section
\tref{xtwop}. It also allowed for a discussion of the anomaly
in the presence of extra pointlike quark interactions, section \tref{anomaly}.

In the case of 3 and higher point functions the number of free functions
in the general approach rapidly grows but still some general
features are visible by the reduction to the basic one-loop functions.

The results also allowed the derivation of relations that are valid in
a large class of Nambu-Jona-Lasinio like models. In particular the
one with the inclusion of gluonic corrections. In fact the final results
are rather insensitive to the inclusion of these extra effects precisely
because the success of the ENJL model rests to a large extent on the
relations we have derived here. These are also well satisfied by the
hadronic experimental quantities we have tried to explain with this
approach.

Especially the presence of several of the short-distance constraints from
QCD in the ENJL model makes it a prime candidate for trying to
estimate semi-analytically hadronic matrix elements. This has been
done at present for the $\pi^+-\pi^0$ mass difference and the $B_K$ parameter.
Both results have been reviewed here shortly.

In conclusion the simple extended nambu-Jona-Lasinio model provides us
with a compact way to describe low-energy hadronic physics and understand a
large body of experimental results. It also provides a framework to
systematically explore effects due to explicit chiral symmetry breaking
beyond chiral perturbation theory as shown here in the discussion about the
$B_K$ parameter.

\section*{Acknowledgements}

I would like to thank my collaborators Ch. Bruno, J. Prades, E. de Rafael
and H. Zheng for a pleasant collaboration.

\appendix

\def\theequation{\Alph{section}.\arabic{equation}}
\setcounter{equation}{0}

\section{Derivation of the Ward identities}
\rlabel{xAppB}

In this appendix we generalize the proof in the appendix of ref. \rcite{BRZ}
to the case with nonzero current quark masses. There a proof was given
of all relevant identities in terms of the heat kernel expansion
(for an excellent
 recent review and definitions see ref. \rcite{Ball}) and some
of them in terms of the Ward identities as well. Here those which can
be derived directly from the Ward identities can also be derived from the
heat kernel expansion but since they involve different masses they
require a resummation of different terms. For these the direct derivation
of the Ward identities is actually simpler. Only for the additional relations
will we give the heat kernel derivation.

At the one-loop level we use as Lagrangian the one in
eq. \rref{LENJL} (with the same definitions as there)
\be
\rlabel{xlagrang}
{\cal L}_{\rm ENJL} = \overline{q}{\cal D}{q}
\ee
where $\cal D$ contains the couplings to the external
fields $l_\mu,r_\mu,s$
and $p$ as well as the effects of the four-quark terms in
${\cal L}^{\rm S,P}_{\rm NJL}$ and ${\cal L}^{\rm V,A}_{\rm
 NJL}$ on the quark currents at the one-loop level.
In particular it contains the constituent quark masses, $M_i$.
However, we shall keep the notation $l_\mu,r_\mu,s$ and $p$
to denote the quark current sources in the presence of these
four-quark NJL operators.
The one-loop current identities derived from this Lagrangian are
\ba
\rlabel{xcurr}
\partial^\mu V^{ij}_\mu &=& -i\left(M_i - M_j\right) S^{ij}
\nonumber\\
\partial^\mu A^{ij}_\mu &=& \left(M_i + M_j\right) P^{ij}\ .
\ea
When the whole series of constituent quark bubbles are summed
these identities are satisfied changing constituent quark masses
by current quark masses.
In addition we use the equal time commutation relations for fermions
\be
\rlabel{xeqtime}
\left\{ q^{i\dag}_\alpha(x) , q^j_{\beta}(y)\right\}_{x^0 = y^0}
=
i\delta_{\alpha\beta}\delta_{ij}\delta^3({\bf x} - {\bf y})\ .
\ee
Here $\alpha$ and $\beta$ are Dirac indices and $\bf x$ means the spatial
components of $x$.
Multiplying the two-point functions with $i q_\mu$
is equivalent to
taking a derivative of the exponential under the integrals in eqs.
\rref{x9} to \rref{x12}. By partial integration we then get several terms,
those due to the time ordering which leads to equal time commutators
and those where the derivative hits one of the currents. The first type
are evaluated using eq. \rref{xeqtime} and the second type are related to
other two-point functions using eq. \rref{xcurr}. This then leads
to the expressions \rref{xward1} to \rref{xward2}.

The derivation of the other two identities is slightly more complicated.
The effective action of the Lagrangian in eq. \rref{xlagrang} can
be obtained in Euclidean space as a heat kernel
expansion (see ref. \rcite{Ball}).
The coefficients of this expansion are
the so-called Seeley-DeWitt coefficients, they
are constructed out of the two quantities $E$ and $R_{\mu\nu}$.
These are defined as
\ba
\rlabel{xER}
{\cal D}^\dagger {\cal D} &\equiv&
-\nabla_\mu\nabla^\mu + E +\overline{M}^2\ ,
\nonumber\\
R_{\mu\nu} &\equiv& \left[\nabla_\mu ,\nabla_\nu\right]\ ,
\nonumber \\
\nabla_\mu \# &\equiv& \partial_\mu \# -i [v_\mu,\#]
-i [a_\mu \gamma_5 ,\#]\, .
\ea
If in eq. \rref{xlagrang} the Dirac operator $\cal D$ contains couplings to
gluons these should not be taken into account in eq. \rref{xER}. The
relevant heat kernel expansion in that case will have different coefficients
depending on vacuum expectation values of gluonic operators, but will
still be constructed out of the quantities in eq. \rref{xER}
(depending now also on the gluon field). The quantity
$\overline{M}$ is the mass that is used in the heat kernel
expansion. The operator $\cal D$ is
\be
i\gamma^\mu(\partial_\mu -i v_\mu -i a_\mu\gamma_5)
-{\cal M}-s+ip\gamma_5\ .
\ee
Here ${\cal M}={\rm diag}(m_u,m_d,m_s)$ is the current quark mass
matrix and we allow for spontaneous chiral symmetry breaking
solution $\langle 0 | s(x) | 0 \rangle \neq 0$.
For the terms relevant to two-point functions we
have
\ba \rlabel{xE}
R_{\mu\nu} &=&-i(v_{\mu\nu}+a_{\mu\nu} \gamma_5) \, , \nonumber
\\ {\rm and} \hspace*{2cm} && \nonumber \\
E &=& -\frac{i}{2}\sigma^{\mu\nu}R_{\mu\nu}
+i\gamma^{\mu} {\rm d}_\mu \left(M + s + i p\gamma_5 \right)
\nonumber \\ &-& \gamma^\mu
\left\{a_\mu \gamma_5,M+s-ip\gamma_5 \right\} +
\left\{M,s\right\}-i\left[M,p\right]\gamma_5
+s^2+p^2
\nonumber\\&+&
 M^2 -\overline{M}^2\ , \nonumber \\
{\rm with} && \nonumber \\
v^{\mu\nu} &\equiv& \partial^\mu v^\nu - \partial^\nu v^\mu
-i [v^\mu,v^\nu] \, , \nonumber \\
a^{\mu\nu} &\equiv& \partial^\mu a^\nu - \partial^\nu a^\mu
-i [a^\mu,a^\nu] \, , \nonumber \\
{\rm d}^\mu \# &\equiv& \partial^\mu \# - i [v^\mu,\#] \, .
\ea
The main difference with ref. \rcite{BRZ} is the occurrence
of the last line in the expression for $E$ in
\rref{xE}. We shall call this last line $E_0$.
In this equation, $M \equiv
{\rm diag}(M_u,M_d,M_s)$,
the diagonal matrix of the constituent quark masses
defined in eq. \rref{gap}. Notice that the scalar field here has
been shifted and we have now $\langle 0 | s(x) | 0 \rangle =0$
(though we use the same notation for it).
When $G_S \to 0$ in eq. \rref{gap}
then ${\overline M} \to 0$ and $M \to {\cal M}$.
 Let us now
systematically go through
all possible types of terms in the expansion. We shall not discuss the mixed
two-point functions here since we only want to prove eqs.
\rref{xward5}-\rref{xward6}.

In the heat kernel expansion, those terms
containing two factors  $R_{\mu\nu}$ only contribute to the
transverse parts, $\ovpi^{(1)}_{V,A}$ and in the same way. Their contributions
hence obviously satisfy eqs. \rref{xward5}-\rref{xward6}. Similarly,
one factor $R_{\mu\nu}$ requires the presence of two covariant
derivatives $\nabla_\mu$.
By commuting derivatives (the extra terms only contribute
to three and higher point functions)
and partial integration these can be brought next to each other so they convert
into a second factor  $R_{\mu\nu}$. This brings us back to the
previous case. Intervening $E$'s can only contribute via $E_0$ but these
do not spoil the above argument.
The first term in $E$, namely $\sigma_{\mu\nu}R^{\mu\nu}$,
requires a 2nd $\sigma_{\mu\nu}
R^{\mu\nu}$ because otherwise the trace over Dirac indices vanishes.
These also behave like terms with two factors $R^{\mu\nu}$. Therefore,
in the remainder we are only concerned with $E$ without this first term.

$E$ can also directly contribute to the scalar and pseudoscalar two-point
function in the same way via $s^2 + p^2$. Extra factors $E$ become again
$E_0$ and extra derivatives also respect the relation \rref{xward6}.
The most complicated case is where both fields come from a different $E$.
This contributes in the form $E_0^n E E_0^m \partial^{2i} E$.
These contribute to all form factors in the form $M_i^n M_j^m q^{2i}$
times the coefficients listed in Table \tref{xtable1}.
\begin{table}[htb]
\begin{center}
\begin{tabular}{|c|c|}
\hline
Function & Contribution \\
\hline
$\ovpi_S $&$ -q^2+(M_i+M_j)^2 $\\
$\ovpi_P $&$ -q^2+(M_i-M_j)^2 $\\
$\ovpi^{(0)}_A $&$ (M_i+M_j)^2/q^2$\\
$\ovpi^{(1)}_A $&$ -(M_i+M_j)^2/q^2$\\
$\ovpi^{(0)}_V $&$ (M_i-M_j)^2/q^2$\\
$\ovpi^{(1)}_V $&$ -(M_i-M_j)^2/q^2$\\
\hline
\end{tabular}
\end{center}
\caption{The contribution of terms of the type $E^{m+n+2}$ to the two-point
functions.\rlabel{xtable1}}
\end{table}
These coefficients obviously satisfy the relations
\rref{xward5}-\rref{xward6}.
The last type of terms is where  the external fields come out of a derivative.
We do not consider the mixed case here, so both the fields have to come
out of a derivative due to the $\gamma_\mu$ that is necessarily present
in the $E$ that would be a candidate for the external field.
So there are those where the external fields are contained in
two factors $\nabla_\mu$.
If the indices of these are different, then there need to be
at least two extra derivatives present that will produce a $q_\mu q_\nu$.
This contributes equally to $\ovpi^{(0)}_V$ and $\ovpi^{(0)}_A$.
If the indices are equal, it will contribute proportional to $g_{\mu\nu}$
and thus to the vector and axial-vector equally with
$\ovpi^{(0+1)}_{V,A} = 0$. This completes the proof of the identities
\rref{xward5}-\rref{xward6}.

Now it remains to prove that these contributions will never produce a
pole in $\ovpi^{(0+1)}$ at $q^2=0$.
Terms that contain two factors $R_{\mu\nu}$ contain
two factors of momenta and hence do not. Terms with
one factor $R_{\mu\nu}$ can be brought in the form with two so do not
produce a pole either. From Table \tref{xtable1} there
is no contribution from that
type of terms to $\ovpi^{(0+1)}_{V,A}$. Then those with
external fields from $\nabla_\mu$
with different derivatives necessarily contain
extra factors  $q_\mu q_\nu$ so do not contribute to a possible
 pole at $q^2=0$ and the last type
of terms does not contribute to $\ovpi^{(0+1)}_{V,A}$ as shown above.
This completes the proof.

\section{Explicit expressions for the barred
two-point functions}
\rlabel{xAppC}
\setcounter{equation}{0}

Here we shall give the one-constituent-quark-loop
expression for the
two-point functions defined in eqs. \rref{x9}-\rref{x14}
in the presence of current quark masses. These two-point
functions are denoted in the text as the $\ovpi$ ones.
They fulfil the same Ward identities as the full-ones in eqs.
\rref{xward1}-\rref{xward4} changing the current quark masses
there by the constituent quark ones. In addition, they also
satisfy the Ward identities in eqs. \rref{xward5}-\rref{xward6}.
Using these identities one can
see that there are only two independent functions
out of $\ovpi^{(1)}_V$, $\ovpi^{(0)}_V$, $\ovpi^{(1)}_A$,
$\ovpi^{(0)}_A$, $\ovpi_S^M$ $\ovpi_P^M$, $\ovpi_S$
and $\ovpi_P$. We shall take $\ovpi_P^M$ and
$\ovpi^{(1)}_V + \ovpi^{(0)}_V$ as these functions.
The explicit expressions are

\ba
\rlabel{xpiv01}
\left(\ovpi^{(1)}_V + \ovpi ^{(0)}_V \right)
(Q^2)_{ij} &=& \frac{\dis N_c}{\dis 16 \pi^2}\,
8 \, {\dis \int^1_0} {\rm d}x \,x(1-x) \Gamma(0,x_{ij})\, , \\
\rlabel{xpimp}
{\overline \Pi}_P^M (Q^2)_{ij} &=& \frac{\dis N_c}{\dis 16 \pi^2}\,
4 {\dis \int^1_0} {\rm d}x (M_i x + M_j (1-x))
\Gamma(0,x_{ij}) \, ,
\ea
where
\ba
x_{ij} &\equiv& \frac{\dis M_i^2 x + M_j^2 (1-x) + Q^2 x(1-x)}
{\dis \Lambda_\chi^2} \,  .
\ea

One can obtain all the others one-loop two-point functions
in function of these two by
using the Ward identities mentioned above.
For instance, for the $\ovpi^{(0)}_V$ one gets

\be
\renewcommand{\arraystretch}{1.5}
\begin{array}{l} \dis
\rlabel{xpiv0}
{\overline \Pi}^{(0)}_V (Q^2)_{ij}
= - \frac{\dis \left(M_i - M_j\right)^2}{\dis M_i + M_j} \,
\frac{\dis \ovpi^M_P (Q^2)_{ij}}{\dis Q^2 (Q^2 + (M_i - M_j)^2)}
\nonumber \\ \hspace*{0.5cm} \times
\left\{ \left(M_i+M_j\right)^2 + g_A(Q^2)_{ij} m_{ij}^2
(Q^2) \left(1 - \left(\frac{\dis m_i - m_j}{\dis m_i + m_j}
\right) \left(\frac{\dis M_i + M_j}{\dis M_i - M_j} \right)
\right) + Q^2 \right\}  \, . \nonumber \\
\end{array}
\renewcommand{\arraystretch}{1}
\ee

\setcounter{equation}{0}
\section{Explicit expression for the one-loop form
factor $\ovpi^+_\mu(p_1,p_2)$}
\rlabel{xAppD}

Here we shall give the one-constituent-quark-loop
expression for the
three-point function $\ovpi^+_\mu (p_1,p_2)$
 defined in eq. \rref{pi+}. We shall give it for $M_i=M_k=M_m$.
The explicit expression is (remember that we have $j=m$),

\ba
\rlabel{xpivpp}
\ovpi^{+\mu}(p_1,p_2) &=& \nonumber \\
&-& \frac{\dis 1}{\dis 2 M_i} \Bigg\{
\ovpi_P^M(-p_1^2)_{ii} \, \nonumber \\
&+&  \frac{\dis p_1\cdot p_2}{\dis p_1^2 p_2^2 -
(p_1 \cdot p_2)^2} \, \left[ p_2^2
\left(\ovpi^M_P(-p_2^2)_{ii} -
\ovpi_P^M(-q^2)_{ii}\right) \right. \nonumber \\
&+& \left.
(p_1 \cdot p_2) \left( \ovpi_P^M(-p_1^2)_{ii}
-  \ovpi_P^M(-q^2)_{ii} \right) \right]  \nonumber \\
&+& \frac{\dis 2}{\dis M_i} I_3(p_1^2,p_2^2,q^2)
p_2^2 \left[ 1 + (p_1 \cdot p_2)
\frac{\dis p_1^2 + (p_1 \cdot p_2)}{\dis p_1^2 p_2^2 -
(p_1 \cdot p_2 )^2} \right]  \Bigg\} p_1^\mu \nonumber \\
&+& (p_1 \leftrightarrow - p_2) \, . \nonumber \\
\ea
Where the two-point function $\ovpi^M_P(-p^2)$ was given in
appendix \ref{xAppC} and the function $I_3(p_1^2,p_2^2,q^2)$ is

\ba
I_3(p_1^2,p_2^2,q^2) &=& \frac{\dis N_c}{\dis 16 \pi^2}
2  M_i^2 {\dis \int^1_0} {\rm d}x x {\dis \int^1_0} {\rm d}y
\frac{\dis \Gamma(1, M^2(x,y)/\Lambda_\chi^2)}{\dis
M^2(x,y)} \,
\ea
with
\ba
M^2(x,y) &\equiv& \nonumber \\
&&M_i^2 -p_1^2 (1-x) - p_2^2 x(1-y)
+ (p_1 (1-x) - p_2 x(1-y))^2 \,  .
\nonumber \\
\ea

\listoftables
\listoffigures
\end{document}